\input psfig
\input harvmac
\newcount\figno
\figno=0
\def\fig#1#2#3{
\par\begingroup\parindent=0pt\leftskip=1cm\rightskip=1cm\parindent=0pt
\global\advance\figno by 1
\midinsert
\epsfxsize=#3
\centerline{\epsfbox{#2}}
\vskip 12pt
{\bf Fig. \the\figno:} #1\par
\endinsert\endgroup\par
}
\def\figlabel#1{\xdef#1{\the\figno}}
\def\encadremath#1{\vbox{\hrule\hbox{\vrule\kern8pt\vbox{\kern8pt
\hbox{$\displaystyle #1$}\kern8pt}
\kern8pt\vrule}\hrule}}
	
\def\WP{{\bf WP}}
\def\C{{\bf C}}
	
\def\N{{\cal N}}
 \overfullrule=0pt

 %
 \def\Lambda{\wedge}
 \def\co{{\cal O}}
 \def\tilde{\widetilde}
 \def\bar{\overline}
 \def\Z{{\bf Z}}
 \def\N{{\cal N}}
 \def\T{{\bf T}}
 \def\S{{\bf S}}
 \def\R{{\bf R}}

 \font\zfont = cmss10 
 \font\litfont = cmr6
 
 \def\bigone{\hbox{1\kern -.23em {\rm l}}}
 \def\ZZ{\hbox{\zfont Z\kern-.4emZ}}
 \def\half{{\litfont {1 \over 2}}}

 \Title{hep-th/9701162, IASSNS-HEP-97-3}
 {\vbox{\centerline{VECTOR BUNDLES AND $F$ THEORY}}}
 
 \smallskip
 \centerline{Robert Friedman\foot{Research supported in part by NSF Grant DMS-96-22681. }
 and John Morgan\foot{Research supported in
part by NSF Grant DMS-94-02988.}}
 \smallskip
 \centerline{\it Department of Mathematics, Columbia University, New York,
NY 10027 }
 \smallskip\centerline{Edward Witten\foot{Research supported in part
by NSF Grant PHY-95-13835.}}
\smallskip
\centerline{\it School of Natural Sciences, Institute For Advanced
Study, Princeton, NJ 08540}
\bigskip

 \medskip

 \noindent
To understand in detail duality between heterotic string and
$F$ theory compactifications, it is important to understand the
construction of holomorphic $G$ bundles over elliptic Calabi-Yau manifolds,
for various groups $G$.  In this paper, we develop techniques to describe
these bundles, and make several detailed comparisons between the
heterotic string and $F$ theory.
\Date{January, 1997}

\newsec{Introduction}

 \nref\vafa{C. Vafa, ``Evidence For $F$-Theory,'' hep-th/9602065, Nucl. Phys.
 {\bf B469} (1996) 403.}
 
 \nref\sen{A. Sen, ``$F$-Theory And Orientifolds,'' hep-th/9605150, Nucl. Phys.
 {\bf B475} (1996) 562.}
 
 \nref\mv{D. Morrison and C. Vafa, ``Compactifications Of $F$-Theory On 
Calabi-Yau
 Threefolds, I, II,'' hep-th/9602114,9603161, Nucl. Phys. {\bf B473} (1996) 
74, 
 {\bf B476} (1996) 437. }

 \nref\witten{E. Witten, ``Phase Transitions In $M$-Theory And $F$-Theory,''
 hep-th/9603150, Nucl. Phys. {\bf B471} (1996) 195.}

 One of the important recent insights 
 about string duality is that the
 compactification of the heterotic string on $\T^2$ is equivalent
 to the compactification of $F$ theory on an elliptically
 fibered K3 with a section \refs{\vafa,\sen}.
 Extending this idea, one then expects that the heterotic
 string compactified on an 
 $n$-fold $Z$ which is elliptically fibered over a base $B$ 
 should be equivalent to $F$ theory compactified
 on an $n+1$-fold $X$ which is  fibered with K3 fibers over
 the same base.  This should follow upon fiberwise application
 of the basic heterotic string/$F$ theory duality on the fibers.

 \def\P{\bf P}
 \def\K3{{\rm K3}}\def\K3{{\rm K3}}
 \nref\nonp{E. Witten, ``Nonperturbative Superpotentials In String Theory,''
 hep-th/9604030, Nucl. Phys. {\bf B474} (1996) 343.} 
 \nref\eeight{R. Donagi, A. Grassi, and E. Witten, ``A Non-Perturbative
 Superpotential With $E_8$ Symmetry,'' hep-th/9607091, Mod. Phys. Lett. {\bf A11} (1996)
 2199.}
 \nref\whos{I. Brunner and R. Schimmrigk, ``$F$-Theory On
 Calabi-Yau Four-Folds,'' hepth/9606148, Phys. Lett. {\bf B387} (1996) 750.}
 \nref\svw{S. Sethi, C. Vafa, and E. Witten, ``Constraints On Low-Dimensional
 String Compactifications,'' Nucl. Phys. {\bf B480} (1996) 213.}
\nref\beckers{K. Becker and M. Becker, ``$M$-Theory On
Eight-Manifolds,'' Nucl. Phys. {\bf B477} (1996) 155.}
 \nref\bianchi{M. Bianchi, S. Ferrara, G. Pradisi, A. Sagnotti, and Ya. S. Stanev,
 ``Twelve-Dimensional Aspects Of Four-Dimensional $N=1$ Type I Vacua,'' Phys. Lett.
 {\bf B387} (1996) 64.}
 \nref\kach{S. Kachru and E. Silverstein, ``Singularities, Gauge Dynamics, and
 Nonperturbative Superpotentials In String Theory,'' hep-th/9608194.}
 \nref\mayr{P. Mayr, ``Mirror Symmetry, $N=1$ Superpotentials, and Tensionless Strings
 On Calabi-Yau Fourfolds,'' hep-th/9610195.}
 \nref\brunn{I. Brunner, M. Lynker, and R. Schimmrigk, ``Unification Of
 $M$- and $F$-Theory Calabi-Yau Fourfold Vacua,'' hep-th/9610195.}
 \nref\kv{S. Katz and C. Vafa, ``Geometric Engineering Of $N=1$ Quantum Field
 Theories,'' hep-th/9611090.}
 \nref\bjp{M. Bershadsky, A. Johansen, T. Pantev, V. Sadov, and C. Vafa,
 ``Geometric Engineering And $N=1$ Dualities,'' hep-th/9612052.}
 \nref\vz{C. Vafa and B. Zwiebach, ``$N=1$ Dualities Of $SO$ and $Sp$ Gauge Theories
 and $T$-Duality Of String Vacua,'' hep-th/9701015.}
 \nref\yau{A. Klemm, B. Lian, S.-S. Roan, and S.-T. Yau, ``Calabi-Yau Fourfolds
 For $M$- and $F$-Theory Compactification,'' hep-th/9701023.}
 \nref\mohri{K. Mohri, ``$F$ Theory Vacua In Four Dimensions And Toric
 Threefolds,'' hep-th/9701147.}
 \nref\ob{M. Bershadsky, A. Johansen, T. Pantev, and V. Sadov,
 ``On Four-Dimensional Compactifications Of $F $-Theory,'' to appear.}
 The first non-trivial case of this fiberwise duality
 is  $n=2$ -- which means in practice that
 $B=\P^1$, $Z=\K3$, and $X$ is a Calabi-Yau three-fold.
 In this case, this duality has been successfully used 
 \mv\ to illuminate many aspects of heterotic
 string dynamics on K3,  including \witten\ aspects of the strong
 coupling singularity.  A successful extension to $n=3$ would be very
 interesting physically and would raise many new issues such as 
the possibility of
a spacetime superpotential.  Several aspects have been discussed
so far \refs{\nonp - \ob}.

 To understand in detail $F$ theory/heterotic duality, for any value
 of $n$, involves understanding and comparing the moduli spaces on the
 two sides.  On the $F$ theory side, the moduli spaces involved
 have been comparatively well understood \nref\many{M. Bershadsky, K. Intriligator,
 S. Kachru, D. R. Morrison, V. Sadov, and C. Vafa, ``Geometric Singularities And
 Enhanced Gauge Symmetries,'' hep-th/9605200, Nucl. Phys. {\bf B481} (1996) 215.}
 \nref\aspin{P. Aspinwall and M. Gross, ``The $SO(32)$ Heterotic String
 On a K3 Surface,'' hep-th/9605131.} \refs{\many,\aspin}, but on the heterotic
 string side there is a major gap.  In compactification of the
 heterotic string on a two-torus or on an elliptically fibered manifold
 of $n>1$, a major ingredient is the choice of a suitable
 $E_8\times E_8$ (or $Spin(32)/\Z_2$) stable holomorphic bundle.
 Only limited information about the relevant bundles has been
 brought to bear so far.

 There is, however, an effective
 framework for understanding stable bundles on elliptically fibered
 manifolds \nref\friedman{R. Friedman, ``Rank Two Vector Bundles Over Regular Elliptic
 Surfaces,'' Invent. Math. {\bf 96} (1989) 283.} 
 \nref\fm{R. Friedman and J. Morgan, ``Smooth Four-Manifolds
 And Complex Surfaces'' (Springer-Verlag, 1994),  Chapter 7.1.} 
 \refs{\friedman,\fm}.  In this approach, which has been
 developed in detail for $SU(2)$  bundles on elliptically fibered
 surfaces (for the purpose of applications to Donaldson theory),
 one describes bundles on an elliptically fibered manifold by
 first describing the bundles on a particular elliptic curve, and
 then working fiberwise.  This approach is not limited to Calabi-Yau manifolds.  Most
 of the present paper is devoted to describing this approach mathematically.  In the
 last part of the paper, we specialize to Calabi-Yau manifolds and make some applications
 to $F$ theory.

\bigskip\noindent
{\it Some Generalities About Bundles}

Before  focussing on our specific problem, we make some general remarks about
bundles (in somewhat more detail than really needed to follow the rest of the
paper).  The bundles of interest, whether over a single elliptic
curve or an elliptically fibered manifold, can be
viewed in either of two ways: (1) as
holomorphic stable bundles (or semistable ones as explained below) 
with structure group the
complexification $G_\C$ of a compact Lie group $G$; (2)  as solutions of
the hermitian-Yang-Mills equations for a $G$-valued connection.\foot{These 
equations say that the $(2,0)$ and $(0,2)$ part of the curvature
vanish, and the $(1,1)$ part is traceless.}  The second
point of view arises most directly in physics; the first point of view
is convenient for analyzing the bundles.  The equivalence of the two viewpoints
is a theorem of Narasimhan and Seshadri 
\ref\nar{M. S. Narasimhan and C. S. Seshadri, ``Deformations Of The Moduli Space
Of Vector Bundles Over An Algebraic Curve,'' Ann. of Math. (2) {\bf 82} (1965) 540.}
\nref\ramanathan{A. Ramanathan, ``Stable Principal Bundles On A Compact Riemann
Surface,'' Math. Annalen 213 (1975) 129.}
\nref\donaldson{S. Donaldson, ``Antiselfdual Yang-Mills Connections
On Complex Algebraic Surfaces And Stable Bundles,'' Proc. London Math. Soc.
{\bf 3} (1985) 1. }
\nref\abott{M. F. Atiyah and R. Bott, ``The Yang-Mills Equations Over Riemann
Surfaces,'' Philos. Trans. Roy. Soc. London Ser. A {\bf 308} (1982) 523.}
for vector  bundles on a Riemann surface,  generalized for arbitrary semi-simple
gauge groups in \refs{\ramanathan - \abott}, and
of Donaldson \donaldson, and Uhlenbeck and Yau \ref\uhlenbeck{K.  Uhlenbeck
and S.-T. Yau, ``On The Existence Of Hermitian Yang-Mills Connections On
Stable Bundles Over Compact K\"ahler Manifolds,'' Comm. Pure App. Math.
{\bf 39} (1986) 257, {\bf 42} (1986) 703.}, in higher dimensions.  

Over a Riemann surface, 
the hermitian-Yang-Mills 
equations for a connection simply say that the connection
is flat, so the Narasimhan-Seshadri theorem identifies the moduli space of
semistable holomorphic $G_{\bf C}$ bundles on a Riemann surface
with the moduli space of flat $G$-valued connections.  
The moduli space of such flat connections has an elementary, explicit
description: a flat connection on the two-torus is given by a pair of commuting
elements in the gauge group $G$.  Two such connections are equivalent if and only
if they are isomorphic, which is the  same thing as the commuting pairs being
conjugate in $G$.  

The description of the same moduli space via
semi-stable holomorphic $G_{\bf C}$ bundles is more subtle in several ways.
First of all, the equivalence relation between semistable bundles that is used
to build the moduli space, called $S$-equivalence, is in general weaker
than   isomorphism.   (For example, ${\cal O}\oplus {\cal O}$ and  the
non-trivial extension of ${\cal O}$ by ${\cal O}$ are $S$-equivalent.
But for the generic semi-stable $G_{\bf C}$ bundle on a torus, $S$-equivalence
is the same as isomorphism.)  
The Narasimhan-Seshadri theorem tells us that every $S$-equivalence class
contains (up to isomorphism) a unique representative that admits a flat
connection.  This preferred representative is not always the one that
arises on the fibers of an elliptic fibration.  In fact, every $S$-equivalence
class has another distinguished representative, a ``regular'' bundle whose
automorphism group has dimension equal to the rank of $G$.  It is the regular representatives
that fit together most naturally in families, as was shown for rank two bundles
over surfaces in \refs{\friedman,\fm}.

When we refer somewhat loosely to a ``$G$ bundle,'' the context should hopefully make clear
whether a given argument is best understood 
in terms of solutions of the hermitian-Yang-Mills equations with a compact gauge
group $G$, or  holomorphic stable (or semistable) $G_\C$ bundles.
Note that in the important case $G=SU(n)$ the complexification $SU(n)_\C$
is customarily called $SL(n,\C)$; the complexifications have no special
names in the other cases.  Hopefully, it will anyway cause no confusion
if we refer loosely to $G$ bundles even for $G=SU(n)$.

 \nref\borel{A. Borel,
``Sous-groupes Commutatifs et Torsion des Groupes de Lie Compacts
Connexes,'' Tohoku Journal Ser. 2 (1961) {\bf 13} 216-240.}
Finally, let us explain the meaning of the term ``semistable'' as opposed to
``stable.''  A stable bundle corresponds to a solution of the 
hermitian-Yang-Mills
equations which is irreducible (the holonomy commutes only with the center of
the gauge group), while a semistable bundle is associated with
a reducible solution of those equations.
In many situations, the generic semistable bundle is actually stable,
but the case of an elliptic curve $E$ is special; as its fundamental group is 
abelian, the flat connections over $E$ have holonomy 
that \borel\ can be conjugated into a maximal torus (if the gauge group is
simply connected and semi-simple)
and so are reducible, and correspond to semistable rather than
stable bundles.  The bundles we will construct on an elliptically fibered
manifold $Z$ of dimension $>1$ are, however, generically stable, if the
K\"ahler class of $Z$ is chosen suitably.  (A sufficient requirement is, as in \fm,
that the fiber is sufficiently small compared to the base, justifying an
adiabatic argument by which stability is proved.) 

\bigskip\noindent
{\it Bundles On An Elliptic Curve}

Now we turn to our specific problem.  In studying semistable 
bundles  on an elliptic curve with general structure group, 
an important role is played
by a  theorem
of Looijenga \ref\looijenga{E. Looijenga, ``Root Systems And
Elliptic Curves,'' Invent.  Math.  {\bf 38} (1977) 17,
``Invariant Theory For Generalized Root Sytems,'' Invent. Math.
{\bf 61} (1980) 1.}
(another proof was given by Bernshtein and Shvartsman \ref\bernstein{I. N. Bernshtein
and O. V. Shvartsman, Func. An. Appl. {\bf 12} (1978) 308.}) which determines the
moduli space ${\cal M}$ of
$G$ bundles on an elliptic curve
$E$ for any simple, connected, and simply-connected group $G$ of
rank $r$.\foot{There is also  a generalization for non-simply-connected
$G$ which can be obtained via the method of section 5 and will be presented
elsewhere.}
  ${\cal M}$ is always a weighted projective space ${\bf WP}^r_{s_0,
s_1,\dots,s_r}$, where the weights $s_0,\dots, s_r$ are 1 and the coefficients
of the highest coroot of $G$.  (In other words, the weights are
the coefficients of the null vector of the {\it dual} of the
untwisted Kac-Moody algebra of $G$.  We will sometimes suppress
the weights from the notation and write just $\WP^r$.)  
The requisite weights, for the various simple groups, are summarized
in figure one.
 
\midinsert
\centerline{\psfig{figure=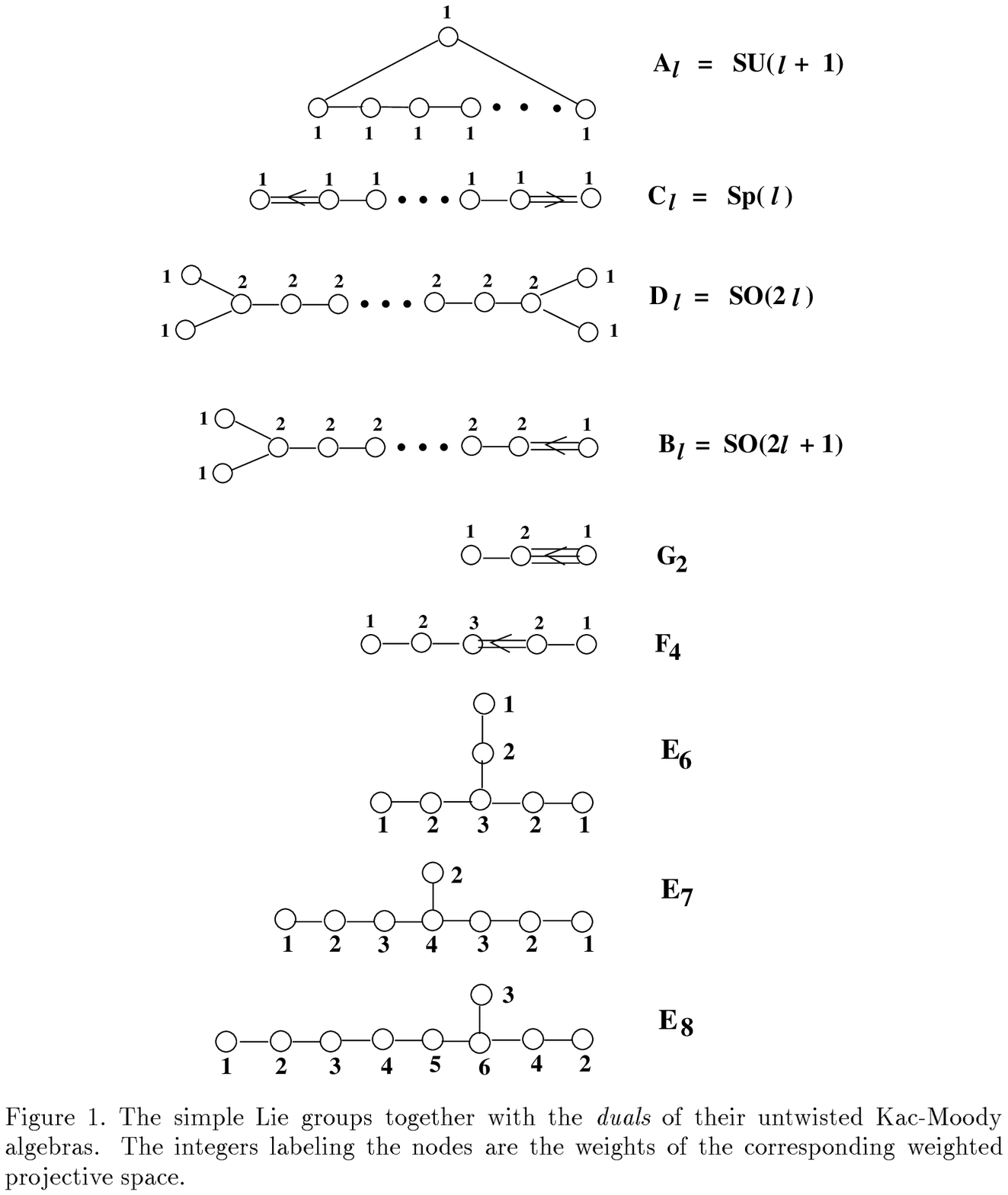,width=5.5in}}
\bigskip
\endinsert

In this paper, we will develop four approaches to understanding
Looijenga's theorem, for different classes of $G$.  

(1) For $G=SU(n)$ or $G=Sp(n)$, the moduli space can be determined
by a completely direct computation that we present in section 2.
  $SU(n)$ and $Sp(n)$
(or $A_{n-1}$ and $C_n$) are the unique cases in which the weights
of the weighted projective space are all 1, so that the moduli
space is actually an ordinary projective space.  In these cases,
a direct treatment along the general lines of \fm\ is possible. 

(2) Every not necessarily simply-laced group $G$ has a canonical
simply-laced subgroup $G'$, generated by the long roots of $G$.
Looijenga's theorem for $G$ is a consequence of Looijenga's theorem
for $G'$, as we will show in section 3.  We also explain another
reduction to the simply-laced case by embedding $G$ in a suitable
simply-laced group.

\nref\othervafa{M. R. Douglas, S. Katz, and C. Vafa, ``Small Instantons, Del Pezzo
Surfaces, And Type $I'$ Superstrings,'' hep-th/9609071.}
\nref\seiberg{D. Morrison and N. Seiberg, ``Extremal Transitions And Five-Dimensional
Supersymmetric Field Theories,'' hep-th/9609070.}
\nref\ganoretal{O. J. Ganor, D. Morrison, and N. Seiberg, ``Branes, Calabi-Yau
Spaces, and Toroidal Compactification Of The $N=1$ Six-Dimensional $E_8$
Theory,'' hep-th/9610251.}

(3) For $E_6,$ $E_7$, $E_8$, and certain subgroups,
 Looijenga's theorem can be proved by relating $G$ bundles
to  del Pezzo surfaces.  This approach, which we will
explore in section 4, is
perhaps closest to Looijenga's original approach. For additional
background see \ref\pinkham{H. C. Pinkham, ``Simple Elliptic Singularities,
Del Pezzo Surfaces, And Cremona Transformations,'' in {\it Several
Complex Variables} (American Mathematical Society) 
Proc. Symp. Pure Math. {\bf 30} (1977) 69.  }.
This method gives an attractive
way to see the relation between groups and singularities (in this case, between
subgroups of $G$ and singularities of the del Pezzo surface) that has
been important in the last few years in studies of string duality.  
The chain of groups related to del Pezzo surfaces is important in applications
of $F$ theory \refs{\othervafa - \ganoretal}.

\nref\fmw{J. Morgan, R. Friedman, and E. Witten, to appear.}
(4) Finally,  we explain in section 5 our most general and powerful
approach.  For any $G$, Looijenga's
theorem can be proved by constructing a distinguished unstable $G$
bundle
on $E$, which has the beautiful property that it can be deformed
in a canonical way to any semistable $G$ bundle.  (This construction always produces
the regular representative of every $S$-equivalence class \fmw.) 

Each of these approaches is most efficient for understanding some
aspects of $F$ theory.  For instance, the first approach, as well as being the
most elementary, gives (at the present level of our understanding) the
most complete information for $SU(n)$ bundles, which enter in most
attempts at using the heterotic string to make models of particle physics.
The last approach is
(at the present level of understanding) the method that enables us
to concretely construct the $E_8$ bundles that are relevant to the
easiest applications of $F$ theory.

\def\W{{\cal W}}
For our applications, we want to understand $G$ bundles not just
on a single elliptic curve $E$, but on a complex manifold $Z$ that is
elliptically fibered over a base $B$.  The basic idea here is to understand
Looijenga's fiberwise fiberwise.  The fiber of $Z$ over a point
$b\in B$ is an elliptic curve $E_b$ (perhaps singular). The moduli
space of $G$ bundles on $E_b$ is a weighted projective space ${\WP}_b$.
The ${\bf WP}_b$ fit together, as $b$ varies, to a bundle $\W$ 
of weighted projective spaces.  Any  $G$ bundle over $E$ that is sufficiently
generic on each fiber determines a section of $\W$, and in many situations
the bundles associated with a given section can be effectively described.

\def\U{{\Omega}}
One of our main goals will therefore be to obtain a description of $\W$.
We will focus on the case that the elliptic manifold $Z\to B$  has a section,
whose normal bundle we call ${\cal L}^{-1}$.  (This is the case
that arises in the simplest applications of $F$ theory.) 
We will see that for every case except $G=E_8$, $\W$ can be described
very simply as the projectivization of a rank $r+1$ vector bundle $\U$ over $B$
which is simply a sum of line bundles.  In fact,  
\eqn\ubbu{\U={\cal O}\oplus
\left(\oplus_{j=1}^{r}
{\cal L}^{-d_j}\right) ,} 
where the  $d_j$ are the degrees of the independent 
Casimir invariants of $G$.
  (This assertion is closely related to
a result of Wirthmuller \ref\wirth{K. Wirthmuller, ``Root Systems And
Jacobi Forms,'' Compositio Math. {\bf 82} (1992) 293, Theorem 3.6.} 
who in particular
discovered the exceptional status of $E_8$.)  In dividing the fibers of
$\U$ by $\C^*$
to make the weighted projective space bundle $\W$, $\C^*$ acts diagonally
on the ${\cal L}^{-d_j}$ with weights $s_j$ introduced above. The matching
of $d_j$ and $s_j$ is described in the table.  This determination of $\W$ 
will serve as the basis in section 6
for an extensive comparison of the moduli
space of $G$ bundles on $Z$ to  appropriate $F$ theory moduli spaces,
in the course of which it will be clear from the $F$ theory point of
view why $E_8$ should be exceptional.

In the decomposition \ubbu, the summand ${\cal O}$ plays a distinguished
role.  The section of ${\cal W}$ coming from the constant section 1 of ${\cal O}$
corresponds to a bundle on $Z$ whose restriction to each fiber is
$S$-equivalent to the trivial $G$ bundle.  The most elementary way to see why Casimir
weights appear is actually to look at the behavior near this section.

Of our four approaches, methods (1) and (4) actually enable us to
construct $G$ bundles over an elliptically fibered manifold $Z$
and not merely to determine  the moduli spaces. When the bundles
can be constructed, one has a starting point for addressing more
detailed question like the computation of Yukawa couplings.  Most such
questions will not be considered in this paper.  However, in section 7,
we    make one important application of the construction of bundles,
which is to compute the basic characteristic class of these bundles
(this is a four-dimensional class which for $G=SU(n)$ is the conventional
second Chern class).  This computation leads to an important comparison
between the heterotic string and $F$ theory; for the case of compactification
of the heterotic string on a Calabi-Yau threefold, we will understand
from the heterotic string point of view the origin of the 
threebranes that appear mysteriously on the $F$ theory side \svw.

$$
\vbox{\tabskip=0pt\offinterlineskip
\def\tablerule{\noalign{\hrule}}
\halign{\strut\tabskip=.8em plus.8em
minus.8em\quad\hfil#\hfil&\vrule#&\hfil#\hfil&\vrule#&\hfil#\hfil&\vrule#&\hfil
#
\hfil&\vrule#&\hfil#\hfil&\vrule#\cr
\omit&height6pt&\omit&&\omit&&\omit&&\omit\cr
&&1&&2&&3&&4\cr
\omit&height6pt&\omit&&\omit&&$\phantom{abcdefgh}$&&$\phantom{abcdefgh}$\cr
\tablerule
\omit&height6pt&\omit&&\omit&&\omit&&\omit\cr
$A_n$&&$0,2,3,\dots,n+1$   &&{}&&{}&&{}\cr
\omit&height6pt&\omit&&\omit&&\omit&&\omit\cr
\tablerule
\omit&height6pt&\omit&&\omit&&\omit&&\omit\cr
$B_n$&&$0,2,4$ &&$6,8,\dots,2n$ && {}&& {}\cr
\omit&height6pt&\omit&&\omit&&\omit&&\omit\cr
\tablerule
\omit&height6pt&\omit&&\omit&&\omit&&\omit\cr
$C_n$&&$0,2,4,\dots,2n$ &&{}&&{}&&{}\cr
\omit&height6pt&\omit&&\omit&&\omit&&\omit\cr
\tablerule
\omit&height6pt&\omit&&\omit&&\omit&&\omit\cr
$D_n$&&$0,2,4,n$&&$6,8,10,\dots,2n-2$ &&{}&&{}\cr
\omit&height6pt&\omit&&\omit&&\omit&&\omit\cr
\tablerule
\omit&height6pt&\omit&&\omit&&\omit&&\omit\cr
$G_2$&&$0,2$&& 6 &&{}&&{}\cr
\omit&height6pt&\omit&&\omit&&\omit&&\omit\cr
\tablerule
\omit&height6pt&\omit&&\omit&&\omit&&\omit\cr
$F_4$&&$0,2$  &&$6,8$  && $12$&&\cr
\omit&height6pt&\omit&&\omit&&\omit&&\omit\cr
\tablerule
\omit&height6pt&\omit&&\omit&&\omit&&\omit\cr
$E_6$&& $0,2,5$  && $6,8,9$  &&$12$&&\cr
\omit&height6pt&\omit&&\omit&&\omit&&\omit\cr
\tablerule
\omit&height6pt&\omit&&\omit&&\omit&&\omit\cr
$E_7$&&$0,2$  && $6,8,10$ && $12,14$ && $18$\cr
\omit&height6pt&\omit&&\omit&&\omit&&\omit\cr
\noalign{\bigskip}
}\vbox{\hsize=13.5cm\noindent\lineskip=\normallineskip
\baselineskip=\normalbaselineskip\lineskiplimit=\normallineskiplimit
This table displays the relation between weights $s_j$ and exponents $d_j$
for the  simple Lie groups (all those other than $E_8$)  for which ${\cal W}$ is
the projectivization of some $\Omega=\oplus_j{\cal L}^{-d_j}$.  Weights are plotted
horizontally and the entries in the table are the exponents $d_j$ for a given weight.
For instance, for the group $G_2$, the exponents are 0 and 2 in weight 1 and 6 in weight
2; no other weights appear for this group. 
\bigskip 
}}
$$

In section 8, we compare the  explicit construction of bundles to what
could be predicted {\it a priori} from index theory.

In this paper, we concentrate on explaining aspects of the problem
that seem likely to be most immediately relevant and useful for physicists.
 A more systematic exposition with full proofs will appear elsewhere \fmw.

 \newsec{Direct Approach For $SU(n)$ and $Sp(n)$}

\subsec{Bundles On An Elliptic Curve}

For the starting point, we consider bundles on a single elliptic
curve $E$ -- that is, a two-torus with a complex structure and
a distinguished point $p$ called the ``origin.''  $p$ is the identity
element in the group law on $E$.

A stable or semistable holomorphic $G$ bundle on a Riemann surface $\Sigma$ in 
general is associated with a representation of the fundamental
group of $\Sigma$ in (the compact form of) $G$.  
For the case that the Riemann surface is a two-torus $E$,
the fundamental group is abelian and generated by two elements, so if $G$ is simply-connected,
a representation of
the fundamental group in $G$ can be conjugated to a representation
in the maximal torus of $G$ \borel.
\def\L{{\cal L}}
\def\O{{\cal O}}
As promised in the introduction, the present section is devoted
to a direct construction of $G$ bundles on $E$ in certain simple cases.
First we take $G=SU(n)$.

In this case, a $G$ bundle determines a rank $n$ 
complex vector bundle $V$, of trivial determinant.
The fact that $V$ can be derived from a representation of the fundamental
group in a maximal torus means that $V=\oplus_{i=1}^n\N_i$,
where the $\N_i$ are holomorphic line bundles.  The fact
that $V$ is an $SU(n)$ (rather than $U(n)$) bundle means
that   $\otimes_{i=1}^n\N_i
=\O$.  ($\O$ is a trivial line bundle over $E$.)  For $V$ to be semistable
means that the $\N_i$ are all of degree zero.
The Weyl group of $SU(n)$ acts by permuting the $\N_i$, and the $\N_i$
are uniquely determined up to this action.

If $\N_i$ is a degree zero line bundle on $E$, there is a unique
point $Q_i$ in $E$ with the following property: $\N_i$ has
a holomorphic section that vanishes only at $Q_i$ and has a pole
only at $p$.  So the decomposition  $V=\oplus_{i=1}^n\N_i$
means that $V$ determines the $n$-tuple
of points $Q_1,\dots,Q_n$ on $E$.  The fact that $\otimes_{i=1}^n\N_i=
\O$ means that (using addition with respect to the group law on $E$)
$\sum_iQ_i=0$.  Conversely, every $Q_i\in E$ determines a degree zero 
line bundle $\N_i=\O(Q_i)\otimes \O(p)^{-1}$ 
(whose sections are functions on $E$ that are
allowed to have  a pole at $Q_i$ and required to have a zero at $p$),
and every $n$-tuple $Q_1,\dots,Q_n$ of points in $E$ with $\sum_iQ_i=0$
determines
the semistable $SU(n)$ bundle $V=\oplus_{i=1}^n\N_i$.
The bundle $V$ determines the $\N_i$ and $Q_i$ up to permutations,
that is up to the action of the Weyl group.

The moduli space of ${\cal M}$ of semistable $SU(n)$ bundles on $E$ is 
therefore
simply the moduli space of unordered
$n$-tuples of points in $E$ that add to zero.  
The space of such $n$-tuples can be conveniently described as follows.
If $Q_1,\dots,Q_n$ is such an $n$-tuple, then there exists
a meromorphic function $w$ which vanishes (to first order) at the $Q_i$
and has poles only at $p$.  (Existence of such a $w$ is equivalent to
the vanishing of the sum of the $Q_i$ in the group law on $E$.)
Such a $w$ is unique up to multiplication by a non-zero complex scalar.
Conversely, let $W=H^0(E,\O(np))$ be the space of meromorphic functions
on $E$ that have a pole of at most $n^{th}$ order at $p$ and no poles
elsewhere.  Such a function $w$ has $n$ zeroes $Q_i$ which add up to
zero (some of these points may
be coincident; 
also, if the pole at $p$ is of order less than $n$, we interpret
this to mean that some of the $Q_i$ coincide with $p$).

\def\P{{\bf P}}
\def\M{{\cal M}}  This correspondence between $n$-tuples and functions
means that  
${\cal M}$ is a copy of complex projective space ${\bf P}^{n-1}$,
obtained by projectivizing $W$:
\eqn\calm{{\cal M}=\P H^0(E,\O(np)).}
Actually, the functions $w\in H^0(E,\O(np))$ can be described very
explicitly.  
If  $E$ is described by a Weierstrass equation
\eqn\algeq{y^2=4x^3-g_2x-g_3}
in $x-y$ space, and $p$ is the point $x=y=\infty$, then a meromorphic
function $w$ with a pole only at $p$ is simply a polynomial in 
$x$ and $y$.  As $x$ has a double pole at $p$ and $y$ has a triple pole, 
$w$ can be written
\eqn\nalm{w=a_0+a_2x+a_3y +a_4x^2+a_5x^2y+\dots,}
where the last term is $a_nx^{n/2}$ for $n$ even, or $a_nx^{(n-3)/2}y$
for $n$ odd.   In other words, $w$  is a general polynomial in $x$ and $y$ with
at most an $n^{th}$ order pole at infinity, and (modulo the Weierstrass equation) 
at most a linear dependence on $y$.   To 
allow for a completely general set of $Q_i$, one restricts
the $a_k$ only by requiring that they are not all identically zero.
(For example, $a_n$ vanishes if and only if one of the $Q_i$ is the
point $p$ at infinity.)
Since the $a_k$ are never identically zero, it makes sense to 
interpret them as homogeneous coordinates of a complex projective
space, and this is the idea behind \calm.

\bigskip\noindent
{\it $Sp(n)$ Bundles}

\def\L{{\cal L}}

The other  case for which $G$ bundles on an elliptic curve
can be described explicitly  with similar methods is the case
$G=Sp(n)$.  Using the $2n$-dimensional representation of
$Sp(n)$, we can think of an $Sp(n)$ bundle  as a rank $2n$ holomorphic
 vector
bundle $V$ equipped with a non-degenerate holomorphic section
$\omega$ of $\wedge^2 V^*$, reducing 
the structure group to $Sp(n)$.  On an elliptic curve, a stable
$Sp(n)$ bundle is simply a direct sum 
$V=\oplus_{i=1}^n(\N_i\oplus\N_i^{-1})$; in this basis, the non-zero
matrix elements of $\omega$ map $\N_i\otimes \N_i^{-1}\to \O$.
We associate each pair $(\N_i,\N_i^{-1})$ with a pair $(Q_i,-Q_i)$
of equal and opposite points in $E$.  The Weyl group acts by
permutation of these pairs and by the interchanges $Q_i\leftrightarrow
-Q_i$.  The moduli space $\M$ of $Sp(n)$ bundles
on $E$ is simply the space of $n$-tuples of unordered pairs
$(Q_i,-Q_i)$, up to permutation.

A point $Q$ on $E$ corresponds to a set of values
 $(x,y)$ obeying the Weierstrass
equation \algeq.
$y$ is determined by $x$ up to sign.  Since the transformation
$Q\to -Q$ is the $\Z_2$ symmetry $y\to -y$ of $E$, being given not
a point $Q$ but a pair $(Q,-Q)$ is tantamount to being given only
the value of $x$.  So an $n$-tuple of pairs $(Q_i,-Q_i)$ is equivalent
to an $n$-tuple of values of $x$, say $x_1,x_2,\dots,x_n$.  Because
of the Weyl action, the $x_i$ are determined by the bundle only up to
permutation.

\def\P{{\bf P}}
\def\WP{{\bf WP}}

As in the discussion of $SU(n)$ bundles, the unordered $n$-tuple
$x_1,\dots,x_n$ is conveniently summarized by giving a polynomial
in $x$ whose zeroes are the $x_i$:
\eqn\jokko{t=c_0+c_1x+c_2x^2+\dots +c_nx^n.}
Once again, to allow for the possibility that some $Q_i$ are equal
to $p$, the $c_i$ are restricted only to not all be zero.
Since a rescaling $t\to\lambda t$ with non-zero complex $\lambda$
does not change the zeroes, the moduli space ${\cal M}$ of $Sp(n)$ bundles
on $E$ is again a projective space, in this case
the projective space ${\P}^n$ whose homogeneous coordinates are the $c$'s.

\def\WP{{\bf WP}}
It should be stressed that what the above constructions determine
is the moduli space of $G$ bundles on $E$ for the {\it simply-connected}
groups $SU(n)$ and $Sp(n)$.  The discussion must be considerably adapted
to describe the $SU(n)/\Z_n$ or $Sp(n)/\Z_2$ moduli spaces
by similar methods.  For example, these moduli spaces have different components
(of different dimension) indexed by the topological type of the bundle.

We conclude by briefly comparing $Sp(n)$ bundles to $SU(2n)$ bundles.
Given the natural embedding of $Sp(n)$ in $SU(2n)$, the moduli space ${\cal M}_{Sp(n)}$
of flat
$Sp(n)$ bundles on $E$ can be embedded as a subspace of the moduli
space ${\cal M}_{SU(n)}$ of flat $SU(2n)$ bundles on $E$.  In fact, according to \nalm,
flat $SU(2n)$ bundles are related to polynomials $w=a_0+a_2x+a_3y+\dots +a_nx^n$.
If we simply set to zero the $a_i$ of odd $i$ (the ones odd under $y\to -y$) such
a polynomial takes the form of the $Sp(n)$ polynomial in \jokko.  By
more carefully examining the above constructions, it can be shown
that this identification of polynomials does give the embedding of ${\cal M}_{Sp(n)}$
in ${\cal M}_{SU(2n)}$.  An analogous relation holds for $Sp(n)$ and $SU(2n)$ bundles
on elliptic manifolds.

\subsec{Bundle Of Projective Spaces}

For our applications, we must understand not vector bundles on
a single elliptic curve $E$, but vector bundles on a family of elliptic
curves, that is on a complex manifold $Z$ which maps to some base
$B$ with the generic fiber being an elliptic curve.
We will assume for simplicity that the map $Z\to B$ has a section
(the case most commonly considered in relation to $F$ theory).  In
that case,   $Z$ can be described by a Weierstrass equation.
The Weierstrass
equation can be  written in a $\P^2$ bundle $W$ over $B$; $W$ is the
projectivization of $\L^2\oplus \L^3\oplus \O$, with  $\L$ being
some line bundle over $B$.  If we describe
$W$ by  homogeneous coordinates $x,y,z$ (which are sections respectively
of $\L^2,\L^3$, and $\O$), then
the Weierstrass equation reads
\eqn\ormal{zy^2=4x^3-g_2xz^2-g_3z^3}
where $g_2$ and $g_3$ are sections of $\L^4$ and $\L^6$, respectively.
Often, we will use affine coordinates with $z=1$.
For $Z$ to be a Calabi-Yau manifold -- our main interest for the
applications in this paper --
one needs $\L=K_B^{-1}$, with $K_B$ the canonical bundle of $B$.  However,
the description of vector bundles over $Z$ does not require this.  

First we consider in some detail $SU(n)$ bundles.
On a {\it single} elliptic curve, we described an $SU(n)$ bundle
by giving an $n$-tuple of points, determined by another equation
\eqn\vormal{a_0+a_2x+a_3y+\dots +a_n x^{n/2}=0.}
(If $n$ is odd, the last term is $x^{(n-3)/2}y$.)
The $a_i$, up to scaling, define a point in a projective
space $\P^{n-1}$ that parametrizes $SU(n)$ bundles on $E$.  
Given that $x$ and $y$ are sections of $\L^2$ and $\L^3$, one
can think of $a_i$ as a section of $\L^{-i}$.

Now if one has a family of elliptic curves, making up an elliptic
manifold $Z\to B$, then over each $b\in B$, there is an elliptic
curve $E_b$ and a moduli space ${\P}^{n-1}_b$ of $SU(n)$ bundles
on $E_b$.  The ${\P}^{n-1}_b$'s fit together into a $\P^{n-1}$ bundle
over $B$ which we will call ${\cal W}$.  
By noting that the $a_i$ can be interpreted as
homogeneous coordinates for this bundle, we see that it can be
constructed by projectivizing the vector bundle over $B$ 
\eqn\bubbu{\Omega=\O\oplus
\L^{-2} \oplus \L^{-3}\oplus \dots\oplus \L^{-n} .}
Note that the exponents here are 0 and $-s_j$, where $s_j=2,3,4.\dots ,n$
are the degrees of the independent Casimir operators of $SU(n)$
(that is, if $\phi$ is a vector in the adjoint representation of
$SU(n)$, regarded as an $n\times n$ hermitian matrix,
then the invariants are  $\Tr \,\phi^k$,
for $k=2,3,\dots,n$; and these have degrees $2,3,4. \dots, n$).  This
is the form for $\Omega$ promised in the introduction.  

The constant section of ${\cal O}$, when embedded as a section of
${ \Omega}={\cal O}\oplus \dots$, gives a section of ${\cal W}$
that can be characterized by the fact that the homogeneous coordinates
$a_i$ all vanish for $i>0$.  This means that 
on each fiber all of the 
$Q_i$ are at infinity; in the description of bundles
by flat unitary connections, such a bundle corresponds to the trivial
flat connection.  This interpretation of the summand $\O$ was promised
in the introduction.

Analogous results for $Sp(n)$ are easily obtained.
We found in section 2.1 that an $Sp(n)$ bundle
over a single elliptic curve $E$ in Weierstrass form
is determined by an equation
\eqn\impu{c_0+c_1x+c_2x^2+\dots +c_nx^n=0.}
The $c_i$ were homogeneous coordinates of a projective space ${\P}^n$
that parametrizes $Sp(n)$ bundles.  
If instead one has a family of elliptic curves, making up an elliptic
manifold $Z\to B$, then we should think of the $c_i$ as homogeneous coordinates
on a $\P^n$ bundle ${\cal W}$
over $B$ (whose fiber over $b\in B$ is the moduli
space of $Sp(n)$ bundles on the elliptic curve $E_b$ that lies over $b$).
One can think of ${\cal W}$ as the projectivization of a vector bundle
$\O\oplus \L^{-2}\oplus \L^{-4}\oplus \dots\oplus \L^{-2n}$ (the exponents
are clear if one recalls that $x$ in equation \impu\ is
a section of $\L^2$).  
Note that the exponents are 0 and  $-k$ 
where $k=2,4,6,\dots,2n$ are the degrees of the Casimir invariants of
$Sp(n)$.  Thus, we obtain again the form
for ${\cal W}$ promised in the introduction.  The section of ${\cal W}$
coming from the summand $\O$ again corresponds on each $E_b$ to a bundle
that is related to the trivial flat connection.

\subsec{ Construction Of Bundles Over Elliptic Manifolds}

Let us begin with a rank $n$ complex vector
bundle over $Z$ with a hermitian-Yang-Mills $SU(n)$
connection.
This determines a holomorphic vector bundle $V$ over $Z$
which can be restricted to give a holomorphic bundle on each fiber.
If the restriction to each fiber is
semistable, it determines a section of the
projective space bundle ${\cal W}\to B$.  The section $s$ is not
the whole story; there is additional data that we will describe
shortly.  But first let us explain in some detail how to construct
a general section $s$.  

\def\M{{\cal M}}
The mapping from ${\Omega}
=\O\oplus \L^{-2}\oplus \L^{-3}\oplus \dots\oplus \L^{-n}$
to ${\cal W}$ (by throwing away the zero section and dividing on each fiber
by $\C^*$)
gives a holomorphic line bundle over ${\cal W}$ that we will call $\O(-1)$ 
(it restricts on each fiber $\P_b^{n-1}$ to the $\C^*$  bundle usually
known by that name).  The homogeneous coordinates $a_k$ ($k=0,2,3,\dots, n$)
are sections
of $\co(1)\otimes \L^{-k}$.  If $s:B\to {\cal W}$ is a section of ${\cal W}\to B$,
then $s^*(O(1))$ is a line bundle on $B$ that we will call $\N$.  Different
$\N$'s can arise; the homotopy class of the section of ${\cal W}$ is
determined by the first Chern class of $\N$.  (We will learn
in section 7 how the first Chern class of $\N$ is related to the
second Chern class of $V$.)

The $a_k$ pull back under $s$ to sections of $\N\otimes \L^{-k}$.
This process can also be read in reverse: if one picks an arbitrary
line bundle $\N$ on $B$ which is sufficiently ample, 
and picks sections $a_k$ of $\N\otimes \L^{-k}$ that
are sufficiently generic as to have no common zeroes, then
$b\to (a_0(b),a_2(b),\dots , a_n(b))$ gives a section of ${\cal W}$.
Two sections coincide if and only if the corresponding $a_j$ are 
proportional, so the space of sections (of given homotopy class) is
itself a projective space ${\bf P}^m$ for some $m$.

So we get an effective way to  construct sections of ${\cal W}$:
pick $\N$ and the $a_k$.  
Now, a suitable
 $SU(n)$ bundle $V$ over $Z$ determines, as we have explained, such
a section $s$.  In particular, it determines an $\N$.  
However, the section may not uniquely determine the bundle, as we
will now explain.

 A section of ${\cal W}$  concretely determines an equation \vormal\
(with the $a_k$ now understood as sections over $B$), and this, together 
with \ormal, determines a hypersurface $C$ in $Z$.  $C$ is an $n$-fold ramified
cover of $B$, since for fixed $b\in B$, the equations \ormal\ and \vormal\
have $n$ solutions.  By analogy with similar structures in the theory of
integrable systems, we call any such hypersurface in $Z$ that projects
to an $n$-fold cover of $B$ a ``spectral cover.''

Although a ``good'' hermitian-Yang-Mills connection on an $SU(n)$ 
bundle over $Z$ determines in this way a unique spectral cover $C$,
many different bundles may give the same spectral cover.
To proceed further, we need to make a digression about the ``Poincar\'e
line bundle.''

\bigskip\noindent
{\it The Poincar\'e Line Bundle}

We have already exploited  the following basic fact.
If $E$ is an elliptic curve, with a distinguished point $p$, then
the degree zero line bundles on $E$ are parametrized by $E$ itself;
a point $Q\in E$ corresponds to the line bundle $\L_Q=\O(Q)\otimes \O(p)^{-1}$.
\def\P{{\cal P}}
Now consider the product $F =E\times E$, and think of the first factor
as parametrizing degree zero line bundles on the second.  Then one can aim
to construct a line bundle $\P$ on $F$, whose restriction to $Q\times E$,
for any $Q\in E$, will be isomorphic to $\L_Q$.  In fact,
one can take $\P$ to be the line bundle $\O(D_0)$, where $D_0$ is the divisor
$D_0=\Delta - E\times p$ (here $\Delta$ is the diagonal in $E\times E$); the idea
here is that $D_0$ intersects $Q\times E$ in the divisor $Q-p$ (or
$Q\times Q-Q\times p$, to be more fastidious), so the restriction of 
$\O(D_0)$ to $Q\times E$ is $\L_Q$.  However, it is more symmetric
to take $D=\Delta-E\times p-p\times E$ and $\P=\O(D)$.  The idea
is now that $\P$ is isomorphic to $\L_Q$ if restricted to either $Q\times
E$ or $E\times Q$.  For our purposes, a line bundle $\P$ with the property
just stated will be called a Poincar\'e line bundle.

We actually want a Poincar\'e line bundle for a family of elliptic curves.
Suppose that one is given an elliptic manifold $\pi:Z\to B$, with a 
section $\sigma$. One forms the ``fiber product'' $Z\times_B Z$ 
which consists of pairs $(z_1,z_2)\in Z\times Z$ such that $\pi(z_1)=
\pi(z_2)$.\foot{In this paper, it will be possible to ignore singularities
of this fiber product.}  The equation $z_1=z_2$ defines a divisor in $Z\times_B Z$ which
we will call $\Delta$.
$Z\times _BZ$ can be mapped to $Z$ in two ways, by forgetting
$z_2$ or $z_1$; the two maps are called $\pi_1$ and $\pi_2$.  One
can also simply project $Z\times_BZ$ to $B$ by $(z_1,z_2)\to \pi(z_1)
$ (which equals $\pi(z_2)$); we will call this map $\tilde \pi$.
For any $b\in B$, $\tilde\pi^{-1}(b)$ is a copy of $E_b\times E_b$,
where $E_b =\pi^{-1}(b)$.

By a Poincar\'e line bundle $\P_B $ over $Z\times _BZ$ we mean a line
bundle which on each $E_b\times E_b$ is a Poincar\'e line bundle in the
previous sense, and which is trivial when restricted to $\sigma\times Z$
or $Z\times \sigma$.  One might think that one should take $\P_B$ to be $\O(D)$,
where $D=\Delta-\sigma\times Z-Z\times \sigma$.  This line bundle certainly
restricts appropriately to each $E_b\times E_b$.  Its restriction to
$\sigma\times Z$ or $Z\times \sigma$ is however non-trivial -- in fact, it is 
isomorphic to
the pullback $\tilde\pi^*(\cal L)$ of $\L\to B$, as we will show in section 7.
For the desired Poincar\'e line bundle, 
we take $\P_B={\cal O}(D)\otimes \tilde\pi^*({\cal L}^{-1})$.

\bigskip\noindent
{\it Bundles From Sections}

Now we want to return to our problem of understanding how a vector bundle
over $Z$ is to be constructed from a section $s:B\to {\cal W}$, or
equivalently from the spectral cover $C$.  We start with 
$Y=C\times _B Z$, which is defined as the subspace of $Z\times_BZ$ with
$z_1\in C$.  The map $\pi_2$ (forgetting $z_1$) maps $Y\to Z$.  $Y$ is
an $n$-fold cover of $Z$, since $C\to B$ was an $n$-fold cover.

\def\R{{\cal R}}
Suppose we are given any line bundle $\R$  
over $Y$.  Away from branch points of the map $\pi_2:Y\to Z$, one can define
a rank $n$ vector bundle $V$ over $Z$ as follows.  Lying above any
given $z\in Z$, there
are $n$ points $y_1,\dots,y_n\in Y$; take the fiber  $V_z$ of $V$ at $z$
to be $\oplus_{i=1}^n\R_{y_i}$ (where ${\cal R}_y$ is the fiber of
${\cal R}$ over $y\in Y$).  The bundle  $V$ so defined can actually
be extended over all of $Z$ by using a more powerful definition based on
the ``push-forward'' operation in algebraic geometry; one defines
a section of $V$ over a small open set $U\subset Z$ to be  a section
of $\R$ over $\pi_2^{-1}(U)$.  The resulting vector bundle over 
$Z$ is denoted $V=\pi_{2*}(\R)$.

Here let us point out a technical fact about this construction.  The bundles
produced in this way have the property that their restrictions to most, but not
all, fibers carry flat $SU(n)$ connections.  If $b\in B$ is such that its
pre-image in the spectral cover $C$ consists of $n$ distinct points, then it is clear
from the construction that the restriction of the resulting vector bundle
to the fiber $E_b$ is a sum of $n$ line bundles of degree zero (given by the
$n$ points in $E_b$) and hence carries a flat $SU(n)$ connection.  At the branch
points of $V$ something entirely different happens \fm.  For example, if the
pre-image of $b$ in the spectral cover $C$ consists of $n-2$ points of
multiplicity one and  a point of multiplicity   two then the restriction of
$V$ to $E_b$ is a direct  sum of $n-2$ line bundles 
 and a rank two bundle that is a non-trivial
extension of a line bundle by a second (isomorphic) line bundle.  This
bundle    admits no flat $SU(n)$ connection.  So, although the section of
${\cal W}$ can be viewed as defining a varying family of holomorphic bundles
with flat connections on the fibers of $Z$ over $B$, to fit these fundles
together to make a holomorphic bundle on $Z$ we must replace some of the flat
bundles by non-isomorphic, $S$-equivalent bundles.  After fitting these bundles
together, we often produce a stable bundle which then carries a 
hermitian-Yang-Mills
connection.  But this connection is not obtained by gluing together the
original flat connections.    In many situations, this construction
yields the generic stable bundle over $Z$.

\bigskip\noindent
{\it Reconstruction Of A Bundle From The Spectral Cover}

Suppose we start with a vector bundle $V$ over $Z$, and use it as 
above to construct a spectral cover $C$ of $B$.  To recover $V$ from
$C$, the basic idea is to start with a suitable line bundle $\R$ 
over $C\times_BZ$, and obtain $V$ as $\pi_{2*}(\R)$.

The instructive first case to consider is that in which $\R=\P_B$,
the Poincar\'e line bundle over $Z\times _BZ$ restricted to $C\times_BZ$.
Recall that to construct the spectral cover $C$ from the vector bundle $V$,
the idea was that the restriction of $V$ to $E_b$ was isomorphic\foot{Or in
general $S$-equivalent.} to
$\L_{Q_1(b)}\oplus \dots\oplus \L_{Q_n(b)}$ for some points $Q_1(b),\dots
,Q_n(b)\in E_b$; we then defined $C$ to be an $n$-sheeted cover of $B$ such
that the points over $b$ are $Q_1(b),\dots,Q_n(b)$.  If we define
$V'=\pi_{2*}(\P_B)$, then from the definitions of $\pi_{2*}$ and $\P_B$,
the restriction of $V'$ to $E_b$ is indeed equivalent to
$\L_{Q_1}\oplus \dots\oplus \L_{Q_n}$.   

\def\S{{\cal S}}
So $V$ and $V'$ are equivalent on each $E_b$.  But this
does not necessarily imply that $V=V'$.  In fact, the above construction
can be generalized as follows.  Let $\S$ be {\it any} line bundle over
$C$, and let $V'=\pi_{2*}(\P_B\otimes \S)$.  Then the isomorphism  class
of the restriction of $V'$ to any $E_b$ is independent of $\S$,
since $\S$  (being trivial locally along $C$) is
trivial when restricted to a neighborhood of the inverse image of
any given $b\in B$.

This is the only ambiguity in the reconstruction of a vector bundle
from its spectral cover in the following sense.
The main theorem of chapter 7 of \fm\ asserts that if the base $B$ is
one-dimensional, then any        generic $V$ can
be reconstructed from its spectral cover $C$ as $V=\pi_{2*}(\P_B\times\S)$
for some $\S$.\foot{The argument in that reference is formulated for rank two
bundles, but that restriction was needed primarily in giving a precise
description of the possible exceptional behavior; in describing
a generic $V$ in the above-stated form, there is no restriction
to rank two.}
For $B$ of dimension bigger than one, it is too much
to expect this to be true for all bundles, but it is true for the
bundles that can be most naturally constructed via spectral covers;
and these suffice to construct (open dense subsets of)
 some components of the moduli space of all
bundles.  To understand those components -- all we will aim for in this
paper --  we ``only'' need to understand spectral covers and line bundles
over them.

To summarize, we have here described in some detail
the construction of bundles from spectral covers for
$G=SU(n)$.  A similar construction should be possible for
$G=Sp(n)$.

\subsec{ A Note On Jacobians}

We will here make a few remarks that are not needed for understanding
most of the paper, but are background for the comparison between the moduli
space of spectral covers and the moduli space of $F$ theory complex structures
that we will make in section six.  These remarks concern the role in the construction of
stable bundles of certain Jacobians and abelian varieties.

In our applications, $B$,  as the base of a Calabi-Yau fibration, 
is simply-connected.

If $B$ is of dimension one, and therefore in practice $B={\bf P}^1$, then
the $n$-sheeted cover $C$ of $B$ is a Riemann surface of higher genus.
$\S$ is then not completely fixed by its first Chern class; any given $\S$ could
be modified by twisting by a flat line bundle on $C$.  Such flat line bundles
are classified by
 the Jacobian $J(C)$ of $C$.   The moduli space of stable bundles over $Z$ is fibered
over the space of $C$'s, with the fiber being this Jacobian.

When $B$ is of dimension one, the Calabi-Yau manifold $Z$ is actually
a K3 surface, and the moduli spaces of bundles are hyper-K\"ahler.
The space of sections of ${\cal W}$ is a K\"ahler manifold
but not hyper-K\"ahler; in fact, it is a projective space ${\bf P}^m$ for
some $m$, as was seen above.  The Jacobian $J(C)$  has the same dimension $m$;
in fact the whole moduli space looks locally, near the zero section of the bundle
of Jacobians, like the cotangent bundle $T^*{\bf P}^m$.  (Indeed, $C$ is
a curve in  $Z$; if $N_C$ is the normal bundle to $C$ in $Z$,
then the tangent space to the space of spectral covers at $C$ is $H^0(C,N_C)$, which
because $Z$ has trivial canonical bundle is dual to $H^1(C,{\cal O}_C)$, which
is the tangent space to the Jacobian of $C$.)
In heterotic string compactification on $Z$, the $m$ chiral superfields
parametrizing the choice of $C$ combine with $m$ chiral superfields 
parametrizing the Jacobian of $C$ to make $m$ hypermultiplets.

Though we have so far considered only $SU(n)$ and $Sp(n)$ bundles,
an analogous picture holds for any $G$.  The moduli space ${\cal M}$ of bundles
is fibered over the space ${\cal Y}$ of sections
of a certain weighted projective space bundle that we will construct; these
sections generalize the notion of the spectral cover.
${\cal Y}$ is itself a weighted projective space, as we will see.  The fiber of the
map from ${\cal M}$ to ${\cal Y}$ is an abelian variety of dimension
equal to that of ${\cal Y}$; the total space ${\cal M}$ is hyper-K\"ahler and
looks locally (near a certain ``zero section'' of ${\cal M}\to{\cal Y}$)
like $T^*{\cal Y}$. 

Duality with $F$ theory relates the heterotic string on the K3 surface
$Z$ to $F$ theory on a Calabi-Yau threefold $X$ that is fibered over
$B$ with K3 fibers.  The part of the
$F$ theory moduli space that is related to the moduli space of bundles
on $Z$ must, if the duality is correct, have a fibration analogous
to 
${\cal M}\to {\cal Y}$, with the fiber being an abelian variety of dimension
equal to the base.  In $F$ theory, abelian varieties (and more general complex
tori) appear in the moduli space of vacua  because
an $F$ theory vacuum is parametrized among other things by the choice
of a point in the complex torus $H^3(X,{\bf R})/H^3(X,\Z)$, which is known
as the intermediate Jacobian $J(X)$ of $X$.  (The appearance of $J(X)$ is
readily seen if one compactifies on another
circle to convert to $M$-theory; in that formulation, $J(X)$ parametrizes
the  periods of the vacuum expectation value of the three-form potential
of eleven-dimensional supergravity.)  The $F$ theory moduli space
is fibered over the    space of complex and K\"ahler structures  on $X$
with fiber $J(X)$.

In duality between the heterotic string and $F$ theory, heterotic string
vacua in which the structure group of the $E_8\times E_8$ 
gauge bundle reduces to $G\times E_8$ for some $G\subset E_8$ correspond
to points in $F$ theory moduli space in which the K3 fibration
$X\to B$ has a section $\theta$ of singularities; the precise nature
of the singularities for arbitrary $G$   was described in \many.
One factor in the heterotic string moduli space is the moduli
space ${\cal M}$ of $G$ bundles over $Z$, with its fibration ${\cal M}\to {\cal Y}$.
In the duality, ${\cal Y}$ should be mapped to the space of certain
data parametrizing the geometry of $X$ near $\theta$; the details of
which parameters should be relevant for given $G$ were worked out
in \refs{\mv,\many}.  The abelian variety that is the fiber of ${\cal M}\to 
{\cal Y}$ should correspond to a certain factor that splits off from
$J(X)$ when $X$ develops the prescribed type of singularity.

In section  six, we will compare the heterotic string to $F$ theory
by comparing the space ${\cal Y}$ (as determined by our analysis of
$G$ bundles) to the
appropriate data describing the behavior of $X$ near $\theta$ (as determined
in \refs{\mv,\many}).  We will not compare the abelian varieties that appear
on the two sides, for two roughly parallel reasons.  (1) On the heterotic
string side, while we will determine the analog of ${\cal Y}$ for general
$G$, we do not have equally good control over the abelian varieties except
for $G=SU(n)$.  (2) On the $F$ theory side, while the complex structure
parameters that should be related to $G$ bundles have been determined
for general $G$ \refs{\mv,\many}, the appropriate factor in $J(X)$
(which presumably involves cycles with a particular behavior near $\theta$) 
has not yet been described.   Identifying the abelian
varieties on the two sides and comparing them is an interesting question.

At the end of section 4, we will state a conjectural description of
the abelian variety for the case $G=E_8$.

\bigskip\noindent
{\it Fibrations Over Surfaces}

Now let us move on to the case that the base $B $ of the elliptic
fibration is of dimension bigger than one.  In practice, the main case
for physics is that $B$ is of dimension  two, so that the elliptic
manifold $Z\to B$ is a threefold and the K3-fibered manifold $X\to B$ is
a fourfold.  Much of the discussion above still applies.  The moduli
spaces ${\cal M}$ of bundles on $Z$ have fibrations ${\cal M}\to {\cal Y}$
where ${\cal Y}$ is a space of spectral covers (in a generalized sense)
 and the fiber is an abelian
variety.  Likewise, on the $F$ theory side, there is a space of complex
moduli of $X$ with fibered over it the abelian variety
$H^3(X,{\bf R})/H^3(X,\Z)$.  Our test of the duality in section six involves
comparing ${\cal Y}$ to the appropriate part of the moduli space of complex
structures on $X$, without trying to compare the abelian varieties.  

For the present
purposes, the main change in going from $B$ of dimension one to $B$ of dimension
two is that the base and fiber of the fibration ${\cal M}\to {\cal Y}$ need
not have equal dimensions, and in particular the fiber vanishes in many
of the simplest examples.  This is so on the heterotic string side
because in many simple examples, the spectral cover $C $ of the base
$B$ of an elliptic Calabi-Yau threefold is simply-connected; when that
is so, a line bundle
$\S\to C$ is completely determined by its first Chern
class, and the choice of $\S$ does not introduce any abelian variety.  
On the $F$ theory side, Calabi-Yau four-folds 
 can very commonly have $H^3=0$, so that the supergravity
three-form has no periods, and there is no Jacobian to consider.
Thus, in many instances, our check of  heterotic string/$F$ theory
duality in section 7 is more complete for $B$ a surface than for $B$ a curve,
in the sense that the abelian varieties over which one does not have good
control are actually trivial.

It would be very interesting, of course, to show that the relevant part of
$H^3(X)$ is non-trivial precisely when $H^1(C)$ is non-trivial,
and to compare the resulting abelian varieties.

\newsec{Reduction To The Simply-Laced Case}

\subsec{Simply-Laced Subgroup}

We now for the moment think of semi-stable holomorphic $G_{\bf C} $ bundles
on an elliptic
curve $E$ in terms of representations of the fundamental group of $E$ in
the compact form of $G$.  Such a representation is determined by two
commuting elements of $G$.  For $G$ simply-connected, these two elements
can be conjugated into a maximal torus $T$ \borel\
 in a way that is unique 
up to the action of the Weyl group $W$.  The moduli space of semi-stable
$G$ bundles on $E$ is thus isomorphic to $(T\times T)/W$, where $W$ acts
in the natural fashion on both factors of $T$.

We propose to use this in the following situation.  Let $G$ be a simple,
connected, and simply-connected group that is not simply-laced.
$G$ then has a canonical simply-laced 
subgroup $G'$ that is generated by the long roots
of $G$.  The embedding of $G'$ in $G$ gives an isomorphism of the maximal
torus of $G'$ with that of $G$.  The Weyl group $W'$ of $G'$ is however
a subgroup of the Weyl group $W$ of $G$.  In fact, it is a normal
subgroup; there is a group homomorphism
\eqn\grp{1\to W'\to W\to \Gamma\to 1}
for some finite group $\Gamma$.    
$\Gamma$ is a group of outer automorphisms of $G'$.
If ${\cal M}=(T\times T)/W$ and ${\cal M}'=(T\times
T)/W'$ are the moduli spaces of $G$ and $G'$ bundles over $E$, then
\eqn\mimmo{{\cal M}={\cal M}'/\Gamma.}

We will use this to describe the moduli space of $G$ bundles given
the moduli space of $G'$ bundles, and thus to reduce the description of the
moduli space to the simply-laced case.

In practice, there are four examples of this construction:

(1) For $G=Sp(n)$, $G'=SU(2)^n$.  $\Gamma$ is the group $S_n$ of permutations
of the $n$ copies of $SU(2)$ in $G'$.

(2) For $G=G_2$, $G'=SU(3)$. $\Gamma$ is the group $\Z_2$ of ``complex
conjugation'' that exchanges
the three-dimensional representation of $SU(3)$ with its dual.

(3) For $G=Spin(2n+1)$, $G'=Spin(2n)$. $\Gamma$ is the group $\Z_2$ generated
by the outer automorphism of $Spin(2n)$ that exchanges
the two spin representations of $Spin(2n)$.

(4) For $G=F_4$, $G'=Spin(8)$.  $\Gamma$ is the triality group
$S_3$ that permutes the three eight-dimensional representations of $Spin(8)$. 

We consider the four examples in turn.

\bigskip\noindent{\it $Sp(n)$ Revisited}

We consider the elliptic curve $E$ to be given by a Weierstrass equation
in the $x-y$ plane.
The moduli space of $SU(2)$ bundles on $E$ is parametrized,
as we learned in the last section, by the choice of a point $x$
which can be regarded as the root of a spectral equation
\eqn\rteqn{a_0+a_2x=0.}
A $G'$ bundle for $G'=SU(2)^n$ is therefore given by an ordered $n$-tuple 
$x_1,x_2,\dots,x_n$.  The group $\Gamma$ acts by permutation of the $x_i$, so
the relation ${\cal M}={\cal M}'/\Gamma$ says in this case that
the moduli space of $Sp(n)$ bundles over $E$ is the space of unordered
$n$-tuples $x_1,x_2,\dots,x_n$.  This is the description that we obtained
``by hand'' in section 2.  As we explained there, the space of unordered
$n$-tuples can be identified with the space of spectral equations of the form
\eqn\yeqn{c_0+c_1x+c_2x^2+\dots +c_nx^n=0.} 
Furthermore, in the case of an elliptically fibered manifold $Z\to B$,
the $c_i$ are homogeneous coordinates for a projective space bundle
${\cal W}\to B$, as we explained in section       2.

\bigskip\noindent{\it $G_2$ Bundles}
 
Now we consider the case that $G=G_2$, $G'=SU(3)$.  From what we learned
in section 2, an $SU(3)$ bundle over $E$ is determined by a spectral equation
\eqn\peqn{a_0+a_2x+a_3y=0}
whose roots are three points $Q_1,Q_2,Q_3\in E$ with $Q_1+Q_2+Q_3=0$.
The moduli space ${\cal M'}$ of $SU(3)$ bundles is thus a copy of ${\bf P}^2$ with
homogeneous coordinates $a_0,a_2,$ and $a_3$.

The exchange of an $SU(3)$ bundle with its dual amounts to $Q_i\to -Q_i$,
or equivalently $y\to -y$.
The moduli space of $G_2$ bundles is therefore ${\cal M}={\cal M}'/\Z_2$,
where $\Z_2$ acts on ${\cal M}'$ by $a_3\to -a_3$.
Thus ${\cal M}$ is a weighted projective space ${\bf WP}^2_{1,1,2}$ with
homogeneous coordinates $a_0,$ $a_2$, and $a_3^2$.  This is Looijenga's
theorem for $G_2$.

In the case of an elliptically fibered manifold $Z\to B$, 
for each $b\in B$        one has a weighted projective space $\WP^2_b$ 
parametrizing $G_2$ bundles on the fiber $E_b$ of $Z$ over $b$.
The $\WP^2_b$ fit together as fibers of a $\WP^2$ bundle ${\cal W}$ over $B$.
The objects $a_0,a_1$, and
$a_2$ must now be interpreted
as sections of $\cal O$, $\L^{-2}$, and $\L^{-3}$ over $B$. So 
the homogeneous coordinates $a_0$, $a_2$, and $a_3^2$ of ${\cal W}$ are
sections of ${\cal O}$, ${\cal L}^{-2}$, and ${\cal L}^{-6}$.
Since the fundamental   Casimir invariants of $G_2$ are of degrees 2 and 6,
this confirms for the case $G=G_2$ the claim made in the introduction
about the structure of ${\cal W}$.

The section of ${\cal W}$ coming from the constant section of the summand
${\cal O}$ corresponds to the trivial $G_2$ bundle on each fiber, since
this was true for $SU(3)$.  (A similar statement holds in the other cases
below and will not be repeated.) 

\bigskip\noindent
{\it $Spin(2n+1)$ Bundles}

In the last two cases, $G'$ is a spin group $Spin(2n)$ or $Spin(8)$.
We have not yet discussed Looijenga's theorem for the spin groups (we will
do so  in section 5),
but we will here show that by analogy with the cases considered above,
Looijenga's theorem for $Spin(2n+1)$ and for $F_4$ follows
from the corresponding statement for the simply-laced groups $Spin(2n)$ and
$Spin(8)$.

The fundamental Casimir invariants of $Spin(2n)$ are of degrees $2,4,6,\dots,2n-2$
and $n$.  If $\phi$ is an element of the adjoint representation, regarded
as an antisymmetric $2n\times 2n$ matrix, then the invariants are
$w_k=\Tr\,\phi^{2k}$ for $1\leq k\leq n-1$, of degree $ 2k$,
and the Pfaffian $w'=
{\rm Pf}(\phi)$, which
is of degree $n$.  The outer automorphism of $Spin(2n)$, which generates
$\Gamma=\Z_2$,  acts trivially
on all Casimir invariants except $w'$, which changes sign.

Looijenga's theorem says that the moduli space ${\cal M}'$ of $Spin(2n)$ bundles 
is a weighted projective space ${\bf WP}^n_{1,1,1,1,2,\dots , 2}$
(four weights one and the rest two).  ${\cal M}'$ has homogeneous coordinates
$s_k$, $k=0,\dots,n-1$, in natural correspondence with the invariants
$w_k$, and $s'$, in correspondence with $w'$.  
The coordinates $s_0,s_1,s_2,$ and $s'$ are of weight 1 and the others of
weight 2.  These statements can be proved by methods of section 5.
The $Spin(2n+1)$ moduli space is hence ${\cal M}={\cal M}'/\Z_2$,
where (in view of its action on the Casimir invariants), the generator
of $\Z_2$ leaves the $s_k$ invariant and maps $s'\to -s'$.  Thus ${\cal M}$
is a weighted projective space ${\bf WP}^n_{1,1,1,2,\dots, 2}$ (three $1$'s
and the rest 2's), with homogeneous coordinates
$s_k$ and $(s')^2$ (the weight one coordinates are $s_0,s_1,s_2$).    
This is Looijenga's theorem for $Spin(2n+1)$.

In the case of an elliptically fibered manifold $Z\to B$, the $s_k$ and $s'$ 
become sections of ${\cal L}^{-2k}$ and ${\cal L}^{-n}$,
respectively.  The usual bundle ${\cal W}'$ of weighted projective spaces
(whose fiber above $b\in B$ is the moduli space of $Spin(2n)$ bundles
on the fiber above $b$) has $s_k$ and $s'$ as homogeneous 
coordinates.  These assertions (which are the
$Spin(2n)$ case of the description of ${\cal W}'$ claimed in the
introduction) can again be proved using the methods of section 5.  The analogous
weighted projective space bundle ${\cal W}$ for $Spin(2n+1)$ 
therefore has homogeneous coordinates $s_0,s_1,s_2,\dots,s_{n-1},$
 and $(s')^2$, of  weights $(1,1,1,2,\dots,2)$
and these homogeneous coordinates
are sections of ${\cal L}^{-2k}$ and ${\cal L}^{-2n}$, respectively.
This is the description promised in the introduction
of the weighted projective space bundle for $Spin(2n+1)$.  Note
that the fundamental Casimir invariants of $Spin(2n+1)$ are of degree
$2,4,6.\dots,2n$.

\bigskip\noindent
{\it $F_4$ Bundles}

For $G=F_4$ the story is similar, but slightly more complicated as
$\Gamma$ is the nonabelian triality group $S_3$.

In this case $G'=Spin(8)$.  The Casimirs are $w_k$, $1\leq k\leq 3$,
of degree $2k$, and $w'$, of degree 4.  $\Gamma$ acts trivially
 on $w_0,w_1$, and $w_3$, but $w_2$ and $w'$ transform in an irreducible
two-dimensional representation.  

The moduli space ${\cal M}'$ of $Spin(8)$ bundles on an elliptic curve $E$
is, according to Looijenga's theorem, a weighted projective space
$\WP^4_{1,1,1,1,2}$ where in notation above the weight one homogeneous
coordinates are $s_0,s_1,s_2$, and $s'$, while $s_3$ has weight 2.
Because of the behavior of the Casimirs,
$\Gamma$ acts trivially on $s_0,s_1$, and $s_3$ while $s_2$ and $s'$
transform in an irreducible two-dimensional representation $\rho$.
The ring of invariants  in the representation $\rho$ is a polynomial
ring generated by a quadratic polynomial $A(s_2,s')$ and a cubic polynomial
$B(s_2,s')$.

The $F_4$ moduli space ${\cal M}={\cal M}'/\Gamma$ is hence a weighted
projective space $\WP^4_{1,1,2,2,3}$ with homogeneous coordinates
$s_0,s_1,s_3$, $A(s_2,s')$, and $B(s_2,s')$ of weights $1,1,2,2,3$.  
This is Looijenga's theorem
for $F_4$.

In the case of an elliptic manifold $Z\to B$,
the usual weighted projective space bundle ${\cal W}'$ for $Spin(8)$ has
homogeneous coordinates $s_0,s_1,s_2,s',s_3$ (of weights $1,1,1,1,2$)
which are sections respectively of ${\cal O},{\cal L}^{-2}, {\cal L}^{-4},
{\cal L}^{-4},$ and ${\cal L}^{-6}$.  (These assertions can again be proved
using the methods of section 5.)  Restricting to the $\Gamma$-invariants,
the weighted projective space bundle ${\cal W}$ for $F_4$ therefore
has homogeneous coordinates $s_0,s_1,s_3,A(s_2,s')$, and $B(s_2,s')$,
of weights $1,1,2,2,3$, which are sections respectively of ${\cal O}$,
${\cal L}^{-2}$, ${\cal L}^{-6}$, ${\cal L}^{-8}$, and ${\cal L}^{-12}$.
This is the promised description of the weighted projective space bundle
for $F_4$. Note that the fundamental Casimir invariants of $F_4$ are of
degrees $2,6,8,$ and $12$.

\subsec{Embedding In A Simply-Laced Group}

We will now more briefly explain another way to reduce Looijenga's theorem to the
simply-laced case.

So far, to understand bundles for a non-simply laced group $G$, we have
compared $G$ bundles to $G'$ bundles, where $G'$ is a canonical simply-laced
subgroup of $G$.  An alternative way to reduce Looijenga's theorem to the simply-laced
case uses the fact that every simple and simply-connnected but
non-simply-laced group $G$ can be, in a unique fashion, embedded in a simply-laced group $G''$
in such a way that               $G$ is the subgroup of $G''$ left fixed
by an outer automorphism $\rho$.  (This construction has been used in understanding the
appearance of non-simply-laced gauge groups in $ F$ theory \refs{\many,\aspin}.)
The automorphism $\rho$ will act on the moduli space ${\cal M}''$ of $G''$ bundles on $E$,
and the desired moduli space ${\cal M}  $ of $G$ bundles is a component of the subspace of
${\cal M}''$ left fixed by $\rho$.  In fact, ${\cal M}$ is the component of the
fixed point set that contains the point in ${\cal M}''$ that corresponds to the trivial
flat connection.

According to Looijenga's theorem for $G''$, ${\cal M}''$ is a weighted projective
space whose homogeneous coordinates are in correspondence with the identity
and the Casimir invariants of $G''$.  The desired component of the fixed point
set of $\rho$ has homogeneous coordinates in correspondence with the identity and the
$\rho$-invariant Casimirs of $G''$.  Looijenga's theorem for $G$ is thus a consequence
of Looijenga's theorem for $G''$ together with an appropriate statement about the
action of $\rho$ on the Casimirs of $G''$.  Here is how things work out in the four
cases:

(1) For $G=Sp(n)$, $G''=SU(2n)$ and $\rho$ is the outer automorphism of $G''$ that
acts by ``complex conjugation.''  The Casimirs of $G''$ are $\Tr\,\phi^k$ for
$k=2,3,4,\dots, 2n$.  $\rho$ acts by multiplication by $(-1)^k$, so the $\rho$-invariant
Casimirs are $\Tr\,\phi^{2m}$ for $m=1,2,\dots,n$.  These are also the Casimirs
of $Sp(n)$, and they appear with weight one for both $SU(2n)$ and $Sp(n)$.  Indeed,
this relation between $Sp(n)$ bundles and $SU(2n)$ bundles was already described
at the end of section 2.1.

(2) For $G=G_2$, $G''=Spin(8)$, and $\rho$ is the triality automorphism.  Of the Casimirs
of $G''$, $\Tr\,\phi^2$ and $\Tr\,\phi^6$ are $\rho$-invariant, and the
quartic Casimirs transform non-trivially.  So
the  
$ \rho$-invariant homogeneous coordinates for $G''$ are associated
with the identity, $\Tr\,\phi^2$, and $\Tr\,\phi^6$, the degrees being
$0,2,6$ and the weights $1,1,2$.   These are the right degrees and weights for $G_2$.

(3) For $G=Spin(2n+1)$, $G''=Spin(2n+2)$, and $\rho$ is a ``reflection of one
coordinate'' that reverses the sign of the Pfaffian and leaves fixed the other
Casimirs.  The $\rho$-invariant homogeneous
coordinates for $G''$ are hence associated with the identity and $\Tr \,\phi^{k}$,
$k=2,4,6,\dots, 2n$, and have weights $1,1,1,2,2,\dots,2$.  These are the correct
degrees and weights for $Spin(2n+1)$.

(4) The final example is $G=F_4$.  For this case, $G''=E_6$, and $\rho$ is the
involution that reverses the sign of the Casimirs of degree 5 and 9 and  leaves
fixed the others.  The surviving homogeneous coordinates -- of weights $1,1,2,2,3$ and
associated with Casimirs of degree $0,2,6,8,12$ -- have the appropriate degrees and
weights for $F_4$.

Note that in this construction based on a simply-laced group $G''$ containing $G$,
we want the $\rho$-invariant Casimirs, which are homogeneous coordinates on
a subspace of ${\cal M}''$, while in the previous construction
based on a simply-laced subgroup $G'$, we wanted the $\Gamma$-invariant functions
of the Casimirs (not only the linear functions), 
which are functions on  ${\cal M}'/\Gamma$.

\newsec{Construction Via Del Pezzo Surfaces}

We here explain how to construct the moduli space of $G$ bundles on an elliptic
curve, for certain $G$, via   del Pezzo surfaces.  We first give
a somewhat abstract account and then proceed to explicit formulas.

A del Pezzo surface $S$ is a two-dimensional 
complex surface whose anticanonical line bundle is ample.
The second Betti number $b_2(S) $ of such a surface
ranges from 1 to 9; we set $k=b_2(S)-1$. 
In practice, a smooth del Pezzo surface (we incorporate singularities later) is isomorphic
 either to ${\bf P}^1\times {\bf P}^1$ or to ${\bf P}^2$ with $k$ general points
blown up for $0\leq k\leq 8$.  We will restrict ourselves to the latter
case. (${\bf P}^1\times  {\bf P}^1$ would be an exception for many of
the statements and is not very useful for the applications.)

The intersection form on $L=H^2(S,\Z)$ is isomorphic
over ${\bf Z}$ to the form
\eqn\jumpy{u_0^2-u_1^2-\dots -u_{k}^2.}  
where we can pick coordinates so that 
$u_0$ generates the second cohomology of an underlying ${\bf P}^2$ and
the $u_i$, $i>0$, are exceptional divisors created by blowing up $k$ points.
Note in particular that this gives a basis for $L$ consisting of the classes
of algebraic cycles, so that $H^{2,0}(S)=0$ and every $y\in L$ is the first
Chern class of a holomorphic line bundle ${\cal L}_y$.

Let $T_S$
be the tangent bundle to $S$ and $x=c_1(T_S)$.  
In the coordinates just described
\eqn\hugy{x=3u_0-u_1-\dots -u_{k}.}
(The anticanonical class of ${\bf P}^2$ is $3u_0$, and all 
exceptional divisors created by blowups enter with coefficient $-1$.)
Evidently $x^2=9-k$ and (as $x^2>0$ follows from ampleness of the anticanonical
divisor) one sees the restriction to $k\leq 8$.
Let $L'$ be the sublattice of $L$ consisting of
points $y$  with $x\cdot y=0$. Then the intersection form on $L'$ is
negative definite
and moreover (since all coefficients in \hugy\ are odd)
is even.  Moreover, as $L$ has a unimodular intersection form,
the discriminant of $L'$ is equal to $x^2=9-k$.  

For $k=8$, the intersection form on $L'$ is thus even and unimodular
and of rank eight and so (after reversing the sign of the quadratic form
to make it positive definite) is the
conventional intersection form of the $E_8$ lattice.  More generally,
for any $k\leq 8$, $L'$ can be similarly identified with the root (or coroot)
lattice of 
a simply-laced simple Lie group $G$ of rank $k$ which we will call $E_k$.
For $k=6,7$, $E_6$ and $E_7$
are the groups usually called by those names, while $E_5=D_5$, $E_4=A_4$, 
etc.  In what follows, we mainly consider $E_6,E_7$, and $E_8$.

One can also see in a   similar way the weight lattice of
$E_n$ (which is defined as the dual of the root lattice).  It is 
$L''=L/x\Z$ (where $x\Z$ is the one-dimensional sublattice of $L$ generated
by $x$).  Notice that the pairing on $L$ induces a perfect pairing $L'\otimes
L''\to{\bf Z}$ identifying $L''$ with the dual of $L'$.

The center of $E_n$ is isomorphic to $L''/L'$. Because $x^2=9-k$, this
is isomorphic to $ \Z/(9-k)\Z$.

\bigskip\noindent
{\it A Note On Flat Connections}

Before explaining how to use del Pezzo surfaces to make bundles on
elliptic curves, we first describe a slightly alternative way 
 to think about semistable $G$ bundles on an elliptic curve $E$,
for simply-connected $G$.

Such a bundle 
is equivalent to a flat connection $A$ with values in the maximal torus
$T$.  Now every point  $\bf w$ in the weight lattice $L''$
of  $G$ determines a representation $\rho_{\bf w}$
of $T$
and, by taking the flat connection 
$A$ in the representation $\rho_{\bf w}$, we get a line bundle $\L_{\bf w}$ over $E$.
This line bundle determines a point on the Jacobian of $E$ (which of course
is isomorphic to $E$ itself).

This correspondence ${\bf w}\to {\cal L}_{\bf w}$ determines a homomorphism
from $L''$ to the Jacobian of $E$.  Conversely, from such a homomorphism one
can recover a  $T$-valued flat connection $A$ and therefore
a $G$ bundle.  (Of course, ${\rm Hom}(L'',E)\cong (L'')^*\otimes  E =L'\otimes E$.)

As $L''=L/x{\bf Z}$, a homomorphism from $L''$ to $E$ is the same as a homomorphism
from $L$ to $E$ that maps $x$ to zero.

A homomorphism to $ E$ from the  root lattice $L'\subset L''$ would
determine the ${\cal L}_{\bf w}$'s for ${\bf w}$ a weight of the adjoint
representation, but not for all weights.  So this would determine 
a flat bundle on $E$ with structure group ${\rm ad}(E_k)$ (which is the quotient
of $E_k$ by its center).

The identifications of
$L'$ and $L''$  with the root and weight lattices of $G=E_k$ 
are natural only up to the action of the Weyl group of $E_k$.  But two $T$
bundles over  $E$  that differ by the action of the Weyl group on $T$
determine isomorphic 
$E_k$ bundles.  So homomorphisms from $L'$ or $L''$ to $E$ do determine
well-defined ${\rm ad}(E_k)$ and $E_k$ bundles, respectively, over $E$.

\bigskip\noindent
\subsec{ Bundles From Del Pezzos}

Now we are in a position to explain  how
to build $E_k$ bundles
over an elliptic curve given the appropriate del Pezzo.

The anticanonical bundle  of a del Pezzo surface $S$ has a non-zero
holomorphic section.  The existence of such a section
can be proved via Riemann-Roch (or seen explicitly, as we do below).  In
general, on an $n$-dimensional complex manifold, a section of the anticanonical
bundle vanishes on an $n-1$-dimensional Calabi-Yau submanifold; in the present
case, $n-1=1$, so this Calabi-Yau submanifold is in fact an elliptic curve $E$.

We have already observed that 
every point $y\in L=H^2(S,\Z)$ is 
 the first Chern class of a holomorphic line bundle $\L_y$. Of course
\eqn\hh{\L_{y+y'}=\L_y\otimes \L_{y'}.}
 
Now fix a particular anticanonical divisor $E$, of genus one.
For $y\in L'$, we have $y\cdot x=0$, and this translates to the fact
that the restriction of $\L_y$ to $E$ (which we will simply denote as $\L_y$)
is of degree zero.  So $\L_y$ defines a point in the Jacobian of $E$.
Because of \hh, the map $y\to \L_y$ is a homomorphism from $L'$ to the Jacobian of 
$E$.  According to the Torelli theorem, the moduli space of such pairs $S,E$ is
isomorphic to the set of homomorphisms $L'\to E$ modulo the action on $L'$ of the
Weyl group of $E_k$.

For $k=8$,  $L''=L'$, and this homomorphism determines an $E_8$ bundle over $E$.
  
For $k<8$, a homomorphism from $L'$ would determine only an ${\rm ad}(E_k)$ bundle.
But suppose we are given a distinguished $(9-k)^{th}$ root ${\cal M}$
of the restriction to $E$ of the anticanonical bundle ${\cal L}_x$  of $S$.
Then we can map $L$ to the Jacobian of $E$ by
$y\to {\cal L}_y{\cal M}^{-y\cdot x}$.  This homomorphism maps $x$ to zero
(since ${\cal M}^{-x\cdot x}\otimes {\cal L}_x$ is trivial), so it induces
a homomorphism  from $L''$ to the Jacobian, which will determine an
$E_k$ bundle.

The basic strategy can now be stated.  We will fix an anticanonical divisor
$E$ in a del Pezzo surface $S$, and let the  complex structure of $S$ vary,
keeping fixed $E$  and the $(9-k)^{th}$ root mentioned above.
Every complex structure on $S$ will determine an
$E_k$ bundle on $E$, and by considering a suitable family
of complex structures, we will get the moduli space of $E_k$ bundles
on $S$.  We will consider this construction in some detail for $k=8,7,6$.

Up to this point, we have tried to be conceptual, but in what follows we will
put more emphasis on being explicit.

\def\P{{\bf P}}
\subsec{Construction Of Bundles For $E_6,E_7,E_8$}

\bigskip\noindent{\it $E_8$ Bundles}

The del Pezzo surface with $k=8$ can be constructed as a hypersurface $S$
of degree six in a weighted
projective space $\WP^3_{1,1,2,3}$, with homogeneous coordinates 
$u,v,x,y$.  $S$ may be defined by an equation such as 
\eqn\kurly{y^2=\alpha x^3+\beta xv^4 + \gamma u^6+\delta u^4x +\dots +\epsilon v^6.}
$S$ is a del Pezzo surface simply because the sum of the weights, namely
$w=1+1+2+3=7$, is bigger than the degree of the hypersurface, which is $d=6$.
That it has $k=8$ can be shown, for instance, by computing the Euler characteristic
of $S$ by standard methods.

The anticanonical divisor of $S$  is of degree equal to the difference $w-d=1$.
So for instance the degree one hypersurface $u=0$ is an anticanonical
divisor.  This divisor is given by an
equation of weighted degree six in $v,x$, and $y$:
\eqn\murly{y^2=\alpha x^3+\beta xv^4 +\epsilon v^6+\dots.}
(Only some representative terms are indicated explicitly.)
This equation defines an elliptic curve $E$ in $\WP^2_{1,2,3}$.
By an automorphism of $\WP^2_{1,2,3}$, this equation can be put in a standard
Weierstrass form
\eqn\urly{y^2=4x^3-g_2xv^4-g_3v^6.}
Note that this curve is really an elliptic curve; there is a distinguished
point on it, namely $v=0$.\foot{If one blows up the point $u=v=0$, one gets
a surface $\widehat S$ which is elliptic (the map that forgets $x$ and $y$ is
a map $\widehat S\to {\bf P}^1$ with elliptic fibers) and has a distinguished section $\sigma$
consisting of the exceptional divisor produced in the blow-up.  Conversely, given
such a rational elliptic surface $\widehat S$ with section $\sigma$, a degree one
del Pezzo surface $S$ can be produced by blowing down $\sigma$.  This gives
a natural isomorphism from degree one del Pezzo surfaces to rational elliptic surfaces with
section.}

As explained above, we want to consider the general deformation of
 the complex structure of $S$ keeping
$E$ fixed.
To construct this general deformation, we add to \kurly\ all possible terms of
degree six that vanish at $u=0$, and divide by automorphisms of
$\WP^3_{1,1,2,3}$ that vanish at $u=0$.  The automorphisms in
question are {\it (i)}  $u$ dependent translations of $x,y,z$ such
as $y\to y+\epsilon ux +\epsilon' u^3+\dots$; and {\it (ii)}
rescaling of $u$, $u\to w^{-1}u$ with $w\in \C^*$.  Dividing by {\it (i)}
can be accomplished by suppressing all $u$-dependent terms divisible by the
$v,x$, and $y$ derivatives of the polynomial
\eqn\purly{P=4x^3-g_2xv^4-g_3 v^6-y^2}
whose vanishing defines $E$.

Assuming that $g_3\not= 0$, dividing by symmetries 
of type {\it (i)} can be accomplished by suppressing all $u$-dependent
terms divisible by  $ y,x^2$, or $v^5$.  (For $g_3=0$, the division by
the type {\it (i)} symmetries must be accomplished in a somewhat different way;
this will have important consequences later.)
The general deformation of interest, modulo the automorphisms of
type {\it (i)}, can thus be described by an equation 
\eqn\zurly{\eqalign{y^2=&4x^3-g_2xv^4-g_3v^6+
(\alpha_6u^6+\alpha_5u^5v+\alpha_4u^4v^2+\alpha_3u^3v^3+\alpha_2u^2v^4)\cr
&+ (\beta_4u^4+\beta_3u^3v+\beta_2u^2v^2+\beta_1uv^3)x
 .\cr}}
Nine complex parameters, namely $\alpha_2,\dots,\alpha_6$
and $\beta_1,\dots,\beta_4$, multiply terms that vanish
at $u=0$.  But to construct the desired space
of $S$'s, we must divide by the symmetries of type {\it (ii)}, that
is by $u\to w^{-1}u$.  The result of this last step is that the $\alpha$'s
and $\beta$'s become homogeneous coordinates of
a weighted projective space 
$\WP^8_{1,2,2,3,3,4,4,5,6}$, 
where the weights come from the fact that $\alpha_j$ and
$\beta_j$ are each of weight $j$. 

Every point in the weighted projective space determines a del Pezzo
surface $S$ (possibly with some singularities of $A-D-E$ type).  
The construction in section 4.1  gives for every point
in $\WP^8_{1,2,2,3,3,4,4,5,6}$ an $E_8$ bundle over $E$.  We thus
get a family of such $E_8$ bundles, parametrized by the weighted
projective space.  Note that the weights that have appeared are the ones
 promised by Looijenga's theorem for $E_8$, which
 is indeed equivalent to the statement that the family of bundles
just constructed is the universal family of $E_8$ bundles over $E$.

The foregoing has the following illuminating 
interpretation.  If we simply set to zero all the $\alpha$'s
and $\beta$'s, we get a hypersurface $C(E)$
\eqn\nurly{y^2=4x^3-g_2xv^4-g_3v^6}
which is a weighted cone over $E$.  This hypersurface has
 at $v=x=y=0$ a singularity that is known as
an elliptic singularity of type $\tilde E_8$.    From
this point of view, the quantity $ g_3^2/g_2^3$ (which is invariant under
rescalings of $v$ and determines the $j$-invariant of $E$) is a modulus
of the singularity.  What is considered in \zurly\ is the general unfolding
of the singularity in which the behavior at infinity  is kept fixed. (Or more informally,
the modulus is kept fixed.)  The parameter space of this
unfolding has a $\C^*$ action induced by the $\C^*$ action on $C(E)$ given by
$(v,x,y)\to (w v,w^{2}x,w^3y)$.  $\C^*$ acts on this parameter
space with all weights of the same sign (the sign is generally taken to be negative
in the literature on singularity theory), and the quotient of the parameter
space by this $\C^*$ is a weighted projective space.

The hypersurface \nurly\ is too singular to define a point on the moduli
space of del Pezzo surfaces.  But if one wishes to understand the
fact that the moduli space of $k=8$ del Pezzo surfaces containing a fixed
$E$ is a weighted projective space with the weights found above,
it is very helpful to start with the singular object and consider
its deformations.  We will see analogous phenomena in section
five in the context of stable bundles.

\bigskip\noindent{\it Reduction Of Structure Group And Singularities}

In this construction, one can see at a classical level
the relation between unbroken gauge symmetry and singularities that
has played an important role in studies of string duality in the last
few years.  Namely,  the bundle induced on an elliptic curve
$E$ by its embedding in a $k=8$ del Pezzo surface $S$ has structure
group that commutes with a semi-simple subgroup $H$ of $E_8$ (which will
always be simply-laced) if and only if $S$ contains a singularity of type $H$.

To make  this argument, it is convenient to work not on $S$ but on a smooth
almost del Pezzo surface $X$ made by resolving singularities of $S$ (replacing
possible $A-D-E$ singularities in $S$ by configurations of rational curves).
One reason that this is convenient is that while the cohomology of $S$ drops
when $S$ acquires a singularity, that of $X$ remains fixed and thus has
the structure we described above for a smooth del Pezzo surface.  In considering
a possibly singular del Pezzo surface $S$, we define $L=H^2(X,\Z)$, $L'$ as
the sublattice orthogonal to the anticanonical divisor $x$ of $X$, and $L''=L/x{\bf Z}$.

We first prove that if $S$ has an $A-D-E$ singularity, then the induced
bundle on $E$ commutes with the corresponding $A-D-E$ subgroup of $E_8$.  Let $L_1$
be the sublattice of $L'$ generated by  
 rational curves in $X$ of self-intersection $-2$.
Let $C$ be such a curve.  Since $E$ is an anticanonical divisor, the cohomology class
of $E$ is $[E]=x$.  So the fact that $C\in L'$ implies that
$C\cdot E=0$, which implies that $C$ and $E$ do not intersect. Hence the line bundle
$\O(C)$ determined by $C$ is trivial if restricted to $E$.  Thus in the map from
$L'$ to the Jacobian of $E$, $L_1$ is mapped to zero.  This means that the induced
bundle on $E$ has a stabilizer of the appropriate $A-D-E$ type.  

To justify the last assertion, recall
first that the automorphism group $H'$ of the $E_8$ bundle $V\to E$ has for its Lie algebra
${\cal H}=H^0(E,V)$.  With $V$ being induced by a homomorphism from $L'$ to $E$, $V$ is a sum
of line bundles of degree zero, and $H^0(E,V)$ is a sum of  one-dimensional contributions
from trivial subbundles in $V$.  From what was seen in the last paragraph,
every length squared $-2$ point in $L_1$ corresponds to a trivial line subbundle of
$V$, and hence to a generator of ${\cal H}$.  So if $S$ has a singularity of type $H$,
then all roots of $H$ appear in $H'$ and so $H\subset H'$.

The proof of the converse is longer.
For ${\cal N}$ a line bundle, let $h^i({\cal N})={\rm dim}\,H^i(X,{\cal N})$.
As will become clear, the main step in the argument is to show that  
if ${\cal L}$ is a holomorphic line bundle over $X$ with $c_1({\cal L})^2=-2$,
then $h^0({\cal L})=h^0({\cal L}^{-1})=0$ implies that the restriction of ${\cal L}$
to $E$ is non-trivial.

\def\L{{\cal L}}
For  such an ${\cal L}$, the index of the $\bar\partial$ operator with
values in ${\cal L}^{\pm 1}$ is zero, so in particular
\eqn\jkl{\eqalign{
                   0 & = h^0(\L^{-1} ) - h^1({\cal L}^{-1})+h^2({\cal L}^{-1}).\cr}} 
By Serre duality,  $h^2(\L^{-1})=h^0(K\otimes
\L)$.  But vanishing of $h^0(\L)$ and existence
of a holomorphic section $s$ of $K^{-1}$ (which vanishes on $E$) imply
vanishing of $h^0(K\otimes \L) $.  (For instance,
multiplication by $s$ would map a non-zero holomorphic section of $K\otimes \L$ to 
a non-zero holomorphic section of $\L$.)  So $h^0(\L)=h^0(\L^{-1})=0$ implies
$h^1({\cal L}^{-1})=0$ and hence by Serre duality $h^1(K\otimes {\cal L})=0$.

Next look at the exact sequence of sheaves
\eqn\hgj{0\to K\otimes \L \to \L \to \L|_E\to 0,}
where the first map is multiplication by $s$ and the second is restriction to 
$E$.  The associated long exact sequence of cohomology groups reads in part
\eqn\grr{\dots \to H^0(X,\L)\to H^0(E,\L) \to H^1(X,K\otimes \L)\to \dots.}
Thus, if $h^0(\L)=h^0(\L^{-1})=0$, then from the above $h^1(K\otimes {\cal L})=0$,
so the exact sequence implies that $H^0(E,{\cal L})=0$.  But this implies that
the restriction of $\L$ to $E$ is non-trivial.

Now, let $L_0$ be the sublattice of $L'$ corresponding to line bundles
whose restriction to $E$ is trivial.  The intersection form on $L_0$ is
even, and the sublattice $L_1$ of $L_0$ generated by the points of length squared
$-2$ is the root lattice of some product of $A-D-E$ groups.  From what we have
just proved, if $y\in L_1$   has $y^2=-2$, then ${\cal L}_y$ or ${\cal L}_y^{-1}$
has a holomorphic section.  Such a section vanishes on a holomorphic curve
$C_y$ with self-intersection number $-2$.  $C_y$ does not meet $E$ (since
$y\cdot x=0$) so the anticanonical bundle of $X$ is trivial when restricted
to $C_y$.  If we go back to $S$, therefore, the $C_y$ are all blown down,
producing the promised singularity of type $L_1$.

\bigskip\noindent{\it $E_7$ Bundles}

Now we consider in a precisely similar way the case $k=7$.  A $k=7$
del Pezzo surface $S$ can be constructed as a hypersurface of degree four
in a weighted projective space ${\bf WP}^3_{1,1,1,2}$, with homogeneous
coordinates $u,v,x,y$.  Such a hypersurface is described
by an equation of the general form
\eqn\dofy{y^2=ax^4+bv^4+cu^4+\dots.}
The difference between the sum of the weights and the degree of the hypersurface
is $1+1+1+2-4=1$, so the degree 1 hypersurface $u=0$ is an anticanonical
divisor.  This divisor is in fact a genus one curve $E$ in a weighted
projective space $\WP^2_{1,1,2}$ with homogeneous coordinates $v,x,y$.
By an automorphism of the weighted projective space, $E$ can be put in the
form 
\eqn\jofy{y^2=4vx^3-g_2xv^3-g_3v^4.}
When put in this form, $E$ is naturally an elliptic curve, with distinguished
point $p$ given by $(v,x,y)=(0,0,1)$, and the line bundle ${\cal O}(p)$ is
a square root of the restriction to $E$ of the anticanonical bundle
of $S$.

If \jofy\ is regarded as defining a hypersurface in ${\bf WP}^3_{1,1,1,2}$,
then that hypersurface is a cone over $E$ and has a singularity of type
$\tilde E_7$ at $x=y=v=0$.  The $j$-invariant of $E$ is a modulus of this
singularity.  The universal unfolding of the $\tilde E_7$ singularity preserving
this modulus (or more precisely the behavior at infinity)
is made by adding to \jofy\ terms that vanish at $u=0$
modulo  $u$-dependent translations of $v,x,y$,
These    translations can be taken into account
by excluding deformations of the equation
divisible by $y,x^3$, or $vx^2$.  (This ``gauge fixing condition''
can be made uniformly, independent of $g_2$ and $g_3$, an important
difference from the $E_8$ case.) The universal deformation thus looks
like
\eqn\bofy{y^2=4vx^3-g_2xv^3-g_3v^4+u(\alpha_1 v^3+\alpha_2xv^2)
+u^2(\beta_1v^2+\beta_2xv+\beta_3x^2) +u^3(\gamma_1v+\gamma_2x)+u^4\delta.}
  
The moduli space of  $S$'s containing the given $E$ is obtained by dividing
by the additional symmetry $u\to w^{-1}u$.  Under this transformation,
the $\alpha$'s have weight one, the $\beta$'s have weight two, the $\gamma$'s
have weight three, and $\delta$ has weight four.
The requisite moduli space of $S$'s is thus a weighted projective space
$\WP^7_{1,1,2,2,2,3,3,4}$.  The construction of  section 4.1 gives
a family of $E_7$ bundles over $E$ parametrized by this weighted projective
space.  The content of Looijenga's theorem for $E_7$ is that this family
is the universal $E_7$ bundle over $E$, so that the moduli space of such $E_7$
bundles is the weighted projective space that we just encountered. 

\bigskip\noindent{\it $E_6$ Bundles}

$E_6$ is treated similarly. A $k=6$
del Pezzo surface $S$ can be constructed as a hypersurface of degree four
in an ordinary projective space  ${\bf P}^3$, with homogeneous
coordinates $u,v,x,y$.  Such a hypersurface is described
by a homogeneous cubic equation in $u,v,x,$ and $y$.
The difference between the sum of the weights and the degree of the hypersurface
is $1+1+1+1-3=1$, so the degree 1 hypersurface $u=0$ is an anticanonical
divisor.  This divisor is in fact a genus one curve $E$ in an ordinary
projective space ${\bf P}^2$, with homogeneous coordinates $v,x,y$.
By an automorphism of the  projective space, $E$ can be put in the
form 
\eqn\jjofy{vy^2=4x^3-g_2xv^2-g_3v^3.}
This way of writing $E$ exhibits it as an elliptic curve with distinguished
point $p$ given by $(v,x,y)=(0,0,1)$, and the line bundle ${\cal O}(p)$ 
is a cube root of the restriction to $E$ of the anticanonical bundle of $S$.

If \jjofy\ is regarded as defining a hypersurface in ${\bf P}^3$,
then that hypersurface is a cone over $E$ and has a singularity of type
$\tilde E_6$ at $x=y=v=0$.  The $j$-invariant of $E$ is a modulus of this
singularity.  The universal unfolding of the $\tilde E_6$ singularity preserving
this modulus (in the sense described earlier) 
is made by adding to \jofy\ terms that vanish at $u=0$
modulo  $u$-dependent translations of $v,x,y$,
These    translations can be taken into account
by excluding deformations of the equation
divisible by $y^2, x^2$ and $vy$.  (It is again significant that this
``gauge fixing'' can be made universally, independent of $g_2$ and $g_3$.) 
The universal deformation thus looks like
\eqn\bofy{vy^2=4x^3-g_2xv^2-g_3v^3+u(\alpha_1 v^2+\alpha_2xv+\alpha_3xy)
+u^2(\beta_1v+\beta_2x+\beta_3y) +u^3\gamma.}

The moduli space of of $S$'s containing the given $E$ is obtained by dividing
by the additional symmetry $u\to w^{-1}u$.  Under this transformation,
the $\alpha$'s have weight one, the $\beta$'s have weight two, and $\gamma$
has weight three.
The requisite moduli space of $S$'s is thus a weighted projective space
$\WP^6_{1,1,1,2,2,2,3}$.  The construction of the last subsection gives
a family of $E_6$ bundles over $E$ parametrized by this weighted projective
space.  The content of Looijenga's theorem for $E_6$ is that this family
is the universal $E_6$ bundle over $E$, so that the moduli space of such $E_6$
bundles is the weighted projective space that we just encountered. 

\subsec{Bundles Over Elliptic Manifolds}

Now we wish to consider bundles over an elliptically fibered manifold
$Z\to B$ with a section $\sigma$ (whose normal bundle we call ${\cal L}^{-1}$).   
For $b\in B$, let $E_b$ be the elliptic curve over $b$.
For each gauge group $G$, there is a weighted projective space
bundle ${\cal W}\to B$ whose fiber over $b\in B$ is the moduli space
of $G$ bundles over $E_b$.  We want to obtain a simple description of 
${\cal W}$ for $G=E_6$ or $E_7$, and to see how the existence of such
a simple description is obstructed for $E_8$.

The basic idea is to make the above construction with parameters.
The only subtlety is that one must give a description of the fibration
$Z\to B$ which is adapted to the choice of $G$.  For instance, for $G=E_6$,
we regard $Z$ as usual as a hypersurface in a $\P^2$ bundle over
$B$, which is obtained by projectivizing ${\cal O}\oplus {\cal L}^2\oplus
{\cal L}^3$, with respective homogeneous coordinates $v,x,y$.
The Weierstrass equation defining $Z$ reads
\eqn\grr{vy^2=4x^3-g_2xv^2-g_3v^3,}
with $g_2$ and $g_3$ being now sections of ${\cal L}^4$ and ${\cal L}^6$.
To obtain the desired ${\cal W}$, we simply make fiberwise the construction
given above.  We embed the $\P^2$ bundle over $B$ in a $\P^3$ bundle, obtained
by projectivizing ${\cal L}^6\oplus{\cal O}\oplus {\cal L}^2\oplus {\cal L}^3$,
with respective homogeneous coordinates $u,v,x,y$ (the choice of exponent $6$
for $u$ is convenient but not essential); we interpret \grr\ as  defining
a singular hypersurface in this bundle (a sort of cone over $Z$).
We consider deformations of this hypersurface of same form as before:
\eqn\bbofy{vy^2=4x^3-g_2xv^2-g_3v^3+u(\alpha_1 v^2+\alpha_2xv+\alpha_3xy)
+u^2(\beta_1v+\beta_2x+\beta_3y) +u^3\gamma.}
The $\alpha_i$, $\beta_j$, and $\gamma$ are now interpreted as homogeneous
coordinates for the desired weighted projective space ${\cal W}$; they are sections
of line bundles which are determined by requiring that \bbofy\ makes sense as an equation
with values in ${\cal L}^6$.
${\cal W}$ is therefore a ${\WP}^6_{1,1,1,2,2,2,3}$ bundle whose
successive homogeneous coordinates are sections   
of ${\cal O},{\cal L}^{-2},{\cal L}^{-5},{\cal L}^{-6},{\cal L}^{-8},{\cal L}^{-9},$
and ${\cal L}^{-12}$.  This is the expected form of ${\cal W}$ for $E_6$.
Note that the Casimir invariants of $E_6$ are of degree $2,5,6,8,9,$ and 12.

$E_7$ is treated similarly.  The only difference is that here we regard
$Z$ as a hypersurface in a $\WP^2_{1,1,2}$ bundle over $B$, obtained by
projectivizing ${\cal O}\oplus {\cal L}^2\oplus {\cal L}^3$, with homogeneous
coordinates $v,x,y$ of weights $1,1,2$, and Weierstrass equation
\eqn\hrr{vy^2=4vx^3-g_2xv^3-g_3v^4.}
We embed the $\WP^2_{1,1,2}$ bundle
 in a $\WP^3_{1,1,1,2}$ bundle over $B$, obtained by
projectivizing ${\cal L}^6\oplus {\cal O}\oplus {\cal L}^2\oplus {\cal L}^3$,
with homogeneous coordinates $u,v,x,$ and $y$.  
\hrr\ describes a singular hypersurface in this larger bundle.  We consider
deformations of this hypersurface of the form
\eqn\bofy{y^2=4vx^3-g_2xv^3-g_3v^4+u(\alpha_1 v^3+\alpha_2xv^2)
+u^2(\beta_1v^2+\beta_2xv+\beta_3x^2) +u^3(\gamma_1v+\gamma_2x)+u^4\delta.}
The $\alpha_i$, $\beta_i$, $\gamma_i$, and $\delta$ are now
interpreted as homogeneous coordinates for the desired weighted projective
space bundle ${\cal W}$, and are again  sections of line bundles that are determined
by requiring that  \bofy\ makes sense as an equation with values in ${\cal L}^6$.
  ${\cal W}$ is thus a $\WP^7_{1,1,2,2,2,3,3,4}$
bundle whose successive homogeneous coordinates are sections of 
${\cal O}\oplus \L^{-2}\oplus \L^{-6}\oplus \L^{-8}\oplus \L^{-10}
\oplus \L^{-12} \oplus \L^{-14} \oplus \L^{-18}.$ Note that the Casimir
invariants of $E_7$ are of degree $2,6,8,10,12,14$, and 18.

One cannot by these methods obtain a description of the weighted projective
space bundle for $E_8$ as the projectivization of a sum of line bundles
(and we believe that there is no such description).  The reason is that
there is no universal way, independent of $g_2$ and $g_3$, to parametrize
the unfolding of the $\tilde E_8$ singularity.  For $g_3\not=0$, a
parametrization is given in \zurly, and one could similarly pick a parametrization
for $g_2\not=0$, but there is no uniform choice. (The natural parametrization
of the weighted projective space bundle for $g_3\not=0$ is related to an
analogous natural parametrization for $g_2\not= 0$ by a nonlinear automorphism
of the weighted projective spaces.)  When one considers elliptic manifolds
over a base $B$ of dimension at least two, one meets cusp fibers with $g_2=g_3=0$; near
such  a fiber the description of bundles is really different.

\subsec{ Relation To  Duality In Eight Dimensions}

We conclude this section with a discussion of the role of the $\tilde E_8$
singularity in string duality.

The basic duality between the heterotic string and $F$ theory maps the
heterotic string on a two-torus $E$ to $F$ theory on an elliptically-fibered
K3.  A heterotic string vacuum on a two-torus $E$ 
is described by a family of conformal field theories depending on 18 complex parameters 
(plus more one real parameter, the string
coupling constant, which determines the K\"ahler class in $F$ theory).  
If one asks for unbroken $E_8\times E_8$ gauge symmetry,
16 complex parameters, which parameterize the flat $E_8\times E_8$ bundles,
are fixed.  The remaining two complex parameters are the complex
structure and complexified K\"ahler class of $E$.

According to Morrison and Vafa (see section  2 of the second paper in \mv),
this two-parameter heterotic string locus corresponds in terms of $F$ theory
to the family of elliptically-fibered K3's described (in affine coordinates)
by the following explicit equation:
\eqn\uhb{y^2=4x^3-g_2 t^4x +t^5-g_3 t^6 +t^7.}
Here $g_2$ and $g_3$ are the two parameters and $t$ is an affine
coordinate on the  base $\P^1$ of the elliptic fibration.  (The fiber over $t=\infty$
should thus be included.)  For given $t,g_2,g_3$, \uhb\ is a Weierstrass
equation defining the elliptic fibration. 

Morrison and Vafa further consider the case in which, on the heterotic
string side, the area $\rho$ of $E$ becomes large, keeping the complex structure
fixed.  They show that this corresponds to $g_2$ and $g_3 $ becoming
large with    fixed $g_2^3/g_3^2$.    
We can enter this region taking $g_2\to  c^2g_2$, $g_3\to c^3g_3$, where $c $ is to
become large.  It is convenient to also rescale $ t$ by $t\to c^{-1}t$. 
In this way, we can actually take the limit as $c\to \infty$.  This
corresponds to decompactification of the heterotic string, with  fixed
complex structure on $E$ but area going to infinity.  Such decompactification
of the heterotic string thus corresponds in $F$ theory to the singular
K3 fibration described in affine coordinates by    the $c=\infty$ limit
of \uhb, or
\eqn\nuhb{y^2=4x^3-g_2t^4x-g_3t^6.}
We see that this has two $\tilde E_8$ singularities, one at 
$x=y=t=0$ and one at $x=y=0$, $t=\infty$.  (To see the latter singularity,
set $t=1/t'$, $x=x'/(t')^4$, $y=y'/(t')^6$.) 

A surface  with these two $\tilde E_8$ singularities does not correspond
to a point on the moduli space of  vacua. This is clear on the heterotic
string side because such a  point can only be reached by  decompactification.
In the natural metric on the moduli space, decompactification is at infinite
distance; one gets a complete metric on the moduli space without including it.
However, the singular surface with
the two $\tilde E_8$ singularities is a convenient
starting point in understanding the part
of the moduli space where classical geometry is a good approximation,
that is, the part where the two-torus has large area.
This is essentially what we have done in using the unfolding of the elliptic
singularity to describe $E_8$ bundles.

\bigskip
\noindent{\it Stable Version And Behavior In Families}

Actually, the degeneration of a K3 surface to produce two
$\tilde E_8$ singularities, as just described, does not correspond
to a stable point on the moduli space of K3 surfaces.  The
stable version is as follows.

Eqn. (4.22) describes a K3 surface $X$ 
that is elliptically fibered
over a base $B'$ which is a copy of ${\bf P}^1$ (parametrized by $t$) with
a section $\sigma$.
There are    24 points $P_i$ on the $t$ plane over
which the fiber degenerates.

To produce the two $\tilde E_8$ singularities, 12 of the
$P_i$ move to $t=0$ and the other 12 to $t=\infty$.  Near this limit,
as explained by Morrison and Vafa,
the hyper-K\"ahler metric on $B'$ (which we identify with the
section of the elliptic K3) looks like a long cigar with 12 of the $P_i$ at
each end; the limit $c\to\infty$ is the limit in which the cigar becomes
infinitely long.  From the point of view of    complex geometry,
the stable version of such a degeneration is that in which
the $B'$ splits into two components $H_1$ and $H_2$  
(each isomorphic to ${\bf P}^1$
and sharing a point $Q$ in common) with  12 $P_i$ in each component.
In this picture,  $X$ degenerates to a union  of two rational elliptic
surfaces $U_1$ and $U_2$ glued together on an elliptic curve $E$.
We write this as $X=U_1\cup_E U_2$.  In terms of the projection
$\pi:X\to B'$, one has $U_1=\pi^{-1}(H_1)$, $U_2=\pi^{-1}(H_2)$, and
$E=\pi^{-1}(Q)$.  
The section $\sigma$ of $X$ splits up into sections $\sigma_1$ and
$\sigma_2$ of $U_1$ and $U_2$; by blowing down the $\sigma_i$
we can map the $U_i$ to del Pezzo surfaces $W_i$, glued together
along $E$.

In view of what has been said by Morrison and
Vafa and above, the correspondence with the heterotic string is simply
that $E$ is the elliptic curve on the heterotic string side, and
the two $E_8$ bundles $V_1$ and $V_2$ over $E$ are coded in the 
complex structures of $W_1$ and $W_2$.  

Now, it is easy to extend this formulation to families.  On the
heterotic string side we replace $E$ by an elliptic $n$-fold
$\pi:Z\to B$.  Assuming that the Kahler metric on $B$ is large so
that we may make fiberwise duality with $F$ theory, this  corresponds in
$F$ theory to an $n+1$-fold that maps to $B$ with K3 fibers,
and maps with elliptic fibers to a certain ${\bf P}^1$ bundle
$B'$ over $B$.   Now if we also take the area of the fibers of $\pi:Z\to B$
to be large, to reduce to classical geometry, and also blow down
the sections, then as was just
seen each K3 fiber of the map $X\to B$ will degenerate to a union
of two del Pezzo surfaces, glued together along an elliptic curve.
Globally, $X$ will degenerate to a union of two $n+1$-folds glued
over an $n$-fold (which is also fibered over $B$); in fact, it degenerates
to  $X=W_1\cup_ZW_2$, where $Z$ is the Calabi-Yau
$n$-fold seen on the heterotic string side, and $W_1$ and $W_2$ are
bundles of del Pezzo surfaces over $B$.\foot{More exactly,
as in the case of a single K3, one gets bundles of rational
elliptic surfaces that can be blown down to make bundles
of del Pezzo surfaces.}  

Fiberwise application of the correspondence between $E_8$ bundles
on an elliptic curve and del Pezzo surfaces shows that the complex
structures of $W_1$ and $W_2$ code the isomorphism classes
of the restriction to each fiber of the $E_8$ bundles $V_1$ and $V_2$ over $Z$.   
                                                   
As we saw in section 2 in the case of $SU(n)$  bundles, given a bundle
$V\to Z$, $V$ is not uniquely determined, in general, by a knowledge
of the isomorphism class of its restriction to each fiber. 
One can make certain twists by a line bundle on the spectral cover.
Similarly for gauge groups other than $SU(n)$, we expect
an abelian variety, generalizing the Jacobian of the spectral cover,
to enter in the parametrization of bundles.
As explained in section   2.4, the additional data should in $F$ theory
show up (along with other things) in the intermediate Jacobian of $X$, 
$J_X=H^3(X,{\bf R})/H^3(X,{\bf Z})$.

When $X$ degenerates to $W_1\cup_ZW_2$, the intermediate
Jacobian of $X$ splits off factors isomorphic to the intermediate
Jacobians $J_{W_i}=H^3(W_i,{\bf R})/H^3(W_i,{\bf Z})$.  
Note that, as $H^{3,0}(W_i)=0$,
the $J_{W_i}$ are abelian varieties (as is $J_X$ when $\dim\,B>1$, but
generally not when $\dim\,B=1$).  As the $E_8$ bundles $V_i\to Z$ are closely
related to the structure of the $W_i$, it is 
natural to believe that the intermediate Jacobians $J_{W_i}$, for $i=1,2$,
contain the additional information  necessary to determine
the $V_i$.

As evidence for this interpretation of the facts, 
we will show that for the case
that $B={\bf P}^1$, the $J_{W_i}$ have the appropriate dimension and in
fact the appropriate tangent space.  

Let $W=W_i$ for $i=1$ or 2.  The tangent space to the space of deformations
of the complex structure
of $W$ that preserve the existence of the divisor $Z$ (other deformations
involve a change in the complex structure of the heterotic string
manifold $Z$ and are not related to bundles on $Z$) is
${\cal T}=H^1(W,T_W\otimes {\cal O}(-Z))$.  Here $T_W$ is the tangent space
to $W$ and ${\cal O}(-Z))$ is the line bundle whose holomorphic sections
are holomorphic functions that vanish on $Z$.  Because ${\cal O}(-Z)$ is
the canonical bundle of $W$, Serre duality says that ${\cal T}^*=
H^2(W,T_W^*)=H^{1,2}(W)$.  But $T_J=H^{1,2}(W)$ is the tangent space
to the intermediate Jacobian $J_W$.  This equality of dimensions between
${\cal T}$ and $T_J$, and in fact the duality between them, is expected
in view of the complex symplectic structure (and hyper-Kahler structure)
of the moduli space of bundles on $Z$, for the case that $B={\bf P}^1$
and $ Z$ is a K3 surface.

Mathematically, it is possible to ``twist'' the intermediate Jacobian
of $X$ by an arbitrary integral class $\alpha\in H^{2,2}(X)$.  (The
twisted intermediate Jacobians are components of the Deligne
cohomology of $X$.)   In physical
terminology, using the language of $M$ theory, making such a twist 
means taking the four-form field strength $G$ of eleven-dimensional
supergravity to represent the cohomology class $\alpha$.   It has been
shown by K. and M. Becker [9] that in the case that $X$ is
a four-fold, introducing $\alpha$ in $M$ theory
(and hence also, with some restriction, in $F$ theory) is compatible
with space-time supersymmetry provided that $\alpha$ is a primitive
element of $H^{2,2}(X)$.  ($\alpha$ is defined to be primitive
if its contraction with the Kahler class vanishes in cohomology.
For $X$ a three-fold, this behavior 
implies that the image of $\alpha$ vanishes
in real cohomology so that $\alpha$ is a torsion class.)  
 This gives physical models, with space-time
supersymmetry, in which twisted
versions of the intermediate Jacobian of $X$ enter.

It is natural to conjecture that twists
by those elements  $\alpha \in H^{2,2}(X)$ that are derived (when $X$ reduces
to $W_1\cup_Z W_2$) from an element $\beta\in H^{2,2}(W)$ have the following
interpretation.  We saw in section two that the moduli space of
$SU(n)$ bundles on $Z\to B$ that have a given restriction to each fiber
is not necessarily connected, but (depending on the Picard group of
the spectral cover $C\to B$) may have different components.   
We conjecture that a similar result holds for $E_8$ bundles and that
the different components of bundles with a fixed restriction to each
fiber correspond in $F$ theory to the twists of $J_W$ by different
primitive elements of $H^{2,2}(W)$ or (more physically) to the
different values of the cohomology class of the $G$ field.

\newsec{Uniform Approach To Construction Of Bundles}

Having reached this far, the reader may yearn for a more uniform approach
to the problem.  In sections 2,3, and 4, we presented different approaches
to understanding the moduli space of $G$ bundles on an elliptic curve $E$;
each approach was effective for a particular class of $G$'s.  Is there
not a more uniform approach?

In this section, we will explain an approach which does in fact work uniformly
for all simple, connected, and simply-connected $G$.\foot{It is even possible
to extend the discussion to non-simply-connected $G$, by using different
parabolic subgroups of $G$, but we will not make this generalization in
the present paper.}  The inspiration for this construction comes in part from
the construction via del Pezzo surfaces in section 4.  We saw there that
to understand the moduli space of ``good'' del Pezzo surfaces, which have at worst
singularities of $A-D-E$ type, it helps to start with a ``bad'' surface
with an $\tilde E_k$ singularity.  The good surfaces are conveniently constructed
as deformations of a bad one.

We will take a similar approach to bundles.  
Though we are interested mainly in semistable bundles, we will find
a distinguished unstable bundle which has the property that the semistable bundles
are conveniently constructed as its deformations.

We explain first the idea
for $G=SU(n)$.  
A  rank $n$ vector bundle $V\to E$ of degree zero is unstable if and only if it contains 
a sub-bundle $U$ of positive degree.   Such a $U$ fits into an exact sequence
\eqn\gruff{0\to U\to V\to U'\to 0} with some $U'$.
To make $V$ just barely unstable, we pick $U$ to be of degree 1 and $U'$ to be
of degree $-1$. We also assume that $U'$ and $U$ are themselves both stable. 

The determinant of $U$ is a rank one bundle of the form ${\cal O}(p)$ for some
point $p\in E$ which we will keep fixed.  The determinant of $U'$ is then
${\cal O}(p)^{-1}$.  Now we specialize to the case that $E$ is of genus one.
In this case, a Riemann-Roch argument (using stability of 
$U^*\otimes U'$)
shows that the sequence \gruff\ splits,
so that in fact $V=U\oplus U'$.  

Also, for $E$ of genus one, $U$ is uniquely determined up to translation on $E$.
In fact, up to isomorphism there is for each $k\geq 1$ 
a unique stable bundle $W_k$ of determinant ${\cal O}(p)$ for any given point
$p\in E$.  For $k=1$,
$W_1={\cal O}(p)$, and if $W_k$ is known then $W_{k+1}$ can be constructed
inductively as the unique non-split extension
\eqn\hux{0\to {\cal O}\to W_{k+1}\to W_k\to 0.}
The $W_k$ will appear extensively in what follows.
The dual of $W_k$, which we write as $W_k^*$, is the unique rank $k$ stable bundle
over $E$ of degree $-1$ and determinant ${\cal O}(p)^{-1}$.

So for our starting point, we take the unstable bundle
\eqn\ubbu{V=W_k\oplus W_{n-k}^*,}
with some  $k$ in the range $1\leq k\leq n-1$.  This is, up to translation
on $E$, the unique minimally unstable bundle with summands of the chosen dimension.

The decomposition \ubbu\ of $ V$  enables one to define a group $H\cong {\bf C}^*$ of
automorphisms of $V$ that acts by scalar multiplication on $W_k$ while
acting trivially on $W_{n-k}^*$.
The structure group of $V$ reduces to the  subgroup of $SU(n)$ consisting
of block diagonal matrices of the form
\eqn\jbn{\left(\matrix{ * & 0 \cr 0 & *\cr}\right),}
where the upper left block is $k\times k$ and the lower right block
is $(m-k)\times (m-k)$.  $H$ acts as a group of automorphisms of
 the Lie algebra of $SU(n)$.  The diagonal blocks transform under $H$ with
 weight  0, the upper right block has weight 1, and the lower left
 block has weight $-1$.
 
 To first order, a deformation of $V$ is determined by an element of 
 $T= H^1(E,{\rm ad}(V))$, where ${\rm ad}(V)$ is the adjoint bundle
 derived from $V$.  Using the facts noted in the last paragraph, $T$
 can be decomposed into pieces of weight $1,0$, and $-1$
 under $H$, as follows:
 
 (1) The weight 1 piece is $H^1(E,{\rm Hom}(W_{n-k}^*,W_k))=H^1(E,W_{n-k}
 \otimes W_k)$.  The bundle $W_{n-k}\otimes W_k$ is  a stable bundle of positive
 degree.  On a curve of genus one, any stable bundle of positive degree has vanishing
 $H^1$, so the weight 1 piece vanishes.
 
 (2) The weight 0 piece is the tangent space  to the space of
 deformations of $W_k$ and $W_{n-k}^*$ that 
 preserve  the decomposition $V=W_k\oplus W_{n-k}^*$ (and the fact that $V$ 
 has trivial determinant).  For reasons already noted above, the only
 such deformation is a motion of the point $p$ such that the determinant
 of $W_k$ is ${\cal O}(p)$.  So the weight 0 piece is one-dimensional and can be viewed
 as the tangent space to $E$ at $p$.
 
 (3) Finally, the weight $-1$ piece is  $H^1(E,{\rm Hom}(W_k,W_{n-k}^*))=H^1(E,
 W_k^*\otimes W_{n-k}^*)$.  This will play the starring role in what follows.
 
 Let us compute the dimension of the weight $-1$ deformation space.  By Riemann-Roch,
and the fact that the bundle is semi-stable,
 this is minus the degree of the bundle $W_k^*\otimes W_{n-k}^*$.  In what follows,
 we will many times have to compute the degrees of such tensor products.
 If the degree $-1$ bundle $W_k^*$ were the sum of a degree $-1$ line bundle and
 $k-1$ line bundles of degree $0$, then in $W_k^*\otimes W_{n-k}^*$, we would
 see one summand of degree $-2$, $n-2$ of degree $-1$, 
 and the rest of degree 0.  The total degree would thus be $-n$.
 Actually, although the $W^*$'s are not such direct sums, the computation just
 performed can be justified using exact sequences such as \hux\ and its dual, so the
 degree is really $-n$.
 
 So the $-1$ part of the weight space is of dimension $n$.  Now, a deformation of
 $V$ by an element $\alpha\in H^1(E,W_k^*\otimes W_{n-k}^*)$ produces a bundle $V'$ with an
 exact sequence 
 \eqn\rogo{0\to W_{n-k}^*\to V'\to W_k\to 0.}
 The existence of such a sequence does not contradict stability of $V'$ since
 $W_{n-k}^*$ is of negative degree.  In fact, a straightforward argument shows that bundles
 on an elliptic curve constructed by non-trivial extensions of the form \rogo\ 
 are all semistable; see \fmw.
 
 We want to consider only deformations of negative weight, suppressing the weight 0 part
 of the deformation space.  This is analogous to
 considering only deformations of fixed $j$ in the construction via del Pezzo
 surfaces.
 
 \bigskip\noindent
 {\it Structure Of The Deformations}
 
 In an explicit description of bundles by an open covering and transition functions,
 the transition functions for the extension $V'$ look like
 \eqn\omigo{\left(\matrix{ * & 0 \cr \alpha & * \cr}\right),}
 where the upper left and lower right blocks are the transition functions
 for $W_k$ and $W_{n-k}^*$.  Because $\alpha$ only appears in the lower
 left block, the $\alpha$-dependent terms
 in the cocycle condition for such transition functions are linear in $\alpha$.  That
 is why, even though in many problems in geometry $H^1$ controls only the
 linearized deformations, the choice of $\alpha\in H^1(E,{\rm Hom}(W_k,W_{n-k}^*))$ produces
 an actual extension $V'$ as described in \rogo, and not just a first order
 approximation to one.

 Closely related is the fact that it does not matter if $\alpha$ is ``big'' or ``small,''
 in the sense that if $\alpha$ is replaced by $t\alpha$ with $t\in {\bf C}^*$, then
 the bundle $V'$ is unchanged, up to isomorphism.  This is so because $t$ can be
 scaled out by using the scaling by $H$.  The point is that the ``unperturbed'' bundle 
 $V$ has an automorphism group $H\cong {\bf C}^*$ that is ``broken'' by the perturbations.
 To construct the moduli space of bundles that can be built by perturbations that
 do not preserve the $\C^*$, one must divide the space of first order deformations
 by $\C^*$.
 
 If $T_-=H^1(E,W_k^*\otimes W_{n-k}^*)$ is the space of negative weight
 first order deformation, then the family of bundles that can be constructed
 via such deformations is naturally parametrized by
 ${ M}=\left(T_--\{0\}\right)
 /{\bf C}^*$
 (which we abbreviate below as $T_-/{\bf C}^*$).
 Since $T_-$ is a copy of ${\bf C}^n$ and the $\C^*$ acts by scalar
 multiplication, $M$ is a copy of $\P^{n-1}$.
 
 We have already seen in section 2 that the moduli space ${\cal M}$ of
 semistable $SU(n)$ bundles
 on $E$ is a copy of $\P^{n-1}$.  This raises the question of whether one can naturally
 identify $M$ with ${\cal M}$.  It will be proved elsewhere that this is so \fmw\
 (for any $k$ in the range from 1 to $n-1$).  In other words, $M$ can be identified
 with the projective space predicted by Looijenga's theorem in the case of $SU(n)$.
 
 \bigskip\noindent
 {\it Framework For Generalizations}
 
 In the rest of this section, we will show how to make an analogous construction
 for any simple, connected and simply-connected Lie group $G$.  In each case,
 we find a distinguished, slightly unstable $G$ bundle $V$ over $E$ with the property that
 the semistable $G$ bundles over $E$ all arise naturally as deformations of $V$.
 Before considering specific examples, we pause for some useful generalities.
  
 A subgroup of $SL(n,{\bf C})$ is called parabolic if (perhaps after conjugation) it
 contains the diagonal and upper triangular matrices:
 \eqn\nippo{\left(\matrix{ * & * &*& \dots& * \cr
                           0 & * & * &\dots &* \cr
                           {}& \dots & {} &{} &{}\cr
                            0& 0 &{} &* &*\cr
                             0 & 0 &\dots & 0 &*\cr }\right).}
 Any such group $P$ has the property that $SL(n,{\bf C})/P$ is compact. 
The existence of the exact sequence \gruff\ is equivalent to a reduction of
the structure group to a group $P_k$ of block upper triangular matrices of the form
\eqn\ippo{\left(\matrix{ * & * \cr 0 & * \cr}\right)}
(the upper left hand block being $k\times k$).
Such a group is certainly parabolic, and in fact it is a maximal parabolic
subgroup of $SL(n,\C)$.  The maximal reductive subgroup\foot{For our purposes,
a reductive group is a group that can be obtained by complexifying a compact
group; equivalently, it is locally a product of simple factors and $U(1)$'s.}
 of $P_k$  is a group $R_k$
of matrices of the form
\eqn\jcb{\left(\matrix{ * & 0 \cr 0 & * \cr}\right).}
The Lie algebra of $R_k$ is that of $SL(k)\times SL(n-k)\times \C^*$.
The $\C^*$ plays an important role.
If one decomposes the $SL(n,\C)$ Lie algebra into eigenspaces of
$\C^*$, then the Lie algebra of $P_k$ is the sum of the spaces of non-negative
eigenvalue, while the Lie algebra of $R_k$ is the subalgebra that commutes with
$\C^*$.  

For any simple Lie group $G$, a subgroup $P$ of $G_\C$ is called parabolic
if $G_\C/P$ is compact.  The maximal parabolic subgroups of $G_\C$, up to conjugation, are in
one to one correspondence with the nodes on the Dynkin diagram of $G$.
Each node determines a $\C^*$  subgroup $U$ of the complexified maximal torus of $G$.  
(The choice of a node in the Dynkin diagram generalizes the fact that for $SU(n)$
one chooses an integer $k$ with $1\leq k\leq n-1$.)  One
decomposes the Lie algebra of $G$ under $U$; the sum of the non-negative
eigenspaces is the Lie algebra of a maximal parabolic $P$, and the 
subalgebra that commutes with $U$ is the Lie algebra of the maximal reductive
subgroup $R$ of $P$.

Let $V$ be a $G$ bundle over $E$.  The structure group of $V$ can be reduced
to a maximal parabolic subgroup $P$ in many possible ways.  
Because of the $\C^*$ factor $U\subset R\subset P$,
any such reduction enables one to define a first Chern class.  The bundle $V$
is unstable if and only if for some reduction to a maximal parabolic subgroup,
the first Chern class is positive.  

We will call an unstable bundle ``minimally unstable''  with respect to
a reduction to $P$ if the first Chern class 
determined by the reduction takes the smallest possible positive value.  
One might think that, for given $P$, there would be many minimally unstable
bundles.  But we will find that for every Lie group $G$, there exists a choice
of $P$ such that a  bundle $V$ that is minimally unstable in a reduction to $P$ 
has the same
degree of uniqueness that we found for $SU(n)$: it is unique up to translation
on $E$, that is up to the choice of a distinguished point $p\in E$.

For $E$ of genus one, a Riemann-Roch argument shows that if the structure group
of $V$ can be reduced to $P$ in such a way that the first Chern class is positive,
then it can be further reduced to $R$.  The importance of this is that
$R$ has the center $\C^*$ (the subgroup $U$ that we started with).
We can therefore decompose $H^1(E,{\rm ad}(V))$ in subspaces of definite weight
under the $\C^*$ action.  As in the case considered above, the subspace of positive
weight vanishes, the subspace of weight zero is one-dimensional, and we want
to consider deformations of negative weight.
If $T_{-}$ is the negative weight deformation space, then 
the family of $G$ bundles built by negative weight deformations of $V$ is
a weighted projective space $T_-/\C^*$.  (It is a weighted projective space in general,
not an ordinary one, because various weights appear in the action of $\C^*$ on $T_-$.)
  The weights turn out to  be just the ones predicted for $G$ by Looijenga's
theorem.   That the family  we make this way is indeed the moduli
space of semistable $G$ bundles over $E$ will be proved elsewhere \fmw.

For general $G$ and $P$, the negative weight part of the Lie algebra is
nilpotent (repeated commutators vanish after finitely many steps) but not abelian.
It therefore takes some additional argument, which we give elsewhere \fmw, to
identify the linearized deformation space
$H^1(E,{\rm ad}(V))$ with a space of actual deformations of $V$.  

A perhaps surprising difference between $SU(n)$ and other groups is that
while for $SU(n)$ we were able to use any maximal parabolic subgroup as the
starting point, for other $G$ there is just a unique choice with the right properties.
The vertices that work are the ones indicated in figure two.

\midinsert
\centerline{\psfig{figure=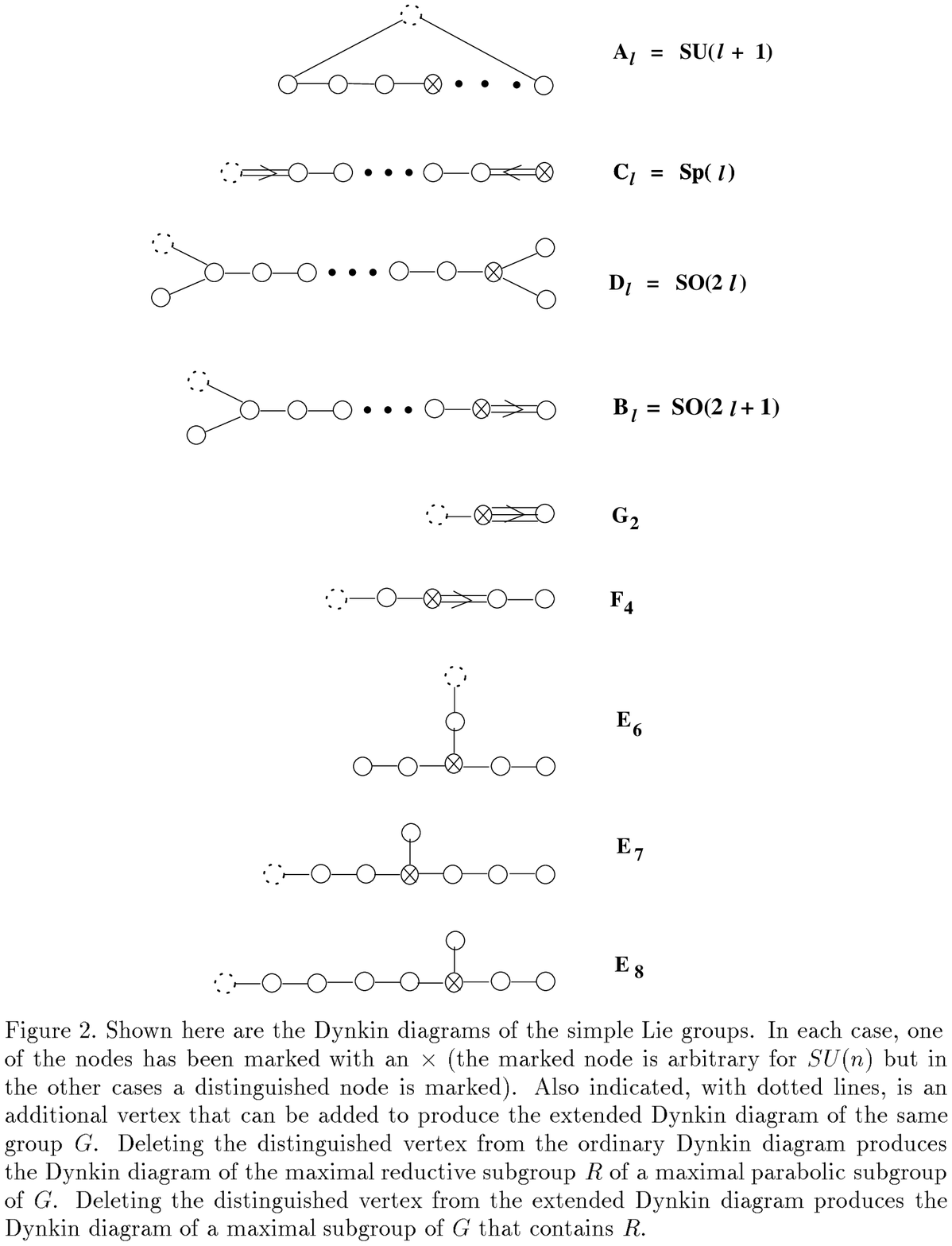,width=5.5in}}
\bigskip
\endinsert

In the remainder of this section, we carry out this program for the various
simple Lie groups.  Then we conclude with a few remarks about bundles
on elliptic manifolds.

\subsec{Fresh Look At $Sp(n)$ }

$Sp(n)$ contains a subgroup $U(n)$ whose complexification is the maximal
reductive subgroup $R$ of a maximal parabolic subgroup $P$ of $Sp(n)_\C$.
If $K_n$ denotes the standard $n$-dimensional representation of $U(n)$, then
the Lie algebra of $Sp(n)$ decomposes under $U(n)$ as
\eqn\ippo{u(n)\oplus {\rm Sym}^2(K_n)\oplus {\rm Sym}^2(K_n^*),}
where $u(n)$ is the adjoint representation of $U(n)$, ${\rm Sym}^2(K_n)$ is the
symmetric part of $K_n\otimes K_n$, and $K_n^*$ is the dual of $K_n$.  We normalize
the $U(1)$ factor in the Lie algebra of $U(n)$ so that the three pieces 
in \ippo\ transform with weights $0,1,$ and $-1$.  The pieces of weight
$0$ and $1$ generate the Lie algebra of a maximal parabolic subgroup $P$ of $Sp(n)$.

An $Sp(n)$ bundle can be represented by a rank $2n$ holomorphic vector bundle
with a symplectic pairing.  A minimally unstable $Sp(n)$ bundle is 
\eqn\urrp{V=W_n\oplus W_n^*}
with $W_n$ as      before the unique stable bundle of determinant ${\cal O}(p)$.  
The symplectic pairing of $V $ comes from the pairing of $W_n$ with $W_n^*$. $V$ is unstable 
because the first Chern class of the summand $W_n$ is positive, and it is minimally
unstable because this first Chern class has the smallest positive value.  $V$ is unique
up to translations on $E$ because $W_n$ has that property.

Now we consider deformations of $V$.  The first order deformations are classified
by $H^1(E,{\rm ad}(V))$.  This can be decomposed using \ippo\ in terms of weights
$1,0$, and $-1$.  The weight one term would be $H^1(E,{\rm Sym}^2W_n)$, and vanishes
because the bundle in question is semistable and
of positive degree.  A similar vanishing for
the deformation space of positive weight holds in all other cases considered
below and will not be mentioned subsequently.   The weight zero deformations
are just the deformations of the bundle $W_n$ in \urrp\ and correspond (in view of
the uniqueness statement about $V$) to translations of $E$.  Again, given the
uniqueness statement about the unstable bundle, this will have an immediate analog
in all the other cases, and will not be repeated.

Thus in subsequent examples we will focus at once on the negative weight part of
the deformation space, which in this $Sp(n)$ example is $H^1(E,{\rm Sym}^2(W_n^*))$.
This equals minus the degree of the semi-stable bundle ${\rm Sym}^2(W_n^*)$.
We compute that degree to give one more illustration of the methods for such
computations; in subsequent examples we will give only the result.
If $W_n^*$ were the sum of a line bundle of degree $-1$ and $n-1$ line bundles
of degree 0, then ${\rm Sym}^2(W_m^*)$ would have one summand of degree $-2$,
$n-1$ of degree $-1$, and the rest of degree 0.  The degree would thus be $-(n+1)$.
Though $W_n^*$ is not actually such a direct sum, this type of computation
can be justified by considering the exact sequences involving $W_n$ and $W_n^*$.

So the negative weight space $T_-$ is of dimension $n+1$.  The unstable bundle
$V$ has a $U(1)$ symmetry (coming from the center of $U(n)$) that is broken
by the deformations.  So the family of bundles that one builds by perturbing
$V$ by a negative weight deformation is parametrized by $M=T_-/{\bf C}^*$
which is a copy of $\P^{n}$.  It will be proved elsewhere that this projective
space is actually the projective space predicted by Looijenga's theorem for $Sp(n)$.

\subsec{$Spin$ Groups}

We will next consider the spin groups.
We work first of all at the Lie algebra level, and thus initially we do not
distinguish $SO$ from $Spin$ or describe the precise global forms of the various
relevant  subgroups of the $Spin$ group.

We begin with $Spin(2n)$.
We consider a maximal parabolic subgroup of $Spin(2n)$ associated
with the ``trivalent'' node of the Dynkin diagram, as in figure two.
The reductive part of the maximal parabolic subgroup associated with the given
vertex is $U(2n-2)\times SO(4)$.  This group is embedded 
in $Spin(2n)$ by the chain $U(2n-2)\times SO(4)\subset SO(2n-4)\times SO(4)
\subset Spin(2n)$.  

An $SO(2k)$ bundle can be regarded as a rank $2k$ bundle with a nondegenerate
holomorphically varying quadratic form.
A minimally unstable $SO(2n-4)$ bundle, with respect to a parabolic subgroup
of reductive part $U(2n-2)$, would be $W_{n-2}\oplus W_{n-2}^*$; a minimally
unstable $SO(2n)$ bundle would be
\eqn\nursko{V=W_{n-2}\oplus W^*_{n-2}\oplus Q_4}
where $Q_4$ is a stable (or semistable) $SO(4)$ bundle.

Now we have to pay some attention to the global forms of the groups, and a crucial
subtlety arises.  The $SO(2n-4)$ bundle $W_{n-2}\oplus W_{n-2}^*$ actually
has a non-zero second Stiefel-Whitney class $w_2$, because the first Chern class of
$W_{n-2}$ is odd.  Since we want $V$ to lift to a $Spin(2n)$
bundle, we must cancel the obstruction by taking for $Q_4$ a stable $SO(4)$ bundle
which likewise has a non-zero $w_2$.

There is a unique such $Q_4$, up to isomorphism.  As a flat $SO(4)$ bundle,
it can be described by saying that the monodromies around two independent one-cycles
in the two-torus $E$ are in a suitable basis
\eqn\imppo{\left(\matrix{1 & 0 & 0 & 0 \cr
                         0 & 1 & 0 & 0 \cr
                         0 & 0 & -1 & 0 \cr
                         0 & 0 & 0 & -1 \cr}\right) ~~{\rm and}~~   
                           \left(\matrix{1 & 0 & 0 & 0 \cr
                         0 & -1 & 0 & 0 \cr
                         0 & 0 & 1 & 0 \cr
                         0 & 0 & 0 & -1 \cr}\right).}
If lifted to $Spin(4)$, these matrices anticommute instead of commuting, so they
define a bundle with non-zero $w_2$.  As a holomorphic bundle, $Q_4$ can be
described as
\eqn\kki{Q_4=\oplus_{\alpha}\L_\alpha}
where the sum runs over the four isomorphism classes of line bundles of order two on $E$.
The quadratic form on $Q_4$ is diagonal with respect to this decomposition and comes
from the isomorphisms ${\L_\alpha\otimes \L_\alpha} \cong {\cal O}$.  
$Q_4$ is unique as any deformation of the $\L_\alpha$ spoils the existence
of this quadratic form.  It is
``fortunate'' that $w_2$ appeared in this way, because such uniqueness would
certainly {\it not} hold for $SO(4)$ bundles with vanishing $w_2$.

Uniqueness of $Q_4$ and uniqueness up to translation of $W_{n-2}$ mean that the
minimally unstable bundle $V$ of equation \nursko\ is unique up to translation.
Now, let us consider its deformations.  $V$ has a $\C^*$ symmetry (coming from the center
of $U(n-2)$) under which $W_{n-2}, Q_4$, and $W_{n-2}^*$ have weights $1,0$, and $-1$.
The deformations we want to consider are the negative weight deformations of $V$.

A novelty relative to the previous cases we considered is that in the decomposition
of the Lie algebra of $Spin(2n)$ under $U(n-2)\times SO(4)$, two different
negative weights appear, not just one.   In fact, the Lie algebra has a piece
of weight $-1$, corresponding to $W_{n-2}^*\otimes Q_4$, and a piece of weight
$-2$ corresponding to $\Lambda^2(W_{n-2}^*)$ (for a bundle $F$, $\Lambda^2(F)$ will
be the antisymmetric part of $F\otimes F$).
The negative weight deformation space of $V$ is thus the sum of two terms:

(1) The weight $-1$ piece is $T_{-1}=H^1(E,W_{n-2}^*\otimes Q_4)$.  As $Q_4$ has rank
four and degree zero, while $W_{n-2}^*$ has degree $-1$, the semi-stable bundle
$W_{n-2}\otimes Q_4$ has degree $-4$, so ${\rm dim}\,T_{-1}=4$.

(2) The weight $-2$ piece is $T_{-2}=H^1(E,\Lambda^2(W_{n-2}^*))$. By methods
explained before, one computes that $\Lambda^2(W_{n-2}^*)$ has degree $-(n-3)$, so
that ${\rm dim}\,T_{-2}=n-3$.

The negative weight deformation space of $V$ is $T=T_{-1}\oplus T_{-2}$.  
We want to consider
the family of $Spin(2n)$ bundles parametrized by $M=T_-/\C^*$, where $\C^*$ is the
symmetry of $V$ broken by the deformations.
Because two different weights appear in $T_-$, this is a weighted projective space
$M={\bf WP}^n_{1,1,1,1,2,2,\dots,2}$ (four $1$'s and the rest 2's) as predicted
for $Spin(2n)$ by Looijenga's theorem.                                   

\bigskip\noindent{\it The Odd Case}

$Spin(2n-1)$ can be considered with almost no change.  The reductive part of a
maximal parabolic (obtained by deleting from the Dynkin diagram the vertex indicated
in figure two) is $U(n-2)\otimes SO(3)$.  A minimally unstable bundle
is now
\eqn\ilpo{V=W_{n-2}\oplus W_{n-2}^*\oplus Q_3}
where now $Q_3$ should be a stable $SO(3)$ bundle with non-zero $w_2$.  There is
a unique $Q_3$, of the form $Q_3=\oplus_{\alpha\not=0}\L_\alpha$ (the sum runs
now over the three {\it non-trivial} line bundles of order two), so $V$ is unique
up to the translations of $E$, acting on $W_{n-2}$.

$V$ has a $\C^*$ symmetry for which the three pieces written in \ilpo\ have weights
$1,-1$, and 0, respectively.  
The negative weight deformation space of $V$ is  the sum of two terms:

(1) The weight $-1$ piece is $T_{-1}=H^1(E,W_{n-2}^*\otimes Q_3)$, and has dimension
three.

(2) The weight $-2$ piece is $T_{-2}=H^1(E,\Lambda^2(W_{n-2}^*))$, and has dimension
$n-3$.

The negative weight deformation space of $V$ is $T_-=T_{-1}\oplus T_{-2}$. 
The family of $Spin(2n)$ bundles parametrized by $M=T_-/\C^*$, where $\C^*$ is the
symmetry of $V$ broken by the deformations,
is a weighted projective space
$M={\bf WP}^n_{1,1,1,2,2,\dots,2}$ (three $1$'s and the rest 2's) as predicted
for $Spin(2n-1)$ by Looijenga's theorem.    

\subsec{$E_8$ Bundles}

What remain are the exceptional groups.  
For these we switch notation slightly.  For the classical groups, we
first described a minimally unstable $G$ bundle $V$ 
using a distinguished representation of $G$, and then we considered the
adjoint bundle ${\rm ad}(V)$.  For the exceptional groups, we will simply
start with the adjoint representation from the beginning.  So the bundle
$V$ will be an adjoint bundle, and the deformation space will be $H^1(E,V)$.

First we consider the simply-laced
groups $E_8$, $E_7$, and $E_6$.  In each case, we consider the parabolic
subgroup associated with the ``trivalent'' vertex, as in figure two.
We first consider $E_8$.

Deleting the indicated vertex from the {\it extended} Dynkin diagram of $E_8$
would give the Dynkin diagram of a maximal subgroup of $E_8$, namely
$H=(SU(6)\times SU(2)\times SU(3))/\Z_6$.  (If one thinks of $\Z_6$ as the group
of sixth roots of unity, then the $\Z_6$ subgroup of $SU(6)\times SU(2)\times SU(3)$
consists in an obvious notation of group elements of the form $\omega\times \omega^3
\times \omega^2$.)  In what follows, 
let $C_n$ be the fundamental $n$-dimensional representation of
$SU(n)$.
The adjoint representation of $E_8$ has the following decomposition
under $H$; it consists of the adjoint representation of $H$ plus the following
pieces:
\eqn\primo{\eqalign{&\Lambda^3C_6\otimes C_2, ~~ \wedge^2C_6^*\otimes C_3,
~~ C_6\otimes C_2\otimes C_3,\cr &
~~ \wedge^2C_6\otimes C_3^*, ~{\rm and}~ C_6^*\otimes C_2\otimes C_3^*.\cr}}
This expansion is easily computed using the chain $SU(6)\times SU(2)\times SU(3)
\subset E_6\times SU(3)\subset E_8$.

If the trivalent vertex is deleted from the {\it unextended} Dynkin diagram of
$E_8$, one is left with the Dynkin diagram of $SU(5)\times SU(2)\times SU(3)\times U(1)$
(where we include a $U(1)$ for the deleted node).  This is the local form of
the reductive part $R$ of the maximal parabolic subgroup of $E_8$ associated
with the given node.  To describe the global form of $R$ and the embedding of $R$
in $H$, note that
$SU(6)$ contains a subgroup $U(5)=(SU(5)\times U(1))/\Z_5$, so $H$
has the subgroup $R=(U(5)\times SU(2)\times SU(3))/\Z_6=
(SU(5)\times SU(2)\times SU(3)\times U(1))/\Z_{30}$. 

A minimally unstable
bundle should have a first Chern class (for the $U(1)$ factor) which is positive
and as small as possible.   Here and in subsequent examples, it is convenient
to work somewhat formally and introduce the ``$SU(n)$ bundle''
\eqn\iggo{B_n={\cal O}(p)^{-1/n}\otimes W_n.}
The fractional exponents will cancel out of all final formulas; if one wishes
one can give a precise meaning to a fractional root of a line bundle in a suitable
formal context.  That the bundle we construct is minimally unstable will be clear
from the fact that its decomposition contains summands of degree 1.

\def\O{{\cal O}}
We consider an $H$ bundle in which the $SU(2)$ factor is $B_2$, the $SU(3)$ factor
is $B_3$, and the $SU(6)$ factor is 
\eqn\pobo{X_6=(W_5\oplus {\cal O})\otimes {\cal O}(p)^{-1/6}.}
Clearly, such a bundle does not make sense as a $SU(6)\times SU(2)\times SU(3)$ bundle,
but it makes sense as an $H$ bundle because the fractional exponents cancel
out for all representations of $SU(6)\times SU(2)\times SU(3)$ in which the $\Z_6$
acts trivially.
\def\ad{{\rm ad}}
For instance, the $E_8$ bundle that we really want can be described as follows.
The part coming from the adjoint representation of $SU(6)$ gives
\eqn\obbo{\ad (W_5) \oplus W_5\oplus W_5^*\oplus {\cal O},}
while the adjoint representation of $SU(2)\times SU(3)$ gives just 
\eqn\robo{\ad (W_2)\oplus \ad(W_3).}
(By $\ad (W_n)$ we mean the {\it traceless} part of $W_n\otimes W_n^*$.)
The part of the $E_8$ bundle coming from \primo\ can be expanded
\eqn\ilpo{
\eqalign{\Lambda^3C_6\otimes C_2& = \Lambda^3 W_5\otimes W_2\otimes \O(p)^{-1} 
\oplus \Lambda^2W_5\otimes W_2 \otimes \O(p)^{-1}\cr
 \wedge^2C_6^*\otimes C_3 & = \Lambda^2W_5^*\otimes W_3\oplus W_5^*\otimes W_3 \cr
 C_6\otimes C_2\otimes C_3 & = W_5\otimes W_2\otimes W_3\otimes \O(p)^{-1}
\oplus W_2\otimes W_3\otimes \O(p)^{-1}\cr
 \wedge^2C_6\otimes C_3^* & = \Lambda^2 W_5\otimes W_3^*\oplus W_5\otimes W_3^* \cr
 C_6^*\otimes C_2\otimes C_3^*&= W_5^*\otimes W_2^*\otimes W_3^*\otimes \O(p)
\oplus W_2^*\otimes W_3^*\otimes \O(p).\cr}}
This $E_8$ bundle $V$ is unique up to translations on the elliptic curve because of the
corresponding statement for the $W_n$'s.

Now we want to consider the negative weight deformations of $V$.  The $\C^*$ in
question is easy to identify because it originated as a subgroup of the $SU(6)$
factor in $H$.
So it acts trivially on $B_2$ and $B_3$, 
while in the decomposition \pobo\ of $X_6$, the  $\C^*$ acts
on $W_5\otimes \O(p)^{-1/6}$ with weight $1$ and on the other summand $\O(-p)^{-1/6}$ with
weight $-5$.  The negative weight deformation space of $V$ can now be analyzed
as follows:

(1) The weight $-1$ summand of $V$ is $V_{-1}=
W_5^*\otimes W_2^*\otimes W_3^* \otimes \O(p)$,
of degree $-1$.  So the weight $-1$  subspace $T$ of $H^1(E,V)$ is
$T_{-1}=H^1(E,V_{-1})$, and has dimension 1.

(2) The weight $-2$ summand of $V$ is $V_{-2}=\Lambda^2W_5^*\otimes W_3$, of degree $-2$.
So the weight $-2$ deformation space  $T_{-2}=H^1(E,V_{-2})$
has dimension 2.

(3) The weight $-3$ summand of $V$ is $V_{-3}=\Lambda^2W_5\otimes W_2\otimes \O(p)^{-1}$, 
of degree $-2$.  So the weight $-3$ deformation space  $T_{-3}=
H^1(E, V_{-3} )$  has dimension 2.

(4)  The weight $-4$ summand of $V$ is $ V_{-4}=W_5\otimes W_3^*$, of degree $-2$.
So the weight $-4$ deformation space  $T_{-4}=H^1(E, V_{-4} )$
 has dimension 2.

(5) The weight $-5$ summand of $V$ is $V_{-5}=
 W_2\otimes W_3\otimes \O(p)^{-1}$, of degree $-1$.
So the weight $-5$ deformation space  $T_{-5}=H^1(E,V_{-5}  )$  has dimension 1.

(6) The weight $-6$ summand of $V$ is $V_{-6}= W_5^*$, of degree $-1$.
So the weight $-6$ deformation space $T_6=H^1(E, V_{-6} )$ has dimension 1.

Putting the pieces together, we can identify the parameter space $M=T_-/\C^*$
of bundles built by a negative weight deformation of $V$.  It is a weighted
projective space ${\bf WP}^8_{1,2,2,3,3,4,4,5,6}$, as predicted by
the $E_8$ case of Looijenga's theorem.

\subsec{$E_7$ Bundles}

We next consider $E_7$ in a similar spirit.
The ``trivalent'' node on the extended $E_7$ Dynkin diagram is associated
with the maximal subgroup $H=(SU(4)\times SU(4)\times SU(2))/\Z_4$ of $E_7$.
The $\Z_4$ consists of elements of $SU(4)\times SU(4)\times SU(2)$ of the form
$\omega\times \omega\times \omega^2$, where $\omega^4=1$.  The Lie algebra of
$E_7$ decomposes under $H$ as the adjoint representation of $H$ plus
\eqn\hojo{C_4\otimes C'_4\otimes C_2\oplus C_4^*\otimes C_4'^*\otimes C_2^*
\oplus \wedge^2C_4\otimes \wedge^2C'_4.}
(Here $C_4$, $C'_4$, and $C_2$ are the basic representations of the three
factors in $H$.)

The reductive part of the maximal parabolic associated with this node is
obtained by restricting to a subgroup of the first $SU(4)$ in $H$ that is
isomorphic to $U(3)=(SU(3)\times U(1))/\Z_3$.
This maximal reductive subgroup is thus $R=(SU(3)\times SU(4)\times SU(2)\times U(1))/
\Z_{12}$.  (The Dynkin diagram of $ R $ is obtained by omitting the trivalent
vertex from the {\it unextended} $E_7$ Dynkin diagram, with the missing node
understood to represent a $U(1)$ factor.)

We will describe the minimally unstable bundle first of all in terms of $SU(4)
\times SU(4)\times SU(2)$.  In the first $SU(4)$ we take 
the bundle to be
\eqn\kormo{X_4=(W_3\oplus \O)\otimes \O(p)^{-1/4}.}
In the second $SU(4)$, we take $B_4=W_4\otimes \O(p)^{-1/4}$, and for the
$SU(2)$ factor we take $B_2=W_2\otimes \O(p)^{-1/2}$.  Just as in the $E_8$
case, the fractional exponents disappear when one constructs the associated
bundle in any representation of $SU(4)\times SU(4)\times SU(2)$ in which
the $\Z_4$ subgroup acts trivially, that is, any representation of $H$.
So we get an $H$ bundle, and therefore an $E_7$ bundle.
The center of $R$ is a $\C^*$ that acts with weight one on the first
summand in \kormo, with weight $-3$ on the second, and trivially on $B_4$ and $B_2$.

It is now straightforward, using \hojo, to describe in detail the minimally
unstable $E_7$ bundle $V$.  Rather than repeating this in as much detail as we did for
$E_8$, we will just write down the pieces of negative weight.

(1) The weight $-1$ subbundle of $V$ is $V_{-1}=W_3^*\otimes W_4^*\otimes W_2^*
\otimes \O(p)$.
This has degree $-2$, so $T_1=H^1(E,V_{-1})$ has dimension 2.

(2) The weight $-2$ subbundle of $ V$ is $V_{-2}=W_3\otimes \Lambda^2W_4\otimes 
\O(p)^{-1}$. This has degree $-3$, so $T_2=H^1(E,V_{-2})$ has dimension 3.

(3) The weight $-3$ subbundle of $V$ is $V_{-3}=W_4\otimes W_2\otimes \O(p)^{-1}$.
This has degree $-2$, so $T_3=H^1(E,V_{-3})$ has dimension 2.

(4) Finally, the weight $-4$ subbundle of $V$ is $V_{-4}= W_3^*$.  This
has degree $-1$. so $T_4=H^1(E,V_{-4})$ has dimension 1.

Putting the pieces together, we see that the parameter space $M=T_-/\C^*$ of
negative weight deformations of $V$ is a weighted projective space
${\bf WP}^7_{1,1,2,2,2,3,3,4}$, as predicted by Looijenga's theorem for $E_7$.

\subsec{$E_6$ Bundles}

Now we consider the last simply-laced group $E_6$.  Removing the trivalent vertex
from the extended Dynkin diagram leaves the Dynkin diagram of the maximal
subgroup $H=(SU(3)\times SU(3)\times SU(3))/\Z_3$
of $E_6$; the $\Z_3$ is the diagonal subgroup of the product of the
centers of the three $SU(3)$'s.  The Lie algebra of $E_6$ consists of the adjoint
representation of $H$ plus
\eqn\ippop{C_3\otimes C_3'\otimes C_3''\oplus C_3^*\otimes {C'}_3^*\otimes {C''}_3^*.}
Here $C_3,$ $C_3'$, and $C_3''$ are the three-dimensional representations of the
three $SU(3)$'s.

The maximal reductive subgroup of the corresponding maximal parabolic is obtained
by replacing the first $SU(3)$ in $H$ by $U(2)=(SU(2)\times U(1))/\Z_2$.
The reductive group is thus $ R=(SU(2)\times SU(3)\times SU(3)\times U(1))/\Z_6$.

We describe a minimally unstable bundle first of all in terms of $H$.
In the first $SU(3)$ we take 
\eqn\pilk{X_3=(W_2\oplus \O)\otimes \O(p)^{-1/3} ,}
and  in the second and third $SU(3)$'s we take $B_3=W_3\otimes \O(p)^{-1/3}$.
Once again, this gives something which makes sense as an $H$ bundle,
and therefore also as an $E_6$ bundle.
The center of $R$ is a $\C^*$ which acts with respective weights $1$ and $-2$
on the two summands in \pilk\ and trivially on factors coming from the other
$SU(3)$'s.

It is straightforward to give a detailed description of the minimally
unstable $E_6$ bundle $V$.  We content ourselves with looking at the pieces
of negative weight:

(1) In weight $-1$, we have $V_{-1}=W_2^*\otimes W_3^*\otimes W_3^*\otimes \O(p)$,
of degree $-3$.  So $T_{-1}=H^1(E,V_{-1})$ has dimension 3.

(2) In weight $-2$, we have $V_{-2}= W_3\otimes W_3\otimes \O(p)^{-1}$, of degree
$-3$.  So $T_{-2}=H^1(E,V_{-2})$ has dimension 3.

(3) In weight $-3$, we have $V_{-3}= W_2*$, of degree $-1$.  So
$T_{-3}=H^1(E,V_{-3})$ has dimension 1.

Putting the pieces together, we see that the space $M=T_-/\C^*$ of negative
weight deformations of $V$ is a weighted projective space
${\bf WP}^6_{1,1,1,2,2,2,3}$, as predicted by Looijenga's theorem for
$E_6$.

\subsec{$G_2$ Bundles}
\def\Sym{{\rm Sym}}
We come now to the two exceptional groups that are not simply
laced.

$G_2$ has a maximal subgroup $H=(SU(2)\times SU(2))/\Z_2$, where the $\Z_2$ is
the diagonal subgroup of the product of the centers of the
two $SU(2)$'s.  (The Dynkin diagram of $H$ is obtained from the extended Dynkin
diagram of $ G_2$ by omitting the vertex indicated in figure two.)
The Lie algebra of $G_2$ decomposes under $H$ as the sum of the adjoint
representation plus
\eqn\koppo{C_2\otimes {\rm Sym}^3C'_2.}
(Here ${\rm Sym}^3C'_2$ denotes the symmetric part of $C'_2\otimes C'_2\otimes
C'_2$.) 

By restricting to a subgroup $U(1)$ of the first $SU(2)$, we get
a group $R=(U(1)\times SU(2))/\Z_2$, which is the maximal reductive subgroup
of a maximal parabolic subgroup of $G_2$.

A minimally unstable $G_2$ bundle $V$ can be described at the level
of $SU(2)\times SU(2)$ by taking  the first $SU(2)$ factor to be 
\eqn\hibbob{X_2 =(\O(p)\oplus \O)\otimes \O(p)^{-1/2}}
and the second to be $B_2=W_2\otimes \O(p)^{-1/2}$.  
$\C^*$ acts with weight $1$ and $-1$ on the two summands in \hibbob\ and
trivially on the second $SU(2)$.

$G_2$ is so small that we can fairly painlessly write down a detailed
description of the minimally unstable bundle $V$.  It is
\eqn\hbc{\O(p)\oplus \O\oplus \O(p)^{-1}\oplus {\rm Sym}^2W_2\otimes \O(p)^{-1}\oplus
\Sym^3W_2\otimes\O(p)^{-1}\oplus\Sym^3W_2\otimes \O(p)^{-2}.}
The first three summands come from the Lie algebra of the first $SU(2)$, the
fourth from the Lie algebra of the second $SU(2)$, and the last two from decomposing
$C_2\otimes {\rm Sym}^3C'_2$ under $R$.

In particular, the subbundles of $V$ of negative weight are as follows.

(1)  $V_{-1}={\rm Sym}^3W_2\otimes \O(p)^{-2}$, of degree $-2$. 
So $T_{-1}=H^1(E,V_{-1})$ is of dimension 2.

(2) $V_{-2}= \O(p)^{-1}$, of degree  $-1$.  So $T_{-2}=H^1(E,V_{-2})$ is of dimension
1.

So the parameter space $M=T_-/\C^*$ of negative weight deformations of $V$ is 
isomorphic to a weighted projective space ${\bf WP}^2_{1,1,2}$, as predicted
by Looijenga's theorem for $G_2$.  

\subsec{$F_4$ Bundles} 

We conclude by examining $F_4$.  

$F_4$ has a maximal subgroup (related to the node of the extended Dynkin diagram
indicated in figure two) isomorphic to $H=(SU(3)\times SU(3))/\Z_3$ 
where $\Z_3$ is the diagonal subgroup of the product of the centers of the
two $SU(3)$'s.
The Lie algebra of $F_4$ decomposes under $F_4$ as the adjoint representation
plus
\eqn\rilpo{C_3\otimes {\rm Sym}^2C'_3\oplus C_3^*\otimes \Sym^2{C'}_3^*,} 
where $C_3$ and $C_3'$ are the basic three-dimensional representations of the
two $SU(3)$'s.

To obtain the reductive subgroup of a maximal parabolic, one restricts
to a $U(2)=(SU(2)\times U(1))/\Z_2$ subgroup of the first $SU(3)$.
So the reductive group in question is $R=(SU(2)\times SU(2)\times U(1))/\Z_6$.

A minimally unstable bundle $V$ can be obtained at the level of the   
$SU(3)\times SU(3)$ by choosing  in the first $SU(3)$
\eqn\helbo{X_3=\left(W_2\oplus \O\right)\otimes \O(p)^{-1/3}}
and $B_3=W_3\otimes \O(p)^{-1/3}$ in the second.
The center $\C^*$ of $R$ 
acts with weights $1$ and $-2$ on the two summands in \helbo, and trivially
on the second $SU(3)$.

The negative weight subbundle of $V$ is explicitly described as follows.

(1) $V_{-1}=W_2^*\otimes {\rm Sym}^2W_3^*\otimes \O(p)$, of degree $-2$.
So $T_{-1}=H^1(E,V_{-1})$ is of dimension 2.

(2) $V_{-2}= {\rm Sym}^2\,W_3\otimes \O(p)^{-1}$, of degree $-2$.  So
$T_{-2}=H^1(E,V_{-2})$ is of dimension 2.

(3) $V_{-3}= W_2^*$, of degree $-1$.  So $T_{-3}=H^1(E,V_{-3})$ is of dimension 1.

So the parameter space $M=T_-/\C^*$ of negative weight deformations of $V$ is 
isomorphic to a weighted projective space ${\bf WP}^2_{1,1,2,2,3}$, as predicted
by Looijenga's theorem for $F_4$.

\subsec{Bundles Over Elliptic Manifolds}

We actually wish to construct $G$ bundles not just over a single elliptic curve but over
an elliptically fibered manifold $\pi:Z\to B$ with a section $\sigma$.  $Z$ is
described by a Weierstrass equation
\eqn\jub{y^2=4x^3-g_2x-g_3,}
where $x$ and $y$ are sections of ${\cal L}^2$ and ${\cal L}^3$, with 
${\cal L}$ being some line bundle over $B$.  The section $\sigma$ is given by
$x=y=\infty$.

To imitate the above construction in this situation, 
 we would like to construct suitable unstable $G$ bundles over $Z$, which
reduce on every fiber of $\pi$ to the minimally unstable bundle constructed
above, and can be deformed to stable $G$ bundles over $V$.

The minimally unstable bundles were all built from tensor products and sums
of the basic building blocks $\O(p)$ and $W_n$.
\foot{In the $Spin$ case, we used distinguished bundles $Q_3$ and $Q_4$.
It can be shown that $Q_4=W_2\otimes W_2^*$, and $Q_3={\rm ad}(W_2)$.} 
So all we need is to generalize those  to an elliptic manifold.

The global version of $\O(p)$ is just $\O(\sigma)$, since $\sigma$ intersects
each fiber $E$ of $\pi$ in a distinguished point $p$.   To construct a global
version of the $W_n$, we must go back to the inductive procedure defining them.
On a single elliptic curve, we had $W_1=\O(p)$, so globally we take
$W_1=\O(\sigma)$. $W_2$ was defined over a single elliptic curve by the existence of
an exact sequence
\eqn\uuu{0\to \O\to W_2\to \O(p)\to 0.}
So globally we ask that $W_2$ should have an exact sequence
\eqn\vuru{0\to {\cal M}\to W_2\to \O(\sigma)\to 0,}
where ${\cal M}$ is the pullback to $Z$ of some line bundle on $B$.  (Thus, ${\cal M}$
is trivial on each fiber.) 
Moreover, we want the extension in \uuu\ to be non-trivial when restricted
to the fiber $E_b$ over any $b\in B$.  This is a strong condition which (up to
isomorphism) uniquely determines ${\cal M}$ and the extension in \vuru.  The condition
is equivalent to the statement that
the line bundle over $B$ whose fiber at $b$ is $H^1(E_b,\O(\sigma)^{-1}
\otimes {\cal M})$ should be trivial, so that it has an everywhere non-zero section.

${\cal M}$ can be determined as follows.  We have $H^1(E_b,\O(\sigma)^{-1}\otimes 
{\cal M})={\cal M}_b\otimes  H^1(E_b,\O(\sigma)^{-1})$.  (${\cal M}_b$ is the
fiber at $b$.)
By Serre duality, $H^1(E_b,\O(\sigma)^{-1})$ is dual to $H^0(E_b,K\otimes \O(\sigma))$
(where $K$ is the  canonical bundle); this is generated by $dx/y$,  so a natural
generator of $H^1(E_b,\O(\sigma)^{-1})$ can be identified with $y(dx)^{-1}$.
For the line bundle whose fiber over $b$ is
${\cal M}_b\otimes H^1(E_b,\O(\sigma)^{-1})$ to be trivial over $B$,
$y(dx)^{-1}$ should make sense globally as a section of ${\cal M}$, so we
need ${\cal M}={\cal L}$.

This type of reasoning can be generalized to get global versions of all the $W_n$'s.
$W_n$ is defined inductively by an exact sequence
\eqn\vvv{0\to {\cal L}^{n-1} \to W_n\to W_{n-1}\to 0.}
The line bundle ${\cal L}^{n-1}$ in \vvv\ is chosen to ensure the existence
of an extension that is non-trivial on each fiber.

\bigskip\noindent{\it Global Version Of Unstable $G$ Bundles}  

Having identified the global versions of  ${\cal O}(p)$ and the $W_n$'s,
we can construct appropriate global versions of the minimally unstable
bundles.  We simply replace in all above formulas ${\cal O}(p)$ and $W_n$
by their global versions.
The only subtlety is that one can twist by additional data coming from $B$.

Instead of trying to be abstract, let us first  write down a concrete example
for $G=Sp(n)$.   The minimally unstable bundle over a single elliptic curve
was $W_n\oplus W_n^*$, and (having explained what we mean by $W_n$) we could
take the same starting point over a general elliptic manifold $Z$.  
However, we can generalize slightly, pick an arbitrary line bundle ${\cal M}$ over
$B$, and consider the $Sp(n)$ bundle
\eqn\pimmob{V=W_n\otimes {\cal M}\oplus W_n^*\otimes {\cal M}^{-1}}
which is isomorphic to the minimally unstable bundle $W_n\oplus W_n^*$ on each fiber.

To express this in a language that is more general, a bundle over $Z$
that is isomorphic on each fiber to $W_n\oplus W_n^*$ is not uniquely
determined because the bundle $W_n\oplus W_n^*$ has automorphisms.
Let $A_b$ be the automorphism
group of the $Sp(n)$ bundle
$W_n\oplus W_n^*$ over $E_b$, and let $A$ be the sheaf of groups over $B$
whose fiber at $b\in B$ is $A_b$.  Then the bundle $W_n\oplus W_n^*$ over
$Z$ can be ``twisted'' by any element of $H^1(B,A)$.

The maximal reductive subgroup of $A_b$ is the  center, $U\cong \C^*$, of
the reductive group $R$ that was used in building the minimally unstable
bundles over the fibers.  What we have done in \pimmob\ is to  twist
by an element of $H^1(B,U)$. 
 
This discussion can be slightly generalized as follows.  If ${\cal M}$ is not
well-defined as a line bundle, but is the square root of a line bundle, then
$V$ is not well-defined as a vector bundle, but associated objects such as
$V\otimes V$ and ${\rm ad}(V)$ are well-defined as vector bundles with structure
group $Sp(n)/\Z_2$.  With this starting point, 
one can use the parabolic construction to construct
$Sp(n)/\Z_2$ bundles over an elliptic manifold $Z$ that can be lifted to    an
$Sp(n)$ bundle on each fiber, but not globally.  A similar construction can be
made for non-simply-connected forms of groups other than $Sp(n)$.

\bigskip\noindent{\it Weighted Projective Space Bundle Over $B$}

For every $G$, there is a bundle  ${\cal W}\to B$ of weighted projective
spaces whose 
fiber over $b\in B$ is the moduli space of semistable $G$ bundles
over $E_b$.  We claimed in the introduction that for arbitrary simple,
connected, and simply-connected $G$ except $E_8$, ${\cal W}$ is a bundle
of weighted projective spaces that can be obtained by projectivizing
a certain sum of line bundles.  In sections 2,3, and 4, we exhibited such
structures for certain classes of $G$.  Here we will briefly point out
a general framework for exhibiting this structure.

Let $\Omega$ be the bundle over $B$ whose fiber at $b\in B$
is the negative weight part of $H^1(E_b,{\rm ad}(V))$.  
For each $b$, the moduli space of $G$ bundles on $E_b$ is simply
$\Omega_b/\C^*$.  So  ${\cal W}$ is obtained by projectivizing
the vector bundle $\Omega$.

Our claim is that for $G$ other than $E_8$, $\Omega$ is a certain
sum of line bundles over $B$, in fact a sum of powers of ${\cal L}$, with
exponents and $\C^*$ weights that were summarized in the table in the introduction.
But we now have a general framework for computing $\Omega$ and verifying that this
is so.  For instance, for $G=Sp(n)$, $\Omega$ is the bundle whose
fiber at $b$ is $H^1(E_b,{\rm Sym}^2(W_n^*))$.  By analyzing this
cohomology group and its analogs for other $G$, 
the decomposition of $\Omega$ as a sum of line bundles
will be exhibited elsewhere \fmw.
 
Note that if it is true that ${\cal W}$ is obtained by projectivizing
a vector bundle $\Omega$ with $\C^*$ action, then $\Omega$ is
 not uniquely determined; one could pick an arbitrary line bundle
${\cal N}\to B$ and twist the weight $k$ subbundle of $\Omega$ by
${\cal N}^k$, without changing the projectivization $B$. 
This freedom is reflected in the fact that ${\cal W}$ can be determined
starting with \pimmob\ (or its analogs for other $G$) for arbitrary ${\cal M}$;
$\Omega$ depends on ${\cal M}$ but ${\cal W}$ does not.

\bigskip\noindent
{\it Deformation To A Stable Bundle}

Stable bundles over $Z$ can (often) be made by deforming the unstable
bundle $V$.  First order deformations are classified by $H^1(Z,{\rm ad(V)}$.  If
(following standard notation in algebraic geometry) we denote the bundles
on $B$ made by taking the $i^{th}$ cohomology of ${\rm ad(V)}$ along the fibers
of $\pi:Z\to B$ as $R^i\pi_*(V)$, then in the situation considered here the
Leray spectral sequence for $\pi:Z\to B$ degenerates to an exact sequence:
\eqn\uppu{0\to H^1(B,R^0\pi_*({\rm ad}(V)))\to H^1(Z,{\rm ad}(V))\to H^0(B,
R^1\pi_*({\rm ad}( V)))\to 0.}  (The Leray spectral sequence of
a mapping always reduces to such an exact sequence when the cohomology
of the fibers is nonzero in only two dimensions.)
So the space of deformations of $V$ maps to $H^0(B,R^1\pi_*({\rm ad}(V)))$,
which is the space whose projectivization is the space of sections of ${\cal W}$.
The fiber of the induced map from the space of bundles
to the space of sections has for its tangent space
$H^1(B,R^0\pi_*({\rm ad}(V)))$. This is the tangent space to an abelian variety
which generalizes the Jacobian found for $G=SU(n)$ in section 2.
\uppu\ thus generalizes part of the structure found in section 2: the moduli
space of bundles maps to the space of sections of ${\cal W}$, the fiber being
an abelian variety.

\newsec{Comparison To $F$ Theory Moduli Spaces}

The remainder of this paper mainly aims at using our results
to make two tests of duality between the heterotic string and $F$ theory.  The first
test, in this section, involves comparison of moduli spaces.

We consider on the heterotic string side an elliptically fibered
manifold $Z\to B$, with a section $\sigma$ whose normal bundle we call ${\cal L}^{-1}$. For each
$G$, there is a weighted projective space bundle ${\cal W}_G\to B$, which
parametrizes $G$ bundles over the fibers of $Z\to B$.

In heterotic string theory, $Z$ is endowed with an $E_8\times E_8$ bundle.
All of our interest will focus on what happens in one of the two $E_8$'s, say
the first one.  We consider the locus of heterotic string vacua on $Z$
in which the structure group of this $E_8$ bundle
reduces to a subgroup $G$, of rank $r$.  To specify a point in this locus requires picking
among other things a $G$ bundle over $Z$.  It will become clear that the
$G$ bundles relevant to comparison with the simplest $F$ theory compactifications
are semistable when restricted to the generic fiber.  Such a $G$ bundle
determines a section $s$ of the bundle ${\cal W}_G$.\foot{$s$ is defined at least over the
dense open subset in $B$ that parametrizes fibers over which the bundle actually
is semistable.  In general, $s$ can be defined everywhere only after some blow-ups
of $B$.  When there is enough ampleness, and the rank of $G $ exceeds the dimension
of $B$, such blowups are generically not necessary.}

Much of the work of this paper can be summarized by saying that ${\cal W}_G$,
for $G\not= E_8$, has homogeneous coordinates $a_j$, $j=0,1,\dots, r$, which are
sections, respectively, of the line bundles
 $\O(1)^{s_j}\otimes \L^{-d_j}$.  Here $\O(1)$ is a line bundle over
 ${\cal W}_G$ which restricts
on each fiber to the basic line bundle on the weighted projective
space; the numbers $s_j$ are the weights appearing in Looijenga's theorem
and the $d_j$ are the degrees of the fundamental Casimir invariants of $G$.  
Under a section $s:B\to {\cal W}_G$, $\O(1)$ pulls back to
a line bundle ${\cal N}$ over $B$, and the $a_j$ pull back to sections
of ${\cal N}^{s_j}\otimes \L^{-d_j}$.
Conversely, sections 
\eqn\hupp{\tilde a_j\in H^0(B,{\cal N}^{s_j}\otimes \L^{-d_j})}
which are sufficiently generic (no common zeroes) determine a section $s$ of
${\cal W}_G$.  The $\tilde a_j$ are uniquely determined by $s$
up to
\eqn\upp{\tilde a_j\to \lambda^{s_j}\tilde a_j, ~~{\rm with}\,\lambda\in\C^*.}

The heterotic string compactified on the elliptic manifold
$Z\to B$ is believed to be dual to 
$F$ theory compactified on a K3-fibered manifold $X\to B$.  The topology
of $X$ depends on the topology of the $E_8\times E_8$ bundle over $Z$, in a way
first analyzed in \mv. 

When the structure group of the $E_8$ bundle reduces to $G$, the heterotic
string acquires an unbroken gauge symmetry $H$, where $H$ is
the commutant of $G$ in $E_8$. If $B$ is a point, then $H$ is necessarily
simply-laced and unbroken $H$ symmetry of
the heterotic string  corresponds to the appearance of a singularity 
of type $H$ in $F$ theory (in a fashion that we analyzed in the del Pezzo
context in section 4). For $B$ of positive dimension,
unbroken gauge symmetry corresponds in $F$ theory
to appearance of
a section $\theta:B\to X$ of singularities.  In general, the singularity
along $\theta$ is not of type $H$ ($H$ may not even
be simply-laced); it is of type $H'$, where $H'\supset H$ is a  simply-laced
group, and the $H'$ symmetry is broken to $H$
\refs{\many,\aspin} by a monodromy corresponding to an automorphism of the Dynkin diagram
of $H'$ whose quotient is the Dynkin diagram of $H$.  (We used this
automorphism in section 3.2 to compare $H$ and $H'$ bundles.) 

In \many, the precise parameters controlling the complex structure near
$\theta$ that should be related to bundle data on the heterotic string side
were identified, for each $H$. 
The correspondence between the two theories was checked
by counting  parameters on the two sides.  Here we will be more
precise and actually exhibit a natural map from complex structure
parameters in $F$ theory to bundle parameters on the heterotic string side.
In fact, we will show that the choice of a section $s:B\to {\cal W}_G$
is in natural correspondence with the data identified in \many\ in $F$ theory.
As we explained in  section 2.4, a more complete comparison of the two
theories would involve also comparing certain abelian varieties.

In comparing \hupp\ and \upp\ to the results of \many, we will actually generalize
the statements of \many\ in a fairly obvious way.  In \many, the case $B={\bf P}^1$ was considered,
and an important role was played by two line bundles over $B$, namely
$K_{{\bf P}^1}^{-1}={\cal O}(2)$ and an additional line bundle ${\cal O}(12+n)$ that
enters in constructing the K3 fibration over $B$.
We will generalize to the case that $B$ may have dimension greater than one,
will write ${\cal L}$ wherever $K_{{\bf P}^1}^{-1}$ appears in \many, and
will replace the line bundle ${\cal O}(12+n)\to{\bf P}^1$ used in \many\ by a general
line bundle ${\cal N}\to B$.  There is no difficulty in adapting the reasoning
and conclusions of \many\ to this more general case.  We will
not attempt here an explanation of the arguments of \many, but will just
cite their answers and compare to \hupp\ and \upp.

We have seen in this paper that the description of $E_8$ bundles on an elliptically
fibered manifolds is rather different from the description of $G$ bundles for
any $G$ other than $E_8$.  In terms of heterotic string/$F$ theory duality,
this is related to the following.
The ability to compare bundle data on the heterotic string side to $F$ theory
in the way we will do below depends on considering heterotic string bundles
whose structure group is a proper subgroup $G$ of $E_8$, so that a singularity
appears on the $F$ theory side; the structure of the $G$ bundle is then coded
in the behavior near the singularity.  The case $G=E_8$ is quite exceptional
as then there is no singularity and no way to ``localize'' the bundle information
on the $F$ theory side.  The comparison of the heterotic string and  $F$ theory
moduli spaces thens involve many additional issues such as heterotic
string $T$ dualities that can mix geometrical and bundle moduli. One way
to turn off the $T$ dualities while looking at $E_8$ bundles is to take
the area of the fibers to be big; this option was explored at the end of
section 4.

\subsec{Comparison Of Moduli Spaces}
\bigskip\noindent{\it $SU(2)$ Bundles}

First we consider the case of $SU(2)$ bundles. For $G=SU(2)$ and all other
groups $G$ considered subsequently, we assume a ``minimal'' embedding of
$SU(2)$ in $E_8$, for which the generator of $H^3(E_8,{\bf Z})$ pulls back to
the smallest possible value.
 Since the commutant of a minimally embedded $SU(2)$
in $E_8$ is $E_7$, reduction of the structure group of an $E_8$ bundle to such
an $SU(2)$  corresponds in $F$ theory to considering
a K3 fibered manifold $X\to B$ with a section $\theta $ of $E_7$ singularities.
($E_7$ has no outer automorphisms, so there is no monodromy breaking $E_7$.)

Only the behavior of $X$ near $\theta$ is relevant, and one can
write a rather explicit local formula describing $X$ as a hypersurface
in a bundle ${\cal M}\oplus (\L^2\otimes{\cal M}^2)\oplus( \L^3\otimes{\cal M}^3)$,
with coordinates $u,x,y$; ${\cal M}$ is a line bundle over $B$.  (In \many, this
is formulated for $B=\P^1$ and ${\cal M}=\O(n)$.)
Taking $\theta$ to be $u=x=y=0$, the behavior near $\theta$ is given
by an equation
\eqn\gug{y^2=4x^3 -fxu^3-gu^5,}
where $f$ and $g$ are sections of the line bundles $\L^4\otimes {\cal M}$
and $\L^6\otimes {\cal M}$ over $B$.  
An obvious rescaling of $u,x,y$ (with weights $1,2$, and 3)
brings the equivalence $f\to\lambda f$,
$g\to \lambda g$.
Setting ${\cal N}=\L^6\otimes {\cal M}$,
and recalling that for $SU(2)$  one has weights $s_j=1$ and exponents $d_j=0,2$, for
$j=0,1$, we see that $f$ and $g$ correspond in a natural way to the sections
$\tilde a_j$ of equations \hupp\ and \upp.

Two remarks  should be made about this:

(1) In \many, it is asserted that, in a heterotic string description
dual to this $F$ theory model,  the instanton number (of the $E_8$ bundle
whose structure group is reducing to $SU(2)$) is $12+n$.  To express this
in a way that does not assume that the base $B$ is one-dimensional, the assertion
is that if $V$ is the $SU(2)$ bundle and $\pi:Z\to B$ the elliptic fibration,
then Chern classes of $V$ and ${\cal N}$ are related by
\eqn\gilg{c_1({\cal N})=\pi_*(c_2(V)).}
This assertion was based in \many\ on qualitative properties of the
heterotic/$F$ theory duality but without sufficient information about the bundles
to actually compute $c_2(V)$ and verify \gilg.
Having constructed the bundles, we are in a position to do so.
In fact, \gilg\ is equivalent to a result of \fm.
We give a proof, together with generalizations to other groups, in section 7.

(2) Generically along $B$, \gug\ describes a singularity of type $E_7$ at
$u=x=y=0$, but the singularity is worse at zeroes of $f$.  It is
proposed  in \many\ that matter fields in the two-dimensional representation
of $SU(2)$ come from zeroes of $f$.  Translated into bundle language,
the assertion amounts to the mathematical
statement that $H^1(Z,V)$ can be computed locally from the behavior at
zeroes of $f$, and that in case $B$ is a curve (so that there are only
finitely many zeroes), $H^1(Z,V)$ receives a one-dimensional contribution from
each zero.  This proposal was originally made on the basis of counting
parameters and was further supported by a study of the $F$ theory singularity
near zeroes of $f$ \ref\katz{S. Katz and C. Vafa, ``Matter From Geometry,''
hep-th/9606086.}.  Having constructed the bundles,
we are in a position to  verify the relation between matter fields and zeroes
of $f$ by computing $H^1(Z,V)$ directly.  We do so at the end of the present
section.

We now compare results of \many\ to \hupp\ and \upp\ for groups  $G$ other
than $SU(2)$, taking the groups in the same order as in \many.
The remarks just made have analogs and will not be repeated.

\bigskip\noindent{\it $SU(3)$ Bundles}
\def\M{{\cal M}}
\def\N{{\cal N}}

$SU(3)$ bundles correspond to unbroken $E_6$.  In the notation of \many,
the structure near $\theta$ to get unbroken  $E_6$ is
\eqn\yut{y^2=4x^3-fxu^3-gu^5-q^2u^4.}
There is a singularity of type $E_6$ at $u=x=y=0$, away from zeroes of $q$.
The fact that the coefficient of $u^4$ is a square prevents a monodromy
that would break $E_6$ to $F_4$.  $g$, $f$ and $q$ are sections of 
$\L^6\otimes {\cal M}$, $\L^4\otimes \M$, and $\L^3\otimes \M$,
that is of $\N$, $\N\otimes \L^{-2}$, and $\N\otimes \L^{-3}$.  
$g$, $f$, and $q$ transform with weight 1 when $u,x,y$ are scaled
with weights $1,2,3$.  

The weights and exponents just obtained agree with \hupp\ and \upp\ for the
case of $SU(3)$.

\bigskip\noindent{\it $G_2$ Bundles}

A reduction of the $E_8$ structure group to $G_2$ leaves unbroken $F_4$.
This 
corresponds in $F$ theory  to a section of $E_6$ singularities with monodromy
allowed.  The analog of \yut\ is 
\eqn\oyut{y^2=4x^3-fxu^3-gu^5-bu^4.}
The only difference from \yut\ is that the coefficient of $u^4$ is not required
to be a perfect square.
$g,f$, and $b$ have weights $1,1,2$ under scalings of $u,x,y$, and 
are sections of $\N$, $\N\otimes \L^{-2}$, and $\N^2\otimes \L^{-6}$, in
agreement with expectations for $G_2$.  In fact, the relation $b=q^2$
between the descriptions for $G_2$ and for $SU(3)$ was already seen in section
3.  The other examples discussed in section three also have analogs in $F$ theory.

In subsequent examples, the precise formulas for behavior along $\theta$
analogous to \oyut\ become more complicated and will not be presented.
Interested readers are referred to \many.  

\bigskip\noindent{\it $Spin(5)$ Bundles}

The commutant of $Spin(5)$ in $E_8$ is $Spin(11)$.  This corresponds in $F$ theory
to having a $D_6$ singularity along $\theta$, with monodromy allowed. 

In comparing to \many\ for $Spin(5)$ and the other examples,
we  use the following conventions.  In \many, various objects are written
as $f_{2n+12}$, $q_{n+6}$, etc.
In general, if the subscript is $a(n+12)-2b$, then in our notation the
corresponding 
object has weight $a$ and is a section of $\N^a\otimes \L^{-b}$.

For instance, according to \many, the $F$ theory locus with $Spin(11)$ gauge symmetry
is  described
by objects $g_{12+n}$, $f_{8+n}$, and $s_{4+n}$.  In our notation,
these objects all have weight one and    are sections of $\N\otimes \L^{-d_j}$
for $d_j=0,2,4$.   These are the expected weights and exponents for $Spin(5)$.

\bigskip\noindent{\it $Spin(6)$ Bundles}

$Spin(6)$ bundles correspond in $E_8$ to unbroken $Spin(10)$. In $F$ theory,
this corresponds to a section of $D_5$  singularities without monodromy.
According to \many, such a section is described by objects
$h_{n+4},\,q_{n+6},\,g_{n+12},\,$ and $f_{n+8}$, that is to say objects
of weight 1 and exponents $d_j=0,2,3,4$.  These are the expected values
for $Spin(6)=SU(4)$.

\bigskip\noindent{\it $Spin(7)$ Bundles}

We conclude with one more example.  (Many more cases are worked out in \many; the
interested reader can verify that in each case, the weights
and exponents are as expected from our analysis of bundles.)  $Spin(7)$ bundles
correspond to unbroken $Spin(9)$ and to a section of $D_5$ singularities
with ${\bf Z}_2$  monodromy.  In \many, the moduli are described by objects
$g_{n+12},f_{n+8},h_{n+4},$ and $g_{2n+12}$, or in other words objects of 
weights $1,1,1,2$ and exponents $d_j=0,2,4,6$.  This is as expected for $Spin(7)$.

\subsec{Localization Of Cohomology}

One of the insights in \many, further explored in \katz, was that if
$V$ is a  $G$ bundle over an elliptic manifold $Z\to B$, then, depending on $G$,
certain cohomology groups of $V$, which in physics determine the spectrum of light
quarks and leptons, appear to be localized on  certain
subvarieties of $B$.  (The case considered in detail was the case that $B$ is a curve
and a subvariety is therefore  a finite set of points.)
As promised above, we will here explain directly from the bundle point of
view why this is so.  We will also explain why other cohomology groups
are {\it not} localized in this way.

We illustrate the idea with the case (particularly important in applications)
in which $G=SU(n)$.  Let $V$ be a rank $n$ complex vector bundle over $Z$,
constructed by a spectral cover as in section 2.
Suppose that one wants to compute $H^1(Z,V)$.  

If we think of the fibers of $\pi:Z\to B$ as being small, the first
step would clearly be to solve the $\bar\partial $ equation along
the fibers, and then solve for the adiabatic motion along the base.
In fact, in complex geometry, there is a systematic procedure (the Leray
spectral sequence) to compute $H^i(Z,V)$ starting with a computation
of $H^j(E_b,V)$, where $E_b$, for $b\in B$, is the fiber of $Z$ over $b$.  
The result is in particular that $H^i(Z,V)$
is localized along those fibers that have the property that $H^j(E_b,V)$
is non-zero for some $j\leq i$.  

In our problem, along a generic fiber, $V$ splits as a sum of line
bundles none of which are trivial.  It is therefore the case that for generic
$b$, $H^j(E_b,V)=0$ for all $j$. 
The computation of $H^i(Z,V)$ will be localized
along the locus in $B$ on which one of the factors of $V$ {\it is} trivial.

We assume as usual that $Z$ is presented in Weierstrass form and that
the spectral cover is defined by an equation of the form familiar from section 2:
\eqn\numo{a_0+a_2x+a_3y+\dots +a_nx^{n/2}=0}
 (if $n$ is odd the last term is slightly different).
We want to find the condition on
$b\in B$ so that when restricted to $E_b$, $V$ does have a trivial factor.
The condition is simply
\eqn\tumo{a_n=0.}
For $a_n=0$ is the condition under which one of the roots of \numo\ is
at $x=\infty$, which is the point on $E_b$ that corresponds to a trivial
line bundle.

So the computation of $H^1(Z,V)$ will be localized on the subvariety of $B$
defined by vanishing of $a_n$.
If $Z$ is a K3 surface and $B$ is a curve, then
\tumo\ defines a finite set of points.  A universal local computation
shows that each simple zero of $a_n$ 
will contribute a one-dimensional subspace to $H^1(Z,V)$.  In a
higher-dimensional case, \tumo\ defines a hypersurface $D$ in $B$,
and $H^1(Z,V)$ must be computed by solving an appropriate $\bar\partial$
equation along $D$.  

Apart from computing the cohomology of $V$, one also wishes to compute
the cohomology of other bundles derived from $V$, such as the second
exterior power  $\wedge^2V$.  The basic idea is similar: on a generic
$E_b$, $\wedge^2 V$ splits as a sum of line bundles, and $H^i(Z,\wedge^2V)$
will be localized along those $E_b$ on which one of the line bundles
is trivial.

If along $E_b$, $V=\oplus_{i=1}^n \L_i$, then $\wedge^2V=\oplus_{1\leq i
<j\leq n}\L_i\otimes \L_j$, so $\wedge^2V$ contains a trivial line bundle
if and only if for some $i<j$, $\L_i=\L_j^{-1}$.  Inverse line bundles
correspond to points on $E_b$ that differ by $y\to -y$, so the 
localization will be on $b$'s such that the spectral equation \numo\
has two solutions that differ by $y\to -y$.  If in other words we write
the spectral equation as
\eqn\ofumo{P(x)+yQ(x)=0,}
where $P$ and $Q$ are polynomials in $x$ only, then the condition is
that $P$ and $Q$ have a common zero; in other words, the resultant $R(P,Q)$
should vanish.  The computation of $H^i(Z,\wedge^2V)$ will be localized
on the hypersurface in $B$ defined by $R(P,Q)=0$.

Let us make this more explicit for small values of $n$ (which are of
particular interest for applications).
The first non-trivial case is
$n=4$.  In this case, vanishing of $Q$ reduces to $a_3=0$, and 
$H^i(Z,\wedge^2V)$ will be localized on the hypersurface defined by that
equation.

The first case in which one really sees the resultant is $n=5$, for which
\eqn\ibbo{\eqalign{P& = a_0+a_2x+a_4x^2 \cr
                   Q & = a_3+a_5x.\cr}}
Solving the second equation for $x$ and substituting in the first, we see
that the condition for a common zero of $P$ and $Q$ is
\eqn\hibbo{a_0a_5^2-a_2a_3a_5+a_4a_3^2=0,}
and this equation defines the hypersurface in $B$ along which 
$H^2(Z,\wedge^2V)$ will be localized.  
The detailed formula for the resultant becomes
increasingly complicated for larger $n$.

It is not true that the cohomology with values in {\it any} representation
has such a localization.  For instance, there is no such localization
for $H^i(Z,{\rm ad}(V))$.  The reason is that ${\rm ad}(V)$ contains
trivial sub-bundles -- associated with the Cartan subalgebra -- on a generic
$E_b$.  In general, the cohomology is localized precisely for those
representations that contain no vector invariant under a maximal torus,
so that on a generic $E_b$ there is no trivial sub-bundle.
This statement holds for arbitrary $G$, not just the case $G=SU(n)$ where
explicit formulas can be worked out  using the spectral covers.  

All of these assertions are in full agreement with what has been guessed or
calculated on the $F$-theory side in \refs{\many,\katz}.

\newsec{Computations Of Characteristic Classes}

\def\T{{\cal T}}
\def\M{{\cal M}}
\def\Z{{\bf Z}}
\def\P{{\cal P}}
\def\R{{\bf R}}
Our goal in this section is to understand better the $G$ bundles
that we have constructed on elliptic manifolds $Z\to B$ by computing their
characteristic classes, and to use this information for another test
of duality.

To be more precise, every simple Lie group $G$
has $H^3(G,\Z)=\Z$, so that a $G$ bundle always has a four-dimensional
characteristic class $\lambda$.  We focus on the case that $G$ is connected
and simply-connected; then the homotopy groups $\pi_i(G)$ vanish for $i<3$,
and $\lambda$ is the first non-trivial characteristic class of a $G$ bundle.

For $G=SU(n)$, $\lambda$ is the usual second
Chern class (of the associated rank $n$ complex vector bundle).
For $G=Spin(n)$, $\lambda $ is one half of the usual  first Pontryagin class
(of the associated rank $n$ vector bundle).  

We have given in this paper two constructions of $G$ bundles that can be
used to compute their characteristic classes -- the constructions via
parabolic subgroups and via spectral covers.  The parabolic construction
gives a simple method to compute characteristic classes for any $G$;
however, it is not completely general at present because we do not understand
the analog of ``twisting'' by a line bundle on the spectral cover.  The spectral
cover construction in the explicit form discussed in section 2
is limited mainly to $SU(n)$ and $Sp(n)$, and leads to much more
complicated computations, but has the virtue that one can incorporate such
twists.

In the next subsection, we compute $\lambda$ via parabolics for two
cases: $G=SU(n)$ and $G=E_8$.  $SU(n)$ was chosen for illustration and 
to permit comparison
with spectral covers, and $E_8$ was chosen because the computation of
$\lambda$ for $E_8$ bundles will make it possible to resolve a longstanding
mystery about $F$ theory, which is the appearance of certain ``three-branes''
in the vacuum.
After settling that issue in section   7.2, we go on in section 7.3
to compute $\lambda$ via spectral covers
for $SU(n)$ bundles.

\subsec{Computation Via Parabolics}

The basic idea of constructing stable $G$ bundles via parabolics is that
one first defines a very simple unstable $G$ bundle and then one deforms
it to become stable.  For the sake of computing characteristic classes, the second
step is unnecessary; the topology of the bundle is in any case invariant
under deformations.  So we can compute directly for the unstable bundle,
and this makes things simple.

For instance, for $G=SU(n)$ the starting point is the unstable bundle
\eqn\ubv{V=W_k\otimes\M\oplus W^*_{n-k}\otimes \M'}
with some $k$ in the range $1\leq k\leq n$, and $\M,\M'$ two line bundles
over $B$. For the
purposes of computing Chern classes, $W_k$ can be replaced by a direct
sum $\O(\sigma)\oplus \L\oplus \L^{2}\oplus\dots \oplus \L^{k-1}$,
and likewise $W_{n-k}^*$ can be replaced by $\O(\sigma)^{-1}\oplus 
\L^{-1}\oplus\L^{-2}\oplus \dots\oplus \L^{-(n-k-1)}$.  (This can be proved using
the exact sequences by which $W$ and $W^*$ are defined.)  $\M$ and $\M'$ should
be constrained so that $V$ has  trivial determinant; this means
that $\M^k\otimes(\M')^{n-k}\otimes \L^{-\half (n-1)(n-2k)}\cong \O$.

It is straightforward to compute the second Chern class of $V$, using the
fact that if $V=\oplus_{i=1}^n\L_i$, then
\eqn\gogn{c_2(V)=\sum_{i<j} c_1(\L_i)c_1(\L_j).}
Even without computation, it is evident that the answer is a polynomial
in $c_1(\O(\sigma))$, $c_1(\O(\L))$, and $c_1(\M)$.
We exhibit the formula only in a comparatively
simple case that we will use later in comparing
to results obtained from spectral covers.  This is the case
that $n$ is even, $k=n/2$, and $\M'=\M^{-1}$.
In this case, if we set $\sigma=c_1(\O(\sigma))$, and
\eqn\juddo{c_1(\M)=-{1\over 2}\left(\eta-c_1(\L)\right),}
then we get\foot{Here and for the $E_8$ calculation given below, one needs to
know that $\sigma^2=-\sigma\cdot c_1({\cal L})$, a relation proved in section
7.2.}
\eqn\ctwo{c_2(V)=\eta\sigma-{1\over 24}c_1(\L)^2(n^3-n) -{n\over 8}\eta\left(\eta-
 nc_1(\L)\right).}
This formula shows that the interpretation of $\eta$ is that
\eqn\pitwo{\eta= \pi_*(c_2(V)).}
Here $\pi_*$ is the operation of ``integrating over the fibers'' of the elliptic
fibration $\pi:Z\to B$.  Clearly, not all values of $\eta$ are possible; one has
\eqn\ipolo{\eta\equiv c_1(\L)~{\rm modulo}~2.}

{}From \ctwo, we see that $c_2(V)$ is of the form
\eqn\xuddo{c_2(V)=\eta\sigma+\pi^*(\omega),}
with $\omega\in H^4(B,\Z)$.  In this way of writing things, $\eta$ and $\omega$
are uniquely determined.  If we fix the elliptic manifold $Z\to B$, so 
that $\sigma$ and $c_1(\L)$ are fixed, then according to \juddo, $\eta$
is arbitrary (apart from the mod two condition) and a choice of $\eta$
fixes $\M$.  There is no additional freedom in the construction;
$\omega$ is uniquely determined in terms of $\eta$ and $Z$ by the formula given in
\ctwo.

In the spectral cover construction, as we see later, this relation
can be modified by twisting by a line bundle on the spectral cover.
As we will show elsewhere \fmw, in special cases this freedom can be seen
in the construction via parabolics by taking $k\not= n/2$.  For other groups,
we do not know the analog of twisting by a line bundle on the spectral cover.

\bigskip\noindent{\it Characteristic Class Of $E_8$ Bundles}

Now we consider the case of $E_8$.

The starting point in building an $E_8$ bundle $V$ via parabolics is to consider
a bundle whose structure group reduces
to a group that is locally $SU(6)\times SU(2)\times SU(3)$ (and even
to a subgroup thereof).  We thus need to describe $SU(6)$, $SU(2)$, and $SU(3)$
bundles over $B$ that we will call $X_6,$ $X_2$, and $X_3$. 

The fundamental characteristic class $\lambda(V)$ of
an $E_8$ bundle whose structure group reduces to $SU(6)\times SU(2)\times SU(3)$
can be described very simply: it is\foot{The characteristic class $\lambda(V)$ is
simply $c_2(V)/60$.  The following formula is computed directly using this fact
and the form of the embedding of $SU(6)\times SU(2)\times SU(3)$ in $E_8$.}
\eqn\plopp{\lambda(V)=c_2(X_6)+c_2(X_2)+c_2(X_3).}

\def\S{{\cal S}}
In section 5, in working on a single elliptic curve, we took $X_2=
W_2\otimes \O(\sigma)^{-1/2}$.  Globally, we must modify this slightly.
In view of the exact sequence
\eqn\pmb{0\to \L\to W_2\to \O(\sigma)\to 0,}
the determinant of $W_2$ is $\O(\sigma)\otimes \L$, so we take
\eqn\nlopp{X_2=W_2\otimes\O(\sigma)^{-1/2}\otimes\L^{-1/2},}
which has trivial determinant.
Likewise, the definition $X_3=W_3\otimes\O(\sigma)^{-1/3}$ used in section  5 must
be modified to
\eqn\ilopp{X_3=W_3\otimes \O(\sigma)^{-1/3}\otimes \L^{-1}.}
Finally, in  working on a single elliptic curve, the $SU(6)$ bundle
was $(W_5\oplus \O)\otimes \O(\sigma)^{-1/6}$.  Here we want to consider
a bundle that is isomorphic to this on each fiber and has trivial determinant.
The most general possibility is 
\eqn\hilopp{X_6=\left(W_5\otimes \S^{-1}
\oplus \S^5\otimes \L^{-1}\right)\otimes \O(\sigma)^{-1/6}
\otimes \L^{-3/2},}
with $\S$ an arbitrary line bundle on $B$.
Because of the fractional exponents, $X_2$, $X_3,$ and $X_6$ do not
really make sense as $SU(n)$ bundles, but the fractions disappear when
one puts together an $E_8$ bundle (or an $(SU(6)\times SU(2)\times SU(3))/\Z_6$
bundle).
The fractions cause no harm in computing Chern classes; one simply uses
\gogn\ formally, setting $c_1(\L^\gamma)=\gamma c_1(\L)$ for $\gamma\in{\bf Q}$.

If we set
\eqn\ruttu{\eta=c_1({\cal S})+4c_1({\cal L}),}
then after a calculation that is only somewhat tedious,
the fundamental characteristic class of the $E_8$ bundle comes out to be
\eqn\puttu{\lambda(V)= \eta\sigma -15\eta^2+135\eta c_1(\L) -310 c_1(\L)^2.}
In particular,
\eqn\ippo{\eta=\pi_*(\lambda(V)).}

We again see that 
\eqn\ipuot{\lambda=\eta\sigma + \pi^*(\omega),}
for some $\omega\in H^4(B,\Z)$.  Moreover, as in the $SU(n)$ case,
while $\eta$ can be adjusted independently, $\omega$ is determined
uniquely in terms of $\eta$ and $Z$.  If one wishes to vary $\eta$ and $\omega$
independently, or at least more independently, one must learn the analog
of twisting by a line bundle on a spectral cover.

In the next subsection, we will give strong evidence that 
the $E_8$ bundles that appear in the simplest applications of $F$ theory
are actually the ones whose characteristic class we have just computed.
This will be done by showing that the formula \puttu, with the strange numbers
$-15$, $135$, and $-310$, agrees with expectations from $F$ theory.
Of course, \puttu\ is mainly interesting if $B$ is of dimension bigger than
one; otherwise, for dimensional reasons, $\omega=0$ and the discussion collapses.
So for this purpose we are interested in the case that $B$ is a surface.

\bigskip\noindent
{\it $\Z_2$ Symmetry}

It might at first seem unexpectedly lucky that our simplest construction agrees
with $F$ theory.  We have definitely not analyzed the most general stable $E_8$
bundle over $Z$, perhaps
not even the most general one that is semistable on the generic
fiber.  It may be possible to construct more general bundles by
 an analog of twisting by a line bundle
on the spectral curve.

It seems that one 
reason for our good fortune has to do with an important bit of physics
that we have not yet exploited in this paper.
An elliptic manifold $Z$ with a section $\sigma$ has a $\Z_2$ symmetry,
generated by an ``involution'' $\tau$ that leaves $\sigma$ invariant and
acts as $-1$ on each fiber.
In terms of a Weierstrass model $y^2=4x^3-g_2x-g_3$, $\tau$ is just the
operation $y\to -y$ with fixed $x$.

What does this correspond to on the $F$ theory side?  The elliptic
manifold  $Z\to B$ corresponds in $F$ theory to a manifold $X$ that is fibered
over $B$ with K3 fibers.  The K3's are themselves elliptic, so there is
an elliptic  fibration with section $\pi':X\to B'$, where $B'$ is a ${\bf P}^1$
bundle over $B$.  On the $F$ theory side, there is therefore a
potential $\Z_2$ symmetry $\tau'$.  If one tracks through the
duality between the heterotic string and $F$ theory, one can see that
$\tau$ is mapped to $\tau'$.

Now, $\tau'$ is automatically a symmetry in $F$ theory on $X$ unless one
turns on modes of the three-form field $C$ of eleven-dimensional supergravity
that are odd under $\tau$.  Such modes involve either the intermediate
Jacobian introduced in section 2.4, or the discrete data discussed 
by K. and M.  Becker \beckers (for which we proposed an interpretation
at the end of section 4).  If we suppress the discrete data
and if $H^3(X)$ vanishes 
so that there are no periods, then $\tau'$ is automatically
a symmetry in $F$ theory; so in the corresponding heterotic string story
we want bundles that are invariant under $\tau$.

The $E_8$ bundles whose characteristic class was computed above are $\tau$ invariant,
while one would expect that a generic twist or perhaps any twist of the sort
that we do not presently know how to make would break the symmetry.
(For instance, when we compute for $SU(n)$, where we do understand the possible
twists, we will see that the additional twists give bundles that are not
$\tau$-invariant.)  So it is natural that the bundles that we know how to
construct are the ones that should be compared to the simplest cases of
$F$ theory.

For an $SU(n)$ bundle understood as a rank $n$ complex vector bundle,
the relevant statement of $\tau$-invariance becomes $\tau^*(V)=V^*$.
(Any semistable rank $n$ bundle on a single elliptic curve obeys this
relation -- since any line bundle does -- so it is natural to look
for a component of the moduli space of bundles in which every bundle obeys
$\tau^*(V)=V^*$.)  In the construction via parabolics, starting with
$W_k\oplus W_{n-k}^*$, duality exchanges $k$ with $n-k$, and so the condition
$\tau^*(V)=V^*$ is most easily implemented by taking $k=n/2$, as we did
in arriving at \ctwo.  When we compute via spectral covers where many twists
are possible, we will compare \ctwo\
to the Chern classes of a bundle constructed via spectral curves and obeying
$\tau^*(V)=V^*$.

\subsec{Origin Of $F$ Theory Threebranes}

\def\K{{\cal K}}
One of the very surprising features about $F$ theory compactification
on a Calabi-Yau four-fold $X$ is that a consistent compactification
requires the presence of threebranes in the vacuum \svw.\foot{$F$ theory
is defined as Type IIB superstring theory with a coupling constant that varies
in space-time.  It reduces  at long distances to ten-dimensional Type IIB supergravity
with (among other things) additional ``threebranes.''  A threebrane is a sort of impurity
near which the supergravity description breaks down; 
its worldvolume is a four-dimensional submanifold of spacetime.  In
the present discussion, spacetime is ${\bf R}^4\times B'$ (where $B'$ is the base
of the elliptic fibration with total space $ X$), and to maintain four-dimensional
Poincar\'e invariance, the four-manifolds in question
are of the form ${\bf R}^4\times p_i$, where the $p_i$ are points in $B'$.  The number
of threebranes, that is the number of points $p_i$, was determined in \svw\ by observing
that the supergravity equations have a solution only if the correct number of impurities
is included.}
  The number of
threebranes is $I=\chi(X)/24$, where $\chi$ is the topological Euler characteristic.
If $X$ has an elliptic fibration $\pi':X\to B'$ with a smooth Weierstrass
model, then one can prove as in \svw\ that 
\eqn\loopo{I=12+ 15\int_{B'}c_1(TB')^3,}
where $ c_1(TB')$ is the first Chern class of the {\it tangent} bundle of
$B'$.  

Under duality with the heterotic string, the threebranes turn into fivebranes
that are wrapped over fibers of $Z\to B$, and the question is why such
fivebranes should be present.  The explanation depends upon  heterotic
string anomaly cancellation.  Perturbative anomaly cancellation without fivebranes
requires an $E_8\times E_8$ bundle $V_1\times V_2$ with
\eqn\otto{\lambda(V_1)+\lambda(V_2) = c_2(TZ).}
($TZ$ is the tangent bundle of $Z$.)  The general anomaly cancellation
condition with fivebranes is 
\eqn\jotto{\lambda(V_1)+\lambda(V_2)+[W] = c_2(TZ).}
where $[W]$ is the cohomology class of the fivebranes.

It has been suspected in the past that the reason that fivebranes appear is
that $\lambda(V_1)$ and $\lambda(V_2)$ cannot be varied freely
and that, after adjusting $\pi_*(\lambda(V_1))$ and $\pi_*(\lambda(V_2))$ to
specified values whose sum equals $\pi_*(TZ)$, \otto\ would be in error
by the pullback of a cohomology class from $B $.  Any such class is of the form $h[p]$,
where $h\in {\bf Z}$ and
$[p]$ is the class of a point on $B$.  Suppose that  \jotto\ is obeyed with
\eqn\ibbo{[W]=h[F],}
where $[F]=\pi^*([p])$ is the class of a fiber of the elliptic
fibration.  In that case, $h$ will be the number of fivebranes, on the heterotic
string side, and should coincide with the number of threebranes seen in  $F$ theory.

By now, we have seen that the $\lambda(V_i)$ obey restrictions of the appropriate
kind, and we have the information in hand to compute $h$ and verify that $h=I$,
finally giving a heterotic string explanation of the number of threebranes.
To do this, we will have to make some computations of Chern classes.

\bigskip\noindent{\it Reduction Of $F$ Theory Formula}

\def\T{{\cal T}}
First, as a preliminary, we need to make a further reduction of the $F$ theory
formula \loopo, for the case relevant to (the simplest versions of)
heterotic string/$F$ theory duality.  This is the case that $B'$ is a ${\bf P}^1$
bundle over $B$, the ${\bf P}^1$ bundle being the projectivization of 
a vector bundle $Y=\O\oplus {\cal T}$, with ${\cal T}$ a line bundle over
$B$.  We endow the ${\bf P}^1$ bundle with homogenous coordinates $a,b$
which are sections of $\O(1)$ and $\O(1)\otimes {\cal T}$, respectively;
here $\O(1)$ is a bundle that restricts on each ${\bf P}^1$ fiber to the line
bundle that usually goes by that name.  If we set $r=c_1(\O(1))$, $t=c_1(\T)$,
then the fact that the sections $a$ and $b$ of the line bundles $\O(1)$ 
and $\O(1)\otimes \T$ over $B'$ have no common zeroes means that
$r(r+t)=0$ in the cohomology ring of $B'$.  

Let $c_1(B)$ and $c_2(B)$ denote the Chern classes of the {\it tangent} bundle
of $B$.  When confusion is unlikely we will call these simply $c_1 $ and $c_2$,
and likewise we write simply $c_i$, rather than $\pi^*(c_i)$, etc., for the
pullbacks of the $c_i$  under the various fibrations such as $\pi$  and $\pi'$.

A standard adjunction formula says that the total Chern class
of the tangent bundle of $B'$ is
\eqn\onormo{c(B')=(1+c_1+c_2)(1+r)(1+r+t).}
Hence, the first Chern class of $B'$ is  $c_1'=c_1+2r+t$.  One can now
evaluate \loopo\ in terms of the geometry of $B$.  Let $\pi'':B'\to B$ be
the projection. Using $r(r+t)=0$ to reduce $(c_1')^3$ to a linear function of $r$
and then using $\pi''_*(r)=1$, $\pi''_*(1)=0$, we can compute $\pi''_*((c_1')^3)$
and thereby get the following formula expressing the number of threebranes in terms of 
data defined on $B$:
\eqn\juggo{ I=12+90\int_Bc_1^2+30\int_Bt^2.}
The base $B$ of a Calabi-Yau elliptic fibration is rational and hence obeys
\eqn\orgo{12=\int_B\left(c_1^2+c_2\right).}
We can combine the last two expressions and write
\eqn\ojuggo{I=\int_B\left(c_2+91c_1^2+30t^2\right).}

\bigskip\noindent{\it Computations On Heterotic String Side}

Now we can compute on the heterotic string side.
First of all, the conjectured duality between $F$ theory and the heterotic
string says that $F$ theory on the fourfold $X\to B'\to B$, with $B'$ as above,
should be compared to the heterotic string on $\pi:Z\to B$ with $E_8\times E_8$ bundles
that are chosen so that
\eqn\upoggo{\eta_1=\pi_*(\lambda(V_1))=6c_1+t,~~\eta_2=\pi_*(\lambda(V_2))=6c_1-t.}
This is the generalization of the more familiar statement \mv\ that $F$ theory
on the Hirzebruch surface ${\bf F}_n$ corresponds to a heterotic string on K3
with $12+n$ instantons in one $E_8$ and $12-n$ in the other.  In other
words, when the base ${\bf P}^1$   
of the Hirzebruch fibration ${\bf F}_n\to {\bf P}^1$
is replaced by a surface $B$ (which is the base of $B'\to B$),
the generalization of the number $12 $ is the cohomology class $6c_1(B)$,
and the generalization of the number $n$ is the cohomology class $t$.

The $V_i$ are uniquely determined (at least within the
class of bundles we are considering) by specifying $t$ and hence the $\eta_i$,
 and then via \puttu\
the $\lambda(V_i)$ are  determined, given $c_1({\cal L})$.  Also, it is appropriate 
now
to impose the Calabi-Yau condition $c_1(\L)=c_1(B)=c_1$. We get 
\eqn\hormo{\lambda(V_1)+\lambda(V_2)= -80c_1^2+12\sigma c_1-30t^2.}

To proceed further we need $c_2(TZ)$, which is the remaining unknown in \jotto.
This can be computed by the same methods we used to arrive at \juggo.
The Weierstrass equation $zy^2=4x^3-g_2xz^2-g_3z^3$
embeds $Z$ in a ${\bf P}^2$ bundle $W\to B$.  $W$ is
the projectivization of a sum of line bundles $\L^2,\L^3$, and $\O$. $W$ has
homogeneous coordinates $x,y$, and $z$ which we interpret as sections of $\O(1)\otimes
\L^2,\O(1)\otimes \L^3$, and $\O(1)$ over $W$; we set $r=c_1(\O(1))$.
The total Chern class $c(Z)=1+c_1(Z)+c_2(Z)+\dots$ of $Z$ is given by adjunction as
\eqn\tutut{c(Z)=c(B){(1+r)(1+r+2c_1({\cal L}))(1+r+3c_1(\L))\over 1+3r+6c_1(\L)}.}
The denominator expresses the fact that the Weierstrass equation is a section
of $\O(1)^3\otimes \L^6$. The fact that $x,y$, and $z$ have no common 
zeroes means that $r(r+2c_1(\L))(r+3c_1(\L))=0$ in the cohomology ring of $W$.
Since multiplication by $3(r+2c_1(\L))$ can be understood as restriction from
$W$ to $Z$ (which is defined as we just said by vanishing of a section of
$\O(1)^3\otimes \L^6$), the relation for $r$ simplifies in the cohohomology
ring of $Z$ to $r(r+3c_1(\L))=0$.
 As the section $z$ of $\O(1)$ vanishes
on $\sigma$ with multiplicity 3 ($\sigma$ can be described in homogeneous
coordinates by $(x,y,z)=(0,1,0)$, and we see that near $\sigma$, $z$ has
a third order zero, being given by $z\sim x^3$), one has $r=3\sigma$ in the
cohomology ring of $Z$.
With a little patience one can use these facts and expand \tutut\ to learn that
\eqn\wutut{c_2(TZ)= c_2+11c_1^2+12\sigma c_1.} 

Everything is now in place to evaluate \jotto.  Using \wutut\ and \hormo\
we see that \jotto\ is obeyed if and only if $W=h[F]$ with
\eqn\hypo{h=c_2+91c_1^2+30t^2.}
Using \ojuggo,    one can see that this amounts to
the statement that $h$, the number of fivebranes required on the heterotic
string side, equals $I$, the number of threebranes required in the $F$ theory
description.  So, as promised, we have obtained from the heterotic string
point of view some understanding of the appearance of threebranes in 
$F$ theory compactification.

For further use, note that because $r(r+3c_1(\L))=0$ and $r=3\sigma$, we have
obtained the relation
\eqn\gnon{\sigma^2=-\sigma c_1(\L)}
which entered at several points in this paper and is further used below.  
 The relation $r=3\sigma$
that we just exploited means that the line bundle $\O(1)$ over
$W\to B$, when restricted to $Z$, obeys $\O(1)\cong \O(\sigma)^3$.  (The assertions
of this paragraph do not require the Calabi-Yau condition and hold for any $\L$.)

\subsec{Computation Via Spectral Covers}

For the rest of this section, 
our goal will be to compute Chern classes using the description of
bundles over   $Z$  by spectral covers.  We do not assume that $Z$ is Calabi-Yau,
so $c_1(\L)$ and $c_1(B)$ are unrelated.  We otherwise use the same notation as above;
$Z$ is embedded in a ${\bf P}^2$ bundle $W\to B$ via a Weierstrass equation, and
has a section $\sigma$.  We will incorporate the twisting by a line bundle that was
explained in section two.

\def\O{{\cal O}}

The spectral cover  $C$ introduced in section 2 is given by an equation
$s=0$ which defines a hypersurface in $Z$.  $s$ is a section
of $\O(\sigma)^n\otimes {\cal M}$, where ${\cal M}$ is an arbitrary
line bundle over $B$; we set $\eta'=c_1(\cal M)$.  (In eqn. (7.70), we will
see that $\eta'$ coincides with $\eta$ as introduced earlier.)
  Concretely in affine coordinates with $z=1$,
\eqn\ocl{s=a_0+a_2x+a_3y+\dots +a_nx^{n/2} }
(the last term is $x^{(n-3)/2}y$ if $n$ is odd).
Here $a_0$ is a section of 
$\M$, and  ($x$ and $y$ being sections of $\L^2$ and $\L^3$ in the Weierstrass
model)
$a_r$ is a section of $\M\otimes \L^{-r}$.  $s$ has a pole of order $n$ at $x=y=
\infty$,  which is why it is a section of $\M\otimes \O(\sigma)^n$.

\def\P{{\cal P}}
\def\L{{\cal L}}
We recall that $SU(n)$ bundles on $Z$ are constructed as follows.
Let $\P_B$  be the Poincar\'e line bundle on $Z\times_B Z$, which
we restrict to $C\times_BZ$, and let $\N$ be an arbitrary line
bundle over $C$.  Let $\pi_2$ be the projection of $C\times_BZ$ to the
second factor.  The vector bundle over $Z$ that we want to study
is then 
\eqn\defby{V=\pi_{2*} (\N\otimes \P_B).}

\def\ch{{\rm ch}}
\def\Td{{\rm Td}}
Now let us explain the basic strategy for computing Chern classes.
If $G$ is any vector bundle on $Z$, then the index of the $\bar\partial$
operator on $C\times _BZ$, with values in $\N\otimes \P_B\otimes \pi_2^*G$, would
be 
\eqn\nefby{\int_{C\times_BZ} e^{c_1(\N\otimes \P_B) } \ch(\pi_2^*G)
\Td(C\times_BZ),}
where $\ch$ is the Chern character, $\Td$ is the Todd class, and
$\pi_2^*G$ is the ``pullback'' of $G$ to $C\times_BZ$.
The index of the $\bar\partial$ operator on $Z$, with values in
$V\otimes G$, is
\eqn\wefby{\int_Z \ch(V)\ch(G) \Td(Z).}
The Hirzebruch-Riemann-Roch or Atiyah-Singer index theorem says that (with $V$ defined in
\defby), these are equal.  But $G$ is arbitrary, and therefore $\ch(G)$
is essentially arbitrary.  For \nefby\ and \wefby\ to be equal
for any $G$ implies a relationship between the integrands that is known
as the Grothendieck-Riemann-Roch theorem (GRR):
\eqn\corby{\pi_{2*} \left(e^{c_1(\N\otimes \P_B)} \Td(C\times_BZ)\right)
  =\ch(V) \Td(Z).}
Here $\pi_{2*}$ is, at the level of differential forms, the operation
of ``integrating over the fibers'' of the map  $\pi_2:C\times_BZ\to Z$.
(As this map is an $n$-fold cover, integration over the fibers is in this
case somewhat akin to taking a finite sum.)
Since everything else in \corby\ can be computed independently,
\corby\ will serve to determine the Chern classes of $V$.

Our main goal is to compute $c_1(V)$ and $c_2(V)$.  With this
in mind, we expand various factors in \corby\ up to the relevant order.
We have
\eqn\moby{\ch(V)=n+c_1(V)+{1\over 2}c_1(V)^2-c_2(V)+\dots}
where $n$ enters because $V$ is of rank $n$.
For any complex manifold $X$,
\eqn\noby{\Td(X)=1+{c_1(X)\over 2}+ {c_2(X)+c_1(X)^2\over 12}+\dots.}
So given a choice of $\N$ and hence of $c_1(\N)$, if we can
compute the Chern classes of $Z$ and of $C\times_BZ$, then
\corby\ will determine the Chern classes of $V$.

\subsec{The First Chern Class}

Our first task is to compute the first Chern class $c_1(V)$.
For this purpose, a few simplifications occur.  The construction
of $V$ in section 2 ensured that $V$ is an $SU(n)$ bundle when
restricted to any fiber of $\pi:Z\to B$.  So $c_1(V)$ vanishes
when restricted to a fiber, and is therefore determined by its
restriction to $B$, that is, to the section $\sigma$ of $\pi$.  

So instead of working on $C\times_BZ$, we can restrict to $C\times_B\sigma
=C$.  Therefore, instead of using GRR 
for $\pi_2:C\times_BZ\to Z$, we can use GRR for the projection
$\pi:C\to B$ (which is just the restriction to $C$ of $\pi:Z\to B$).
In writing GRR for $\pi:C\to B$, we can moreover set $c_1(\P_B)$ to zero,
since $\P_B$ was defined to be trivial when restricted to $Z\times_B\sigma$.
So we get 
\eqn\orby{\pi_*\left(e^{c_1(\N)}\Td(C)\right) = \ch(V)\Td(B).}
With $\Td(B)=1+c_1(B)/2+\dots$, along with 
$e^{c_1(\N)}=1+c_1(\N)+\dots$, and ${\rm ch}(V)=n+c_1(V)+
\dots$,  the formula
for the first Chern class becomes
\eqn\forby{\pi_*\left(c_1(\N)+{1\over 2}c_1(C)\right) ={n\over 2}
c_1(B)+c_1(V).}

We want to determine the $\N$'s for which $c_1(V)=0$.  The condition
for this is
\eqn\jorby{\pi_*(c_1(\N))= -{1\over 2}\pi_*\left(c_1(C)-\pi^*c_1(B)\right).}
($\pi^*$ is the operation of ``pullback,'' and because the map $\pi:C\to B$
is an $n$-fold cover, $\pi_*\pi^*c_1(B)=nc_1(B)$.)  This
says that
\eqn\ha{c_1(\N)=-{1\over 2} \left(c_1(C)-\pi^*c_1(B)\right)+\gamma,}
where  $\gamma$ is a class such that
\eqn\ga{\pi_*\gamma=0.}

Now, let us discuss the significance of this for the two basic
cases: $B$ a curve, and $B$ of dimension greater then one.

\bigskip\noindent{\it $B$ A Curve}

If $B$ is a curve, then $C$ is also a curve, and the condition
$\pi_*\gamma=0$ implies that $\gamma=0$.  
Hence $c_1(\N)$ is determined uniquely by \ha.

If $K_B$ and $K_C$ are the canonical bundles of $B$ and $C$,
then \ha\ means that 
\eqn\ba{
\N=K_C^{1/2}\otimes K_B^{-1/2}\otimes {\cal F}}
where ${\cal F}$ is
 a flat line bundle over $C$.  This possibility of tensoring with a flat
line bundle means that the Jacobian of $C$ enters the story,
as we saw in section 2.   Note that a curve is a spin manifold,
so that  square roots $K_C^{1/2}$ and $K_B^{1/2}$ do exist.

\bigskip\noindent{\it $B$ Of Higher Dimension}

Now, consider the case that $B$ is of dimension greater than one.
In this case, in many interesting examples $H^{1,0}(C)=0$, and the classification of
line bundles on $C$ is discrete; $\N$ is then uniquely determined up
to isomorphism by its first Chern class.

Now, let us ask what $\gamma$ can be.  $c_1(\N)$ will necessarily
be an integral class of type $(1,1)$.  Such classes on $C$ are relatively
scarce for the following reason.  Although $H^{1,0}(C)=0$,
$H^{2,0}(C)$ is generically non-zero for the $C$'s of interest (with
possible exceptions for small $\eta'$ and $n$);
this tends to  prevent  the existence of many $(1,1)$ integral classes.

The only obvious such classes on $C$ are the cohomology class of
the section $\sigma $, and the pullbacks $\pi^*\beta$ of integral $(1,1)$ classes
$\beta$ on $B$.  We will compute presently (eqn. (7.56)) that
\eqn\kum{\pi_*\sigma = \eta'-nc_1(\cal L).}
Also,
\eqn\um{\pi_*\pi^*\beta = n\beta}
because $\pi:C\to B$ is an $n$-sheeted cover.
So
\eqn\pum{\pi_*(n\sigma-\pi^*\eta'+n\pi^*c_1({\cal L}))=0.}
And this is the only general construction of a class annihilated by $\pi_*$.

So we can take
\eqn\prum{\N=K_C^{1/2}\otimes K_B^{-1/2}\otimes
\left(\O(\sigma)^n\otimes \M^{-1}\otimes \L^{n}\right)^\lambda}
for suitable $\lambda$.

Actually, there is a subtlety here.  The square root $(K_C\otimes K_B^{-1})
^{1/2}$ may not exist, and if it does not one may not take $\lambda=0$; in fact,
$\lambda$ must be half-integral, 
and there is a restriction on $\M$ so that ${\cal N}$ actually exists as a line bundle.
However, for a reason that we will now explain, $\lambda=0$ is
the appropriate case for the simplest tests of the duality between
$F$-theory and the heterotic string. 

\bigskip\noindent
{\it Involution And Duality}

We recall that the elliptic fibration $Z\to B$ has a $\Z_2$ symmetry $\tau$ which
acts on the Weierstrass model by $y\to -y$, while leaving $x$ and $z$
unchanged.  ($\tau$ multiplies each fiber of the elliptic fibration by $-1$.)
As we already explained at the end of section 7.1, 
the most easily seen $F$ theory moduli
are all invariant under $\tau$, so in comparison with $F$ theory there
is particular interest to construct components of the moduli space of bundles
that are entirely $\tau$-invariant.  In the case of a rank $n$ complex
vector bundle, we want $\tau$-invariance in the sense that $\tau^*(V)=V^*$.

In fact, \prum\ with $\lambda=0$, that is the existence of an isomorphism
\eqn\brum{\N^2=K_C\otimes K_B^{-1},}
is the condition for $\tau^*V=V^*$.  To see this, note first that
the condition $\tau^*V=V^*$
says that there is a non-degenerate map $s:V\otimes \tau^*V\to \O$.  This
 is equivalent to the following.  If $\omega$
is a meromorphic one-form on $Z$ with poles on a divisor $D$,
there should be a non-degenerate residue map $\phi_\omega:
V\otimes \tau^*V\to \O_D $ obeying certain standard axioms.  One simply
defines $\phi_\omega(v,w)={\rm Res}(s(v,w)\omega)$ where ${\rm Res}$ is
the residue operation.
\brum\ lets us
construct $\phi_\omega$ as follows.

We let $\tau$ act on $C\times_B Z$ through its action on the second
factor.   Then $\tau^*(\N)=\N$ (since $\N$ is pulled back from $C$, on which
$\tau$ acts trivially), and $\tau^*(\P_B)=\P_B^{-1}$ (a basic property of the
Poincar\'e line bundle).  Hence $(\N\otimes\P_B)\otimes \tau^*(\N\otimes\P_B)
=(\N\otimes\P_B)\otimes(\N\otimes\P_B^{-1})=\N^2$,
and an isomorphism as in \brum\ gives a map of this to $K_C\otimes K_B^{-1}$.
We let $\theta$ be
the composite map $(\N\otimes\P_B)\otimes \tau^*(\N\otimes\P_B)\to K_C\otimes
K_B^{-1}$.  Given sections $v'$ and $w'$ 
of $\N\otimes \P_B$ and $\tau^*(\N\otimes \P_B)
$, $\theta(v',w')\otimes \omega$ 
is a meromorphic section of $K_C$, and 
 we define $\phi'_\omega(v',w')={\rm Res}(\theta(v',w')\otimes\omega)$.
$\phi'_\omega$ 
takes values in $\O_{D'}$, where $D' $ is the divisor in $C\times_BZ$
that lies over $D$.  We then define the desired object $\phi_\omega$
such that if $v=\pi_{2*}(v')$, and $w=\pi_{2*}(w')$, then $\phi_\omega(v,w)
=\pi_{2*}\phi'_{\omega}(v',w')$.  This has the necessary properties to
establish the desired duality between $V$ and $\tau^*V$.  The main point
is to verify non-degeneracy of $\phi_\omega$, which can be checked
by a standard local computation, the interesting detail being that this
works near branch points of $C\to B$.

Thus, to achieve $\tau^*V=V^*$,  we must define $\N$ by \prum, with $\gamma=0$.
This, however, is possible only if the line bundle $K_C\otimes K_B^{-1}$ has
a square root.  A computation we will perform presently shows that 
$c_1(C)=-\eta'-n\sigma+c_1(B)-c_1({\cal L})$, so $c_1(C)-c_1(B)=-\eta' -c_1({\cal L})-
n\sigma$.
The only obvious circumstance  in which this is divisible by two is
that 
\eqn\popp{\eqalign{n & \equiv 0 ~{\rm modulo}~2 \cr
                   \eta' & \equiv c_1({\cal L})~{\rm modulo }~2.\cr}}
These conditions have already been encountered before.  In the construction
of bundles via parabolics, to achieve $\tau$ invariance, we needed
$n$ even (so that we could take $W_n\oplus W_n^*$ as the starting point),
and we found in \ipolo\ that we needed $\eta\equiv c_1({\cal L})$ mod two.
So, except that we have not yet proved that $\eta=\eta'$, we have found the
same mod two conditions in the two approaches.

\bigskip\noindent{\it Computation Of $c_1(C)$}

To justify eqn. \popp, and for later use,
 let us calculate $c_1(C)$.  We already used in section 7.2 the
 fact that the cohomology
of the ${\bf P}^2$ bundle $W\to B$ is generated by $r=c_1(\O(1))$, with
the relation
\eqn\reln{r(r+2c_1({\cal L}))(r+3c_1({\cal L}))=0}
which simplifies in the cohomology ring of $Z$
to 
\eqn\umeln{r(r+3c_1({\cal L}))=0.}
  $C$ is defined inside $W$ by the vanishing of the
Weierstrass equation, which is a section of $\L^{6}\otimes \O(3)$,
and of the section $s$   introduced in \ocl, which is a section of $\M\otimes \O(n/3)$.
(There is a small sleight of hand here: $\O(1)$ has the cube root
$\O(1/3)={\cal O}(\sigma)$ only when restricted to $Z$, but for the computation
that we are about to perform, this is of no moment.)  
It follows from  adjunction that
the total Chern classes $c(C)$ and $c(B)$ of $C$ and $B$ are related
by
\eqn\beln{c(C)=c(B){(1+r)(1+r+2c_1(\L))(1+r+3c_1(\L))\over
                  (1+3r+6c_1(\L))(1+\eta'+{n\over 3}r)}.}
To eliminate the fraction from the denominator, simply recall
that when restricted to $Z$ (and therefore also when restricted to $C$),
$r$ is divisible by 3 and $r/3=\sigma$.
{}From \beln, we get
\eqn\keln{c_1(C) = -\eta'-n\sigma+c_1(B)-c_1(\L).}
(To be more precise, $\eta'$ here could be written as $\pi^*\eta'$,
but we will henceforth not be so fastidious on this and similar points.)
\keln\ is a relation upstairs on $C$.
So the equation \ha\ for $c_1(\N)$ becomes
\eqn\meln{c_1(\N)={1\over 2}\left(n\sigma+\eta'+c_1(\L)\right)
+\gamma.}
For future use, we note also that by expanding \beln\ to higher order one
gets
\eqn\feln{c_2(C)=c_2(B)
+12\sigma c_1(\L)+11c_1(\L)^2+(\eta'+n\sigma)^2-(c_1(B)-c_1(\L))(\eta'
+n\sigma+c_1(\L)).}

Finally, let us by similar methods compute $\pi_*\sigma$, where
$\pi$ is the map $\pi:C\to B$.  $\sigma$ extends over $Z$, and $C$ 
represents in $Z$ the cohomology class $ \eta'+n\sigma$.  So pushing down
$\sigma$ from $C$ to $B$ is the same as pushing down $\sigma(\eta'+n\sigma)$
from $Z$ to $B$.  We saw earlier that $\sigma^2=-\sigma c_1(\L)$,
so $\sigma(\eta'+n\sigma)=\sigma(\eta'-nc_1(\L))$.  The pushdown from $Z$ to $B$
is now easy: $\eta'-nc_1(\L)$ is a pullback from $B$, and integration over the
fiber of $Z\to B$ maps $\sigma$ to 1.  So
finally
\eqn\kimmo{\pi_*(\sigma)=\eta'-nc_1(\L),}
as promised in \kum.

\subsec{The Poincar\'e Line Bundle}

Before trying to compute $c_2(V)$, we need a digression concerning
the Poincar\'e line bundle $\P_B$ over $Z\times_BZ$.  Let $\sigma_1=\sigma\times
_BZ$, $\sigma_2=Z\times_B\sigma$.  We recall that $\P_B$ is defined
by saying that it is trivial when restricted to $\sigma_1$ or $\sigma_2$,
plus the following condition.  Given $b\in B$, the inverse image of $b$
in $Z$ is an elliptic curve $E_b$, and the inverse image in $Z\times_BZ$ is
a copy of $E_b\times E_b$.  The second defining property of $\P_B$ is that
its restriction to each $E_b\times E_b$ is the Poincar\'e line bundle
in the standard sense (which was explained in section 2.3).

$\P_B$ can be described very explicitly as follows.  Its first Chern
class is
\eqn\huft{c_1(\P_B)=\Delta-\sigma_1-\sigma_2-c_1(\L),}
where $\Delta$ is the diagonal in $Z\times_BZ$.  To be more precise,
 one can take
\eqn\nuft{\P_B=\O(\Delta-\sigma_1-\sigma_2)\otimes \L^{-1}.}
This $\P_B$ has the correct restriction to each $E_b\times E_b$, since
(being a pullback from $B$) $\L^{-1}$ is trivial on each $E_b\times E_b$,
and $\O(\Delta-\sigma_1-\sigma_2)$ is the standard Poincar\'e line bundle
on $E_b\times E_b$.  To show that $\P_B$ is trivial when restricted to 
$\sigma_1$ (or equivalently to $\sigma_2$), it suffices to show that
$\sigma_1\cdot c_1(\P_B)=0$.  But we showed above that $\sigma_1\cdot(\sigma_1
+c_1(\L))=0$ in the cohomology ring of $Z$, and 
 $\sigma_1\cdot(\Delta-\sigma_2)
=0$ for a simple geometrical reason (if one is at the standard section
in the first factor of $Z\times_BZ$, then being on the diagonal is equivalent
to being at the standard  section of the second).

For future use, let us note the fact just exploited:
\eqn\futureuse{\sigma_1\cdot\sigma_2=-\sigma_1\cdot \Delta=-\sigma_2\cdot 
\Delta.}
Likewise, one has
\eqn\future{\Delta\cdot \Delta =-\Delta\cdot c_1(\L).}
This can be proved by constructing explicitly a section of the normal bundle
to $\Delta$ with divisor $-c_1(\L)$.  The idea is that if $u$ is a meromorphic
section
of $\L^{-1}$, and $x$ and $y$ are Weierstrass coordinates,
then $\psi=u y \,\,d/dx$ is a meromorphic 
vector field on $Z$ tangent to the elliptic
fibers, whose divisor (being that of $u$) has first Chern class $-c_1(\L)$.
On the other hand, this vector field, taken in, say, the second factor
of $Z\times_BZ$ gives a section of the normal bundle to $\Delta$, and
so $\Delta\cdot \Delta$ is $\Delta$ times the divisor of $\psi$.

\subsec{The Second Chern Class}

Now we will compute $c_2(V)$.  One can see from the GRR formula without any
detailed computation that $c_2(V)$ has  the general form
\eqn\tyro{
c_2(V)=\sigma \cdot \pi^*(\eta)+\pi^*(\omega),} where $\eta$ and $\omega$
are some classes on $B$.

We will first compute the first term in \tyro, which for dimensional reasons is the
only term present if $B$ is a curve.  In any event, the first term
can be detected by restricting to an arbitrary curve $B'\subset B$.
We restrict the elliptic fibration $\pi:Z\to B$ to $Z'=\pi^{-1}(B'),$
and we let $C'=C\cap Z'$.  We will compute the restriction of $c_2(V)$ to
$Z'$ -- which sees the $\sigma\cdot\pi^*\eta$ term --
by using the GRR theorem for the projection $\pi_2:C'\times_{B'}Z'\to Z'$:
\eqn\normo{\pi_{2*}\left(e^{c_1(\N)+c_1(\P_B)}\Td(C'\times_{B'}Z')\right)
 =\ch(V)\Td(Z').}

Note that, even if $Z$ is Calabi-Yau, $c_1(Z')$ is generally non-zero.
However, because the restriction of $c_1(Z')$ to the elliptic fibers
vanishes, $c_1(Z')$ is a pullback from $B'$.

A drastic simplification occurs in evaluating the left hand side of \normo\
because $C'$ and $B'$ are curves.  If $\alpha,\beta$ are any two
two-dimensional classes pulled back from $C'$ or $B'$ -- such as $\sigma,
\eta', c_1(\N), $ or $c_1(Z')$ -- then $\alpha\beta=0$.   Also,
for such an $\alpha$, 
\eqn\foggy{\alpha\cdot (\Delta-\sigma_2)=0}
 in cohomology,
because the left hand side 
is annihilated by the projection from $C'\times_{B'}Z'$ to $C'$
(both $\Delta$ and $\sigma_2$ pick out one point on the fiber of this
projection; these contributions cancel).   With these simplifications,
the left hand side of \normo\ collapses to
\eqn\ormo{\pi_{2*}\left(e^{c_1(\P_B)}\left(1+{c_2(Z')\over 12}\right)\right),}
and the four-dimensional class obtained by expanding this is
\eqn\gormo{\pi_{2*}\left({c_1(\P_B)^2\over 2}\right)+{nc_2(Z')\over 12}.}
The right hand side of \normo\ gives on the other hand (since we assume
$c_1(V)=0$) the four-dimensional
class
\eqn\zormo{{nc_2(Z')\over 12} -c_2(V).}
So \normo\ reduces to
\eqn\opormo{c_2(V)=-{1\over 2}\pi_{2*}\left(c_1(\P_B)^2\right),}
and as we are working on the  four-manifold $Z'$, no information
will be lost if we integrate and write
\eqn\upormo{\int_{Z'}c_2(V)=-{1\over 2}\int_{C'\times_{B'}Z'}c_1(\P_B)^2.}

If we write $c_1(\P_B)=u+v$, with $u=-\sigma_1-c_1(\L)$ and $v=\Delta-\sigma_2$,
then $u^2=0$ because $u$ is pulled back from $B'$, and $uv=0$ because
of \foggy.  So we reduce to computing $v^2$.
We found above $\Delta^2=-\Delta c_1(\L)$, 
so $\int_{C'\times_{B'}Z'}\Delta^2=-\int_{C'\times_{B'}Z'}\Delta c_1(\L)$.
Integration over the second factor maps this to $-\int_{C'}c_1(\L)$,
which, because the map $C'\to B'$ is a $n$-fold cover, equals $-n\int_{B'}c_1(\L)$.
Likewise, we had $\sigma_2^2=-\sigma_2c_1(\L)$, so by similar steps 
$\int_{C'\times B'}\sigma_2^2
=-n\int_{B'}c_1({\cal L})$.  Finally, as $\Delta\sigma_2=\sigma_1\sigma_2$,
we need $\int_{C'\times_{B'}Z'}\sigma_1\sigma_2$; integration over the second
factor maps this to $\int_{C'}\sigma_1=\int_{B'}\pi_*\sigma_1
= \int_{B'}(\eta'-nc_1(\L))$.  So upon putting the pieces together, \upormo\
is equivalent to
\eqn\ipormo{\int_{Z'}c_2(V) =\int_{B'}\eta'.}

Since $B'$ was arbitrary, this is equivalent to
\eqn\ripormo{\pi_*(c_2(V))=\eta',}
or equivalently 
\eqn\kipormo{c_2(V)=\sigma\eta' + \omega,}
where the class $\omega$ is annihilated by $\pi_*$ and is thus a pullback
from $B$.
Equation \ripormo\  (which for $n=2$ was proved in \fm)
was asserted in section 6 as part of the comparison
of $F$ theory and the heterotic string.  Also, we have now confirmed
that $\eta'$ as we have defined it here should be identified with
$\eta$ as introduced in section 7.1 in the computations via parabolics.

\bigskip\noindent{\it Evaluation Of $\omega$}

It remains, then, to evaluate $\omega$.  Since $\omega$ is a pullback
from $B$, it is determined by its restriction to the canonical section
$\sigma $ of $\pi:Z\to B$, and we will make this restriction.
  Since the Poincar\'e line
bundle $\P_B$ is trivial when restricted to $\sigma$, it can be dropped.
As $C\times_B\sigma=C $, the projection $\pi_2:C\times _BZ\to Z$,
when restricted to $C\times_B\sigma$, reduces to $\pi:C\to B$.
Thus, we will use GRR for $\pi:C\to B$ to compute the restriction of
$c_2(V)$ to $\sigma=B$.

We have, therefore,
\eqn\oformo{\pi_*\left(e^{c_1(\N)}\Td(C)\right)=\ch(V)\Td(B).}
Everything needed to make this explicit has already been given:
the Chern classes of $C$ are in \keln\ and \feln, the expansion of the
Todd class is in \noby, and $c_1(\N)$ is in \meln.  Also we have
 $\ch(V)=n-c_2(V)$,
and to calculate $\pi_*$ one only needs to know that $\pi_*(1)=n$,
$\pi_*(\sigma)=\eta'-nc_1({\cal L})$, $\pi_*\gamma=0$, and
finally that  $\sigma^2=-\sigma c_1(\L)$.  The explicit
evaluation of \oformo\ then gives, after some algebra,
\eqn\expeval{c_2(V)|_\sigma=-{c_1(\L)^2(n^3-n)\over 24}-\eta' c_1(\L)-{n\eta'(\eta'
-nc_1(\L))\over 8}-{\pi_*(\gamma^2)\over 2}.}

Note that by the Hodge index theorem, $\gamma^2$ is always negative,
so that $c_2(V)|_\sigma$ is minimized for $\gamma=0$, which we recall
is the unique case in which the involution $\tau$ acts by
$\tau^*V=V^*$.  
For $\gamma=\lambda(n\sigma-\eta'+nc_1(\L))$ (the only general solution of
$\pi_*\gamma=0$, as we explained above), one computes from formulas
summarized above
\eqn\neval{\pi_*(\gamma^2)=-\lambda^2n\eta'(\eta'-nc_1(\L)).}

In combining \kipormo\ and \expeval\ into a general formula for $c_2(V)$,
one must recall that $\sigma^2=-\sigma c_1(\L)$, so that $\sigma|_\sigma
=-c_1(\L)|_\sigma$.  Hence a term $-\eta' c_1(\L)$ in \expeval\ is the
result, in a sense, of restricting the $\eta'\sigma$ term in $c_2(V)$ to
$\sigma$.  The final formula for $c_2(V)$ is thus
\eqn\finalform{c_2(V)=\eta'\sigma -{c_1(\L)^2(n^3-n)\over 24}-{n\eta'(\eta'
-nc_1(\L))\over 8}-{\pi_*(\gamma^2)\over 2}.}
If we set $\gamma=0$ to ensure $\tau^*V=V^*$, then this is in happy agreement
with \ctwo.

\bigskip\noindent{\it Change In $n$}

Before leaving the subject of spectral covers,
we will point out an important consequence of the formulas above
that determine $c_2(V)|_\sigma$ in terms of $\eta=\pi_*c_2(V)$. The
relation between them depends on $n$, the rank of the bundle.  This
means that if $V$ is a rank $n$ bundle of the sort we have been constructing,
then $V$ cannot degenerate by varying parameters to $\O\oplus V'$,
where $V'$ is a rank $n-1$ bundle built from the same type of construction.
(Such a degeneration would of course correspond in physical terms to
restoration 
of some gauge 
symmetry.) In fact, $V$ and $\O\oplus V'$ have different Chern classes.

Let us see concretely where the obstruction is.  Rank $n$ bundles
were constructed using a spectral cover $C$ defined by
\eqn\nocl{a_0+a_2x+a_3y+\dots +a_nx^{n/2}=0}
(or a slightly different formula if $n$ is odd).
For a rank $n-1$ bundle, the equation of the spectral cover is similar
except that the last term is absent.  So one might hope to reduce 
$V$ to $\O\oplus V'$ by setting $a_n=0$.  When this is done,
$C$ becomes a reducible variety with two branches, one a copy of $\sigma$
and one an $n-1$-sheeted spectral cover $C'$.  ($\sigma$ appears because
as $a_n\to 0$, one root of \nocl\ goes to $x=\infty$, that is, to $\sigma$.)
By itself
$\sigma$, regarded as the trivial spectral cover, would correspond to $\O$,
and $C'$ would likewise correspond to a rank $n-1$ bundle $V'$.  However,
$\sigma$ and $C'$ intersect, and because of this intersection one gets
not a direct sum $\O\oplus V'$ but a more complicated extension of bundles
-- with, of course, the same Chern classes as $V$!  ($V$ can in fact be obtained
as a so-called elementary modification of $\O\oplus V'$.)

The interpretation of gauge theory singularities via string theory 
five-branes gives a hint of what one might
be able to do.  Though $\tilde \O\oplus V'$ cannot be deformed to an irreducible
rank $n$ bundle $V$ if $\tilde \O$ is a trivial bundle and $V'$ is a rank
$n-1$ bundle, such a deformation of $\tilde \O\oplus V'$ may be possible
if $\tilde \O$ is a rank one torsion-free sheaf of $c_1=0$ and appropriate
non-zero $c_2$. In other words, $\tilde \O$ would be
the ideal sheaf $I_\Gamma$ for some codimension two subvariety $\Gamma$.  
The codimension two submanifold $\pi^{-1}(\Gamma)$ would be interpreted
physically as the world-volume of a set of  heterotic string fivebranes, 
generalizing the adventures that we had with fivebranes for somewhat analogous
reasons in section 7.2 above.

\def\co{{\cal O}}

\newsec{$\Z_2$ Index Theorem}

We conclude this paper by working out some general information about
bundles on elliptic Calabi-Yau threefolds that can be deduced from index theory.
(The index computation can be performed for other elliptic threefolds but gives a
particularly neat result in the Calabi-Yau case.) 
It is perhaps slightly anticlimactic to conclude with such generalities after
having made detailed constructions of bundles.  However, it is still
interesting to compare the detailed constructions to general theory.

Given any gauge group $G$ and complex manifold $Z$,
 one would hope to at least be able to
determine the dimensions of the moduli spaces of $G$ bundles on
$Z$.   This can be done via index theory if $Z$ is a surface, but
if $Z$ is a Calabi-Yau threefold
one runs into difficulty; in general there is no
index that determines the dimension of the moduli space of bundles.  
One can contemplate an index theorem for the alternating sum
\eqn\indo{\sum_{i=0}^3(-1)^i{\rm dim}\,H^i(Z,{\rm ad}(V)),}
but this sum is zero because of Serre duality, which asserts
that in this problem (because the bundle ${\rm ad}(V)$ is real and
the canonical bundle of $Z$ is trivial) $H^i$ is dual to $H^{3-i}$,
and so has the same dimension.

Physically, the absence of an index theorem reflects the fact
that it can be hard to predict whether an approximately massless
chiral superfield (coming from an apparent modulus of $V$) is really
exactly massless.  One can learn something about the superpotential
of the string theory by using the involution $\tau$ of $Z$ that
was discussed in section 7. Because $\tau$ acts as ``multiplication
by $-1$'' on the fibers of $Z\to B$, while acting trivially on $B$,
it multiplies the canonical bundle of $Z$ by $-1$.  According to the standard
arguments about Calabi-Yau compactification of string theory, $\tau$ is therefore
 observed in heterotic string  compactification on $Z$ as a $\Z_2$ $R$ symmetry;
so the superpotential is  odd under $\tau$. 

\def\ad{{\rm ad}}
Suppose then that we work at a $\tau$-invariant point in the moduli space,
by which we mean that the action of $\tau$ on $Z$ lifts to an action
on the adjoint bundle $\ad(V)$; we fix such a lifting.
If $A$ and $B$ are chiral superfields that are respectively
even and odd under $\tau$, then $A^2$ and $B^2$ terms in the 
superpotential are forbidden by the symmetry. A mass term would
necessarily be of the form $AB$, and gives mass to one even and one
odd superfield.  If then $n_e$ and $n_o$ are the numbers of even and
odd massless chiral superfields, the ``index'' $I=n_e-n_o$ is
invariant under $\Z_2$-invariant perturbations of the superpotential
and may be much easier to calculate than $n_e$ and $n_o$ separately.

In fact, this difference is governed by an index theorem.  
We  project onto
the $\tau$-invariant part of the index problem and consider
\eqn\jindo{I=-\sum_{i=0}^3(-1)^i\Tr_{ H^i(Z,\ad(V))}
{1+\tau\over 2} .}
If we write $H^i_e$ and $H^i_o$ for the   subspaces of $H^i$ that
are even or odd under $\tau$, then 
\eqn\ikko{\Tr_{H^i} {1+\tau\over 2}={\rm dim}\,H^i_e} 
so
\eqn\hilko{I=-\sum_{i=0}^3(-1)^i{\rm dim}H^i_e.}
The dimension of $H^1_e$ is what we called $n_e$, the number of
$\tau$-invariant chiral superfields.  
On the other hand, Serre duality says in this situation
that $H^i_e$ is dual to $H^{3-i}_o$.  (The duality exchanges the
even and odd subspaces because it involves multiplying by a 
holomorphic three-form, which is odd.)    
So the dimension of $H^2_e$ is $n_o$, the number of  odd chiral
superfields.  
Moreover, $H^0_e$ and $H^3_e$ (the latter is dual to $H^0_o$), are the
number of unbroken gauge generators that are even or odd under
$\tau$.  If for example the gauge symmetry is completely broken, these
numbers vanish and we have simply
\eqn\kilko{I=n_e-n_o.}
Even when the gauge symmetry is not completely broken, the correction
to \kilko\ would be known (if one knows the unbroken gauge group) so that
$n_e-n_o$ can always be effectively related to $I$.

As we noted above, in many examples (those which on the $F$-theory
side are described by four-folds $X$ with $H^3(X)=0$) one expects
$n_o=0$, so \kilko\ will reduce to $I=n_e$.  But in any case,
it is the difference $n_e-n_o$ that is calculable from index theory.

Since the ordinary index vanishes, the definition of $I$ is equivalent to
\eqn\kimmo{I=-{1\over 2}\sum_{i=0}^3(-1)^i\Tr_{H^i(Z,\ad(V))}\tau.}
Such a ``character-valued index'' can be effectively computed by a fixed 
point theorem
(originally obtained in
 \ref\abss{M. F. Atiyah and R. Bott, ``A Lefschetz Fixed Point Formula
For Elliptic Differential Operators,'' Bull. Am. Math. Soc. {\bf 72} (1966) 245,
M. F. Atiyah and G. B. Segal, ``The Index Of Elliptic Operators, II,''
Ann. Math. {\bf 87} (1968) 531.}; see \ref\goodman{M. Goodman,
``Proof Of Character Valued Index Theorems,'' Commun. Math. Phys.
{\bf 107} (1986) 391.} 
for a derivation based on path integrals).  
In a case such as this one, in which (as we will see) the components
of the fixed point set are all orientable and
of codimension two, the fixed point
theorem can be stated as follows.  Let $U_i$ be the components
of the fixed point set, and let $N_i$ be the normal bundle to $U_i$
in $Z$, regarded as a complex line bundle.  Let $F_i$ be the
restriction of $\ad(V)$ to $U_i$, and let $F_{i,e}$ and $F_{i,o}$ be
the subbundles of ${\rm ad}(V)$ on which $\tau $ acts by $1$ or $-1$; let
$s_{i,e}$ and $s_{i,o}$ be the rank of $F_{i,e}$ and $F_{i,o}$.  
And let  ${\rm ch}$ denote the Chern character
and ${\rm Td}$ the Todd class.  Then 
\eqn\holf{\sum_{i=0}^3(-1)^i\Tr_{ H^i(Z,{\rm ad}(V))}\tau
=\sum_i\int_{U_i} {{\rm ch}(F_{i,e})-{\rm ch}(F_{i,o})\over
           1+e^{c_1(N_i)}}{\rm Td}(U_i).}

This can be evaluated more explicitly as follows.  In evaluating
the Chern characters, we can stop at four-forms because the $U_i$
have dimension four.  The bundles $F_{i,e}$ and  $F_{i,o}$, being
real, have vanishing first Chern class.  So we get
\eqn\polg{\eqalign{{\rm ch}(F_{i,e}) &=s_{i,e} - c_2(F_{i,e}) \cr
{\rm ch}(F_{i,o}) &=s_{i,o} - c_2(F_{i,o}) .\cr}}
In our actual application, the numbers $s_{i,e}$ and $s_{i,o}$ will
be independent of $i$ - let us call them $s_e$ and $s_o$.  That being
so, the part of the contribution to \holf\ that
is proportional to $s_e$ or $s_o$ can be equated, using the fixed
point theorem, with the value of $s_e-s_0$ times
$ \sum_{i=0}^3(-1)^i\Tr_{H^i(Z,{\cal O})}\tau$.
(In other words, that contribution would be unchanged if $F_{i,e}$ and
$F_{i,o}$ were replaced by trivial bundles of the same rank.)
  That last expression
is two (as $H^0(Z,\co)$ is one-dimensional and even under $\tau$,
$H^3(Z,\co) $ is one-dimensional and odd, and $H^1(Z,\co)=H^2(Z,\co)=0$).
So this part of \holf\ is simply $ 2(s_e-s_o)$.  In evaluating the
rest of \holf, we can replace $c_1(N_i)$ in the denominator by
zero, since the numerator already contains four-forms, and we can
likewise replace the Todd class by 1.  So \holf\
becomes
\eqn\uholf{\sum_{i=0}^3(-1)^i\Tr_{ H^i(Z,\ad(V))}\tau
=2(s_e-s_o) -{1\over 2}\sum_i\int_{U_i} 
\left( c_2(F_{i,e})-c_2(F_{i,o})\right) .}

\bigskip\noindent
{\it Action At Fixed Points}

The index formula depends, of course, on how $\tau$ acts on the fibers of ${\rm ad}(V)$
over the fixed point set.  The  case of most direct interest for comparing to the constructions of
bundles that we have given in this paper is the case in which $\tau$ acts as
the involution of the Lie algebra of $G$ that is induced
from the involution $-1$ of the root lattice of $G$.  We will call
this involution of the Lie algebra $\rho$.
In fact, on a single elliptic curve $E$,
every flat $G$ bundle  $V$ has the property that the involution
$\tau$ of $E$ lifts to ${\rm ad}(V) $ in such a way as to act on fibers over fixed points
as $\rho$.  (This is therefore also true for every semistable $G$ bundle up to
$S$-equivalence.) 
So, given a $G$ bundle $V$ over $Z$ that is semistable on each fiber, it is natural
to look for a lifting of $\tau$ to ${\rm ad}(V)$ so as to act by $\rho$ on fibers over
fixed points.

More physically, a lifting with this property is natural because
in duality between $F$ theory and the heterotic
string, what in $F$ theory is the involution $\tau$ corresponds on the heterotic
string side to multiplication of the full Narain lattice by $-1$, and (in a limit
where classical geometry applies) this certainly induces the involution $\rho$
of the root lattice.

With this choice of lifting of $\tau$, $\uholf$ can be made more explicit.
The  difference $s_e-s_0$ is just the trace of $\rho$ in the adjoint representation
of $G$.  In taking this trace, the non-zero weights (which are exchanged by $\rho$ in pairs)
do not contribute, so the complete contribution comes from the maximal torus,
and is equal to $-r$ ($r$ being the rank of $G$).

One can also make a reduction  of the second Chern classes that appear in \uholf.
We want to express the difference
$c_2(F_{i,e})-c_2(F_{i,o})$ in terms of the fundamental characteristic class $\lambda(V)$ of
$V$.  We claim that in fact
\eqn\poolp{c_2(F_{i,e})-c_2(F_{i,o})=-4\lambda(V)|_{U_i}.}
For $SU(n)$, this means that 
\eqn\hoolp{c_2(F_{i,e})-c_2(F_{i,o})=-4c_2(V)|_{U_i}.}
This is proved using the explicit form of $\rho$ to compute traces.

So we can rewrite \kimmo\ and \uholf\ to say
that the index $I=n_e-n_o$ of bundle moduli is
\eqn\indix{I=r-\sum_i\int_{U_i}\lambda(V).}

\subsec{Comparison With Construction Of Bundles}

We want to compare this index formula to the actual number of moduli found
in our construction of bundles (for the $\tau$-invariant components of the moduli space).
For this, we must
describe explicitly the $U_i$ and determine
the quantities $\lambda(V)|_{U_i}$.  Recall that the manifold
$Z$ is described by a Weierstrass equation
\eqn\ikko{zy^2=4x^3-g_2xz^2-g_3z^3.}
Moreover, $\tau$ is the transformation $y\to -y$, leaving other
coordinates fixed.  A fixed point is a point at which the homogeneous
coordinates $x,y,$ and $z$ are
left fixed up to overall scaling.
There are thus two components of the fixed point set.  One component,
$U_1$, is given by $x=z=0$, $y\not=0$, and is the section
$\sigma$ of $Z\to B$ that figured extensively
above.   The other component, $U_2$, is given
by $y=0$.  $U_2$ is thus a triple cover of $B$, given by the
equation
\eqn\jikko{0=4x^3-g_2xz^2-g_3z^3}
in a certain ${\bf P}^1$ bundle $W'$ over $B$.

We will compare explicitly the index formula \indix\ to our actual construction of
bundles only for $G=SU(n)$.  The index formula predicts that
\eqn\urpu{I=n-1-\int_{U_1}c_2(V)-\int_{U_2}c_2(V).}
 The cohomology class of $U_1$
is simply the class of the section $\sigma$, while (as $y$ is a section of
$\O(1)\otimes {\cal L}^3=\O(\sigma)^3\otimes \L^3$) 
the cohomology class of $U_2$ is $3\sigma+3c_1(\L)$.
So \urpu\ can be written
\eqn\turpu{I=n-1-4\int_\sigma c_2(V)|_\sigma-3\int_Zc_1(\L)c_2(V).}
 So, using \finalform\ (and of course setting $\gamma=0$ to ensure
$\tau$-invariance), we get
\eqn\jutty{I=n-1+\int_B\left({(n^3-n)c_1({\cal L})^2\over 6}+{n\eta(\eta-nc_1({\cal L}))
\over 2}
+\eta c_1({\cal L})\right).} 

Now let us count the parameters in our construction of the bundle.
For this, we simply count the parameters in the equation \ocl\ that
defines the spectral cover, and subtract 1 for overall scaling of that
equation.  As $a_r$ is a section of $\M\otimes {\cal L}^{-r}$, the number of
parameters, assuming a suitable amount of ampleness so that an index
theorem can be used to compute the dimension of $H^0(B,\M\otimes {\cal L}^{-r})$, is
\eqn\otherform{\eqalign{\tilde I
=&-1+\int_Be^\eta\left(1+e^{-2c_1({\cal L})}+e^{-3c_1({\cal L})}+\dots
+e^{- nc_1({\cal L})}\right) \Td(B).   \cr}}
With a small amount of
algebra (and using the Calabi-Yau     condition $c_1(B)=c_1({\cal L})$ since we have
assumed this in doing the index theory), one finds $\tilde I=I$, as expected.
\bigskip

We would like to thank E. Sharpe for comments on the manuscript.
\listrefs
\end